\documentclass[12pt,preprint]{aastex}
\usepackage{color}
\makeatletter

\newcounter{reaction}
\newdimen\reactionindent
\newdimen\reactionarrspc
\def\@reactionnum{\hbox{\reset@font\rm(\thereaction)}}
\newcommand\chem[1]{$\rm #1$}
{%
  \@beginparpenalty\predisplaypenalty%
  \@endparpenalty\postdisplaypenalty%
  \refstepcounter{reaction}%
  \trivlist \item[]\leavevmode%
    \hb@xt@\linewidth\bgroup $\m@th% $
    \displaystyle%
    \hskip\reactionindent%
    \rm
}{%
    $\hfil % $%
    \displaywidth\linewidth\hbox{\@reactionnum}%
  \egroup%
  \endtrivlist%
}

\newcommand\reactionlabel[1]{%
  $\m@th$\hfil%
  \refstepcounter{reaction}%
  \hbox{\@reactionnnum}%
  \label{#1}
}
\newdimen\autotop  \newdimen\autobottom  \newdimen\autosize 
\newcommand\reactionarrow[2]{\ensuremath{%
  \rule{\reactionarrspc}{0pt}%
    \setbox0=\hbox{\textrm\scriptsize #1}\autotop=\wd0\setbox0=\hbox{\textrm\scriptsize #2}\autobottom=\wd0%
    \ifdim\autobottom>\autotop\autosize=\autobottom\else\autosize=\autotop\fi%
    \ifdim\autosize<1em\autosize=1em\fi%
    \advance\autosize by 1em%
    \mathop{\raisebox{0.495ex}{\rule{\autosize}{0.4pt}}}\limits%
    ^{\textrm{\scriptsize #1} \rule{1em}{0pt}}%
    _{\textrm{\scriptsize #2} \rule{1em}{0pt}}%
    \rule{-8.5pt}{0pt}\raisebox{-0.2pt}{$>$}%
  \rule{\reactionarrspc}{0pt}%
}}
\newcommand\reactionrevarrow[2]{\ensuremath{%
  \rule{\reactionarrspc}{0pt}%
    \setbox0=\hbox{\textrm\scriptsize #1}\autotop=\wd0\setbox0=\hbox{\textrm\scriptsize #2}\autobottom=\wd0%
    \ifdim\autobottom>\autotop\autosize=\autobottom\else\autosize=\autotop\fi%
    \ifdim\autosize<2em\autosize=2em\fi%
    \advance\autosize by 3em%
    \raisebox{-1.9pt}{$\smallsetminus$}\rule{-9.3pt}{0pt}%
    \mathop{
     \raisebox{0.490ex}{\rule\autosize{0.4pt}\rule{-\autosize}{0pt}%
     \raisebox{0.500ex}{\rule\autosize{0.4pt}}%
    }}\limits%
    ^{\textrm\scriptsize #1 \rule{1em}{0pt}}%
    _{\textrm\scriptsize #2 \rule{1em}{0pt}}%
    \rule{-9.0pt}{0pt}\raisebox{5pt}{$\smallsetminus$}%
  \rule{\reactionarrspc}{0pt}%
}}
\makeatother
\def\spose#1{\hbox to 0pt{#1\hss}}
\def\simlt{\mathrel{\spose{\lower 3pt\hbox{$\mathchar"218$}}
     \raise 2.0pt\hbox{$\mathchar"13C$}}}
\def\simgt{\mathrel{\spose{\lower 3pt\hbox{$\mathchar"218$}}
     \raise 2.0pt\hbox{$\mathchar"13E$}}}

\begin{document}
\newcommand{\hh}{H$_2$}
\newcommand{\hhh}{H$_3^+$}%
\newcommand{\hhd}{H$_2$D$^+$}
\newcommand{\hdd}{HD$_2^+$}
\newcommand{\ddd}{D$_3^+$}
\newcommand{\dd}{D$_2$}
\newcommand{\lrarrow}{$\leftrightarrow$}
\newcommand{\x}{$\times$}
\bibliographystyle{apj}
\title{Deuterium chemistry in protoplanetary disks. II. The inner 30 AU}
\author{K.~Willacy }
\affil{Jet Propulsion Laboratory, California Institute of Technology, MS 169--506, Pasadena, CA 91109}
\email{Karen.Willacy@jpl.nasa.gov}
\and
\author{Paul~M.~Woods}
\affil{Jodrell Bank Centre for Astrophysics,
The Alan Turing Building,
School of Physics and Astronomy,
The University of Manchester,
Oxford Road,
Manchester,
M13 9PL,
UK}
\email{Paul.Woods@manchester.ac.uk}

\begin{abstract}
We present the results of models of the chemistry, including
deuterium, in the inner regions of protostellar disks.  We find good
agreement with recent gas phase observations of several
(non--deuterated) species.  We also compare our results with
observations of comets and find that in the absence of other
processing e.g.\ in the accretion shock at the surface of the disk, or
by mixing in the disk, the calculated D/H ratios in ices are higher
than measured and reflect the D/H ratio set in the molecular cloud
phase.  Our models give quite different abundances and molecular
distributions to other inner disk models because of the differences in
physical conditions in the model disk.  This emphasizes how
changes in the assumptions about the density and temperature
distribution can radically affect the results of chemical models.
\end{abstract}


\section{Introduction}

Observations and models of the deuterium chemistry in protostellar
disks can make important contributions to our understanding of the chemistry
of the early solar system.  They can be directly related to the
chemistry of the formation of cometary ices, which provides a link
between the current and primitive solar nebula.  Observations
of deuterated molecules can be used to trace the physical and 
thermal history of a protostellar disk
and, combined with models, can determine the relative contribution
of the different phases of star formation (e.g. molecular cloud,
collapse, protostellar disk) to the molecular deuteration ratios
observed in protostellar disks and in comets.  Although the
elemental abundance of deuterium in the Galaxy is only a few \x 10$^{-5}$
\citep{linsky06} the relative abundance of deuterated to
non--deuterated molecules can be much higher and depends sensitively
on the temperature at which the molecules formed, and, at high densities,
 on the degree of molecular depletion.  Models
of the deuterium chemistry have demonstrated the importance of
gas phase reactions of deuterated isotopologues of \chem{H_3^+},
\chem{CH_3^+} and \chem{C_2H_2^+} in transferring deuterium atoms
to molecules, as well as the critical role played by grain surface chemistry,
where reactions of deuterium atoms are able to efficiently form
molecules such as deuterated formaldehyde and methanol.

A few deuterated molecules have now been observed in protostellar disks.
\cite{cec04} observed \chem{H_2D^+} in TW Hya and DM Tau and argued that
the emission originates in the midplane and that
therefore \chem{H_2D^+} provides a means of tracing the ionization
level in this high density, low temperature region where
few other molecules exist in the gas phase. A later paper \citep{cec05}
reported a detection of HDO in DM Tau where the emission is from
a region with a temperature of $\sim$ 25 K, far below that at
which this molecule can be thermally desorbed.  This indicates that
there is an efficient non--thermal desorption mechanism
acting e.g.\ photodesorption \citep{dom05,wl00}.  (Note
that the observations of both \chem{H_2D^+} and HDO have
been disputed by \cite{guilloteau06}.)  DCO$^+$ has been observed
in TW Hya \citep{qi08,vd03} and in DM Tau \citep{guilloteau06}.
\cite{vd03} used JCMT and found a beam--averaged ratio of
DCO$^+$/HCO$^+$ = 0.035.  \cite{qi08} using the SMA found a more
complex distribution of DCO$^+$ which does not follow that
of HCO$^+$.  In this case DCO$^+$/HCO$^+$ increases with radius
between 30 and 70 AU from 0.01 to 0.1, and DCO$^+$ disappears
rapidly at $R$ $>$ 90 AU.  In contrast HCO$^+$ and CO
are present out to 200 AU.  In DM Tau \cite{guilloteau06} find
DCO$^+$/HCO$^+$ $\sim$ 4 \x 10$^{-3}$ in the outer disk.  \cite{qi08} also
observe DCN in TW Hya and determine DCN/HCN to be 0.017.

These observations are all for the outer regions of
the disk ($R$ $\simgt$ 30 -- 50 AU).  Here our focus is
on $R$ $\leq$ 30 AU.  The inner regions of disks are of great 
interest because they cover
the region where planet formation took place in the solar nebula.  
They are now becoming
accessible to observational investigation e.g.\ with the Spitzer
Space Telescope \citep{cn08, salyk08, lahuis06}
or the Keck Interferometer \citep{gibb07,gibb04}.  In the future,
ALMA will further open up this region for study.  

solar system data provides an additional constraint on the models.
Observations of deuerated molecules in comets are limited to two
molecules (HDO and DCN) in four comets -- Halley, Hale-Bopp, Hyakutake
and C/2002 T7 LINEAR.  The ratios are similar in all of the observed
comets with HDO/H$_2$O $\sim$ 5 -- 7 \x 10$^{-4}$ \citep{bm98, eberhardt95,
balsiger95,hut08} and DCN/HCN $\sim$ 2 \x 10$^{-3}$ \citep{meier98a}.  These
values are higher than the solar elemental D/H ratio.
The similarities between
molecular D/H ratios observed in the interstellar medium
and those seen in comets have led to the suggestion
that cometary ices have their origin in the interstellar medium.
However, there is sufficient difference between the
two data sets for this to still be an open question e.g.\
\cite{bergin07} and to suggest that processing during 
star formation, or in the early solar nebula could affect the
ratios.  Once comets formed they
are thought to 
have undergone relatively little subsequent processing 
and therefore
their abundances reflect the composition of material at the time
of their formation.

Enhanced D/H ratios are seen elsewhere in the solar system.
For example, interplanetary dust particles are enriched in D with respect to the
terrestrial value \citep[e.g.][]{messenger03,messenger00}.  They show an 
extremely wide range of D/H ratios \citep{robert00} but the carriers
have not yet been identified.  Enhancements are also seen in 
primitive meteorites (LL3 and carbonaceous chondrites, as well
as meteoritic water), but the anomalies are generally smaller. 
The observations do not yet allow for the origin of the 
deuterated material to be definitively identified, and both
interstellar \citep[e.g.][]{ph05} and protostellar \citep[e.g.][]{remusat}
origins have been claimed.

The deuterium content of the giant planets Jupiter and Saturn 
is considered to be a relic of the early protosolar nebula, since deuterium 
is neither formed nor destroyed in these planets, and their gravity
is sufficient to prevent hydrogen and deuterium from escaping \citep{owen86}.
Accordingly measuring their molecular D/H ratios can provide information about
the deuteration levels  in the protosolar nebula at their time of formation.
Measurements in Jupiter suggest that the D/H ratio is enhanced over
the interstellar value elemental ratio by a factor of a few.  \cite{mahaffy98} used
the Galileo Probe Mass Spectrometer to derive (D/H)$_{H_2}$ = 2.6 $\pm$ 0.7 $\times$
10$^{-5}$, in close agreement with the ISO-LWS value of 2.2 $\times$ 10$^{-5}$
\citep{lellouch99}.
Both these values depend on the models used to analyze the data.
A direct measurement of D/H in the limb of Jupiter using Lyman--$\alpha$
emission was made by \cite{jaffel98} who found (D/H)$_{H_2}$ = 5.9 $\times$ 10$^{-5}$.
Deuterated methane has also been observed with a ratio (D/H)$_{CH_4}$ = 1.8 \x 10$^{-5}$ 
-- 2.9 \x 10$^{-5}$ \citep{feuchtgruber99,encrenaz99}.
Saturn has similar D/H ratios to Jupiter (1.7 $\times$ 10$^{-5}$; 
\cite{lellouch99}).  In Neptune and Uranus 
the ratios are higher -- 6.5 $\times$ 10$^{-5}$ \citep{bezard97} or 
5.6 \x 10$^{-5}$ \citep{orton92} for Neptune
and 5.5 $\times$ 10$^{-5}$ \citep{feuchtgruber99} in Uranus, possibly as
a consequence of the inclusion of highly deuterated icy planetesimals into the
planets as they formed.

In our previous paper \citep{willacy07} we considered the
deuterium chemistry in the outer regions ($R$ $>$ 50 AU) of a protostellar
disk.  We showed that although the gas phase molecular D/H ratios
can be considerably altered by chemical processing in the disk,
the grain mantle ratios reflect those set in the parent molecular cloud,
supporting the idea that comets may contain interstellar material
that survived the formation of the Sun and planets relatively
unchanged.  Here we extend our previous work to cover the inner
disk from 0.5 -- 30 AU. This radius range covers the comet and planet
forming region, and is also appropriate for comparison to infrared
observations of the inner regions of protostellar disks.

We present the results for a typical T Tauri star disk and for
a more massive disk with a higher surface density similar
to the minimum mass solar nebula (MMSN).  Our models include multiply 
deuterated molecules since in high density regions these can become
important \citep{cd05,roberts02,roberts03} although their effects in the warm
inner disk are expected to be less than in the colder outer disk.
We compare our results with the available observations of the
inner disk (non--deuterated molecules) and with observations
of deuterated molecules in comets.

\section{\label{sec:physical_model}The model}
To fully model a disk self-consistently requires many
chemical and physical processes to be taken into account.
To simplify this process and to reduce the required computing
time we separate the calculation of the physical structure
of the disk from the chemical network.  We obtain the
disk structure (density and grain temperature) from a pre-existing
hydrodynamical model \citep{dalessio99,dalessio01}.  These models
use the $\alpha$ prescription to determine the density and temperature
distribution throughout the disk.  

We consider two of d'Alessio's models with different mass
accretion rates.  The first (Model 1) has $\dot M$ = 10$^{-8}$
$M_\odot$ yr$^{-1}$, a central star of mass M$_*$ = 0.7 M$_\odot$,
radius R$_*$ = 2.5 R$_\odot$ luminosity L$_*$ = 0.9 $L_\odot$ and
temperature T$_*$ = 4000K. The viscosity parameter $\alpha$ is 0.01.
The dust distribution is described by a power law $n(a)$ = $n_0
a^{-p}$ where $a$ is the grain radius and $p$ = 3.5.  The maximum
grain size is 0.25 \micron~ and the dust is assumed to be well mixed
with the gas.  The surface density at 5 AU is
38.8 g cm$^{-2}$.  This model is the same as 
used in \cite{willacy07} (hereafter W07) and so represents a continuation of
the model to smaller radii.

The mass of the disk in this model is somewhat lower than that
expected in the early solar system.  We therefore also consider
a more massive disk (Model 2) where the surface density is closer to that
of the
minimum mass solar nebula \citep{hayashi81}.  In this model
$\dot M$ = 10$^{-8}$ M$_\odot$ yr$^{-1}$ and \mbox{$\alpha$ = 0.025}.
All other parameters remain the same
as in Model 1.  The surface density at 5 AU for Model 2 is 150 gcm$^{-2}$.

Although the hydrodynamical models assume that the gas and grain
temperature are equal, this is not the case in the surface layers of
the disk e.g.\ \cite{kd04,gh04}, where the gas temperature can be much
higher than the grain temperature.  Since the gas temperature can
critically affect the chemical (and especially the isotopic)
abundances we elect to calculate the gas temperature separately.  This
calculation is based on the work of \citet{kd04} and is described in
more detail in \cite{ww09}.  We summarize here the processes that are
included.  Heating is provided by the photoelectric effect,
photodissociation of H$_2$, the formation of H$_2$ on the surfaces of
dust grains, the ionization of carbon atoms, 
cosmic rays and stellar X-rays.  Gas-grain
collisions can also act as heating processes in regions where the grains are
hotter than the gas, but this is not an important process in the inner
disk, where this process is more likely to cool the gas.  Cooling occurs
via the emission lines of atomic oxygen and carbon, C$^+$, CO and CH,
and by Lyman-$\alpha$ cooling.  The rates for these processes are given in
\cite{ww09}.

\begin{figure}[p]
\caption{\label{fig:physical_0.01}The physical parameters used in (a) Model 1 ($\alpha$ = 0.01) and (b) Model 2 ($\alpha$ = 0.025).
{\it Left:} The gas density; {\it Center:} the grain temperature; {\it Right:} the
gas temperature.  Gas density and grain temperature are supplied by the 
models of d'Alessio, and the gas temperature is calculated separately by balancing
the heating and cooling processes.}

\includegraphics{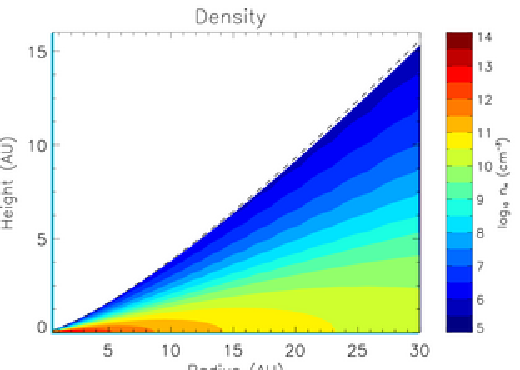}
\includegraphics{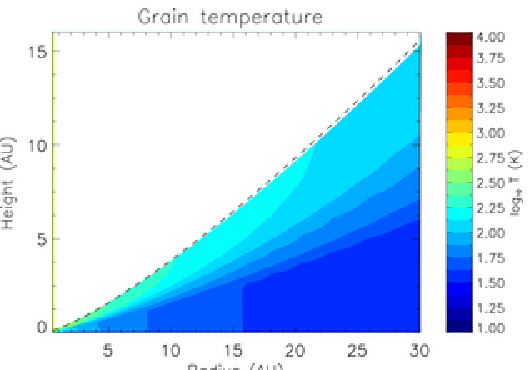}
\includegraphics{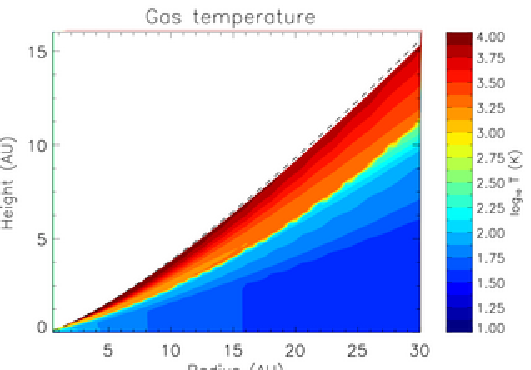}
\end{figure}

For H$_2$ formation heating we use the H$_2$ formation rate 
from \cite{ct02}, which assumes that H atoms can
both chemisorb and physisorb to bare grains.  Chemisorption
only occurs on bare grains and this rate is therefore only
valid at the top of our disk.  Below this where the grains
are covered in ices, we assume that only physisorption occurs
with the binding energy given in \citet{ct02}.  
This results in cooler temperatures than were found by
\cite{kd04} in regions 
just below the disk surface where ices have begun to form. 
Figure~\ref{fig:physical_0.01} shows the 
density and temperature distributions for the two models used.

\addtocounter{figure}{-1}

\begin{figure}[p]
\caption{\label{fig:physical_0.025}(b) Model 2}

\includegraphics{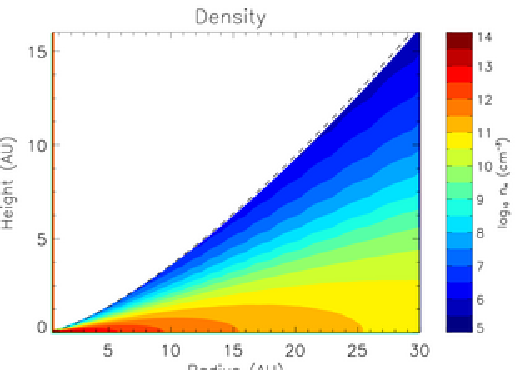}
\includegraphics{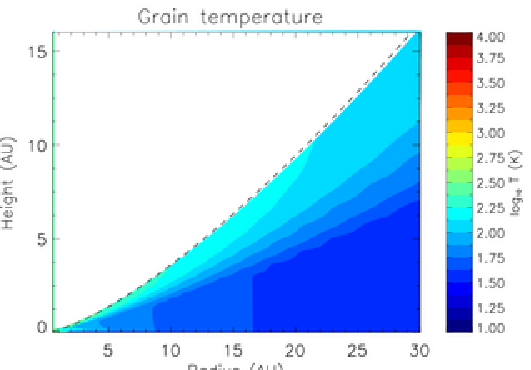}
\includegraphics{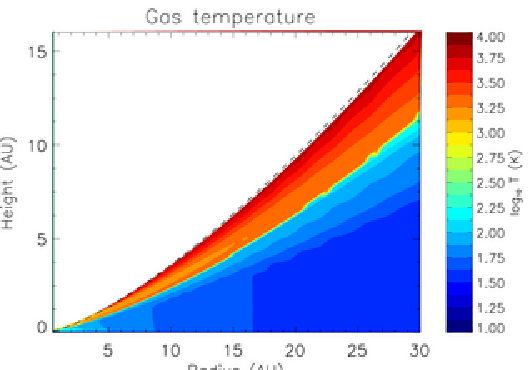}
\end{figure}

The final parameter
required is the UV field.  
In order to estimate the UV field in the disk
we adopt the model of \cite{ry00} which uses a ray tracing method
and the flux limited diffusion approximation to determine the
field at each point in the disk
This allows us to model the effects of UV irradiation
by the interstellar UV field, and by the central protostar
and to include scattering
of the UV photons by dust grains.  The grain absorption and
scattering coefficients are based on \cite{dl84} and \cite{preibisch93}
who assumed a mixture of silicates, amorphous carbon and dirty ice-coated
silicate grains.  The disk is divided into square cells ($\Delta$z = $\Delta$R
= 0.02 AU) and the radiative transfer equation is solved separately for
both sources of radiation and for the direct and diffuse (scattered) components
of each field \citep[see][for details]{ry00}.  The total 
UV flux at a given point is the sum of the contributions from
the stellar and interstellar fields.
In general, this results in a total UV field that is higher than
that estimated from our previous slab method: see for
example W07.  A similar effect was found by
\cite{zadelhoff03} for a more complex radiative transfer model 
in the outer disk.

Using the density and temperature distributions provided by the
d'Alessio disk model we
calculate the UV field for both the interstellar and stellar fields.
For the latter we assume that
the UV field emitted by the T Tauri star is 500 G$_0$ at 100 AU
where G$_0$ is the standard interstellar radiation field (ISRF).  This
field strength is based on observations
of T Tauri stars by \cite{bergin03}, but we do
not take into account the possibility that the
T Tauri spectrum has a different shape to the ISRF
and that it may be dominated by strong emission features such
as Ly$\alpha$.  This can significantly impact the
chemistry of the surface layers of the disk since Ly$\alpha$ can
dissociate some molecules such as OH and CH$_4$, but not others.  
\cite{bergin03} found that this effect can account for high CN/HCN
ratios observed in some disks \citep{dutrey97, kastner97}.

Following previous authors \citep[e.g.][]{ww09,markwick02, willacy98} 
we assume
that material advects inwards towards the star.
We do not have a completely integrated dynamical and
chemical model and instead follow parcels of gas, starting
at the outer edge of the disk (in this case 35 AU) which gradually
move inwards.  The parcels move inwards in discrete radial steps of 
0.5 AU and travel along lines of constant gas pressure scale height.  The time
that a gas parcel remains at a given grid position can
be determined from the radial velocity:
\begin{equation}
v_r = \frac{\dot M}{2 \pi \Sigma R}
\end{equation}
where $\Sigma$ is the surface density of the disk at radius, $R$.  The
timescale for accretion, $\Delta t$ $\approx$ $\Delta R$/$v_r$ and
therefore the time that a parcel of gas spends at each radius is given
by
\begin{equation}
\Delta t \sim \frac{2 \pi \Sigma R \Delta R}{\dot M} \label{eq:dt}
\end{equation}
$\Sigma$ scales roughly with $1/R$ for $R$ $>$ 10 AU \citep{dalessio01}
so equation~\ref{eq:dt} has only a small dependence on R.  Starting
at 35 AU, a parcel of gas will migrate into the star in approximately
0.41 million years.  In our grid, the parcel spends roughly 6000 years
at each radius before moving inwards to the next gridpoint.  We have
70 radial and 56 vertical grid positions.

\subsection{The chemical network}

Our chemical network is the same as used by W07.  We take the basic
gas phase reaction set from the UMIST database RATE95 \citep{rate95}
and add deuterated reactions using the techniques of \citet{millar89}
and \citet{rm96}.  The reaction rates for the multiply deuterated
isotopologues of H$_3^+$ and CH$_3^+$ are taken from \cite{roberts04}.
For details of the grain surface chemistry see W07.
The reaction network links 227 gas phase species (of which 115 are
deuterated) and 91 grain species (44 of which are deuterated) by 9489
reactions.  

We also include freezeout, thermal desorption and cosmic ray heating
of grains.  The rate at which a species, $X$, will freezeout onto dust
grains is given by
\begin{equation}
k_{freeze} = S_X C_X < \pi a^2 n_g > v_X  \mbox{~~~s$^{-1}$}
\end{equation}
where $S_X$ is the sticking coefficient (= 0.3 for all species), 
$a$ is the grain radius, $n_g$ is the number density of grains
and $v_X$ is the gas phase velocity of $X$. 
$C_X$ is a factor to take into account the
increase in freezeout rate for positive ions encountering
a negatively charged grain \citep{un80}. $C_X$ is 1 for
neutral species, and  $1 + 16.71 \times 10^{-4}/(a T_{gr})$ for
positive ions, where $T_{gr}$ is the grain temperature.  
Ions are assumed to recombine on the grain surface in the same
way as they do when reacting with electrons in the gas.
All species are assumed to freezeout with the exception of He and He$^+$
which have a very low binding energy and are easily thermally desorbed
even at low temperatures.  Any He$^+$ that hits a grain is immediately
returned to the gas as neutral He.

The thermal desorption rate is given by
\begin{equation}
k_{therm} = \nu_0 e^{-E_D/T_{gr}} \mbox{~~~s$^{-1}$}
\end{equation}
$\nu_0$ is the frequency of oscillation between the absorbate and
the grain surface and is given by
\begin{equation}
\nu_0 = \sqrt (2 n_s E_D/ \pi^2 m)
\end{equation}
where $n_s$ is the surface density of sites ($\sim$ 1.5 \x 10$^{15}$
cm$^{-2}$), $m$ is the mass of the accreting species and $E_D$ is its
binding energy.  The binding energies used are summarized in
Table~\ref{tab:be}.  For deuterated species, the binding energies are
assumed to be 21 K more than their undeuterated equivalents (21 K is
the difference in zero point energy between the hydrogen and deuterium
atoms) \citep{caselli02}. For example, $E_D$(H$_2$O) = 5770 K, 
$E_D$(HDO) = 5791 K and $E_D$(D$_2$O) = 5812 K. 
Binding energies for species not listed in Table~\ref{tab:be} are
taken from \cite{hh93} and \cite{hhl92}.

Cosmic ray heating can remove weakly bound molecules
from grains in regions where the temperatures are low, and where
the surface density of the disk is low enough that cosmic rays can
penetrate. The desorption rates due to cosmic ray heating are
taken from \cite{hh93} using updated binding energies for some species
e.g.\ CO \citep{oberg05}.

\begin{deluxetable}{lrcclrc}
\tablecolumns{7}
\tablewidth{0pt}
\tablecaption{\label{tab:be}
The binding energies ($E_D$) used
to determine the thermal desorption rates of the abundant
mantle species. }
\tablehead{
\colhead{Species} & \colhead{$E_D$ (K)} & \colhead{Reference} &
\colhead{} & \colhead{Species} & \colhead{$E_D$ (K)} & \colhead{Reference}
}
\startdata
H        & 600  & 1 & & D        & 621  & 2 \\
H$_2$    & 315  & 8 & & C        & 800  & 3 \\
CH       & 645  & 9 & & CH$_2$   & 956  & 9 \\
CH$_3$   & 1158 & 9 & & CO       & 855  & 4 \\
CO$_2$   & 2860 & 5 & & H$_2$CO  & 1760 & 3 \\
CH$_3$OH & 4240 & 7 & & O        & 800  & 3 \\
O$_2$    & 1210 & 3 & & OH       & 1259 & 9 \\
H$_2$O   & 5770 & 6 & & N        & 800  & 3 \\
N$_2$    & 790  & 4 & & NH       & 604  & 9 \\
NH$_2$   & 856  & 9 & & NH$_3$   & 3080 & 5 \\
\enddata
\tablerefs{(1) \citet{ct02,ct04}, (2) \citet{caselli02}, (3) \citet{ta87},
(4) \citet{oberg05}, (5) \citet{sa90}, (6) \citet{fraser01}, (7) \citet{sa93},
(8) \citet{rh00}, (9) \citet{ar77}}
\end{deluxetable}

In our outer disk models (W07)
we included photodesorption
since at $R >$ 50 AU the  grain temperatures are low enough that
thermal desorption is not efficient for most species.  
Photodesorption was found to remove molecules in the surface
layers of the disk even when the grain temperature was low.
In the inner disk model the grain temperatures in the
surface layers are warm enough that even the molecules
with the highest binding energies can be removed efficiently by thermal
desorption and therefore photodesorption will have little effect on the
molecular abundances.  Hence we have chosen to exclude it from 
our current models.

\subsubsection{\label{sec:uv}Ionization processes}
Ionization in the disk can arise from several processes
including irradiation by UV photons (both interstellar
and stellar), protostellar X-rays and cosmic rays.  There is also
a contribution from the decay of radioactive isotopes. 

The UV field at each disk position is calculated as described in Section~\ref{sec:physical_model}.
The stellar field is taken to be 500 $G_0$ at 100 AU 
\citep{bergin03} where $G_0$ is the
standard interstellar radiation field.
We assume that the stellar field does not dissociate CO and H$_2$
\citep{aikawa02},
although these molecules are dissociated by the interstellar field.
We calculate the self--shielding of CO and H$_2$ using the
slab model of \cite{lee96}.  This model was developed for
molecular clouds and provides data for the shielding
due to H$_2$, CO and dust as a function of column density.  The
linewidths assumed are 3 kms$^{-1}$ which is much larger than
observed in disks, where the velocity dispersion in the outer disk 
at least, is almost thermal \citep{gd98}.
To take account of this we have followed \cite{ah99} in
scaling the column densities in Table 10 of \citeauthor{lee96}\ by
$c_s$/3 kms$^{-1}$ where $c_s$ is the local sound speed in the disk.
This scaling factor is only required for H$_2$ since the CO dissociation
lines are broader due to predissociation.

HD does not self-shield, but some of its lines do overlap with those
of H$_2$.  This means that the H$_2$ can provide some shielding.  
\cite{barsuhn77} estimated this would reduce the HD photodissociation
rate by 1/3, assuming that the overlapped HD lines are totally shielded
by H$_2$.  Here we modify the HD dissociation rate accordingly.

Cosmic rays can also cause ionization, both directly and indirectly
(by interaction with H$_2$ producing UV photons).  The
rates for both  processes are taken from the UMIST ratefile,
with an assumed cosmic ray ionization rate $\zeta_0$ =  1.3 $\times$
10$^{-17}$ s$^{-1}$.  Cosmic rays can produce ionization in the
disk only if the surface density ($\Sigma$) is low enough for them
to penetrate i.e.\ $\Sigma$ is less than 150 g cm$^{-2}$ \citep{un81}.
We assume that cosmic rays only penetrate the disk
vertically, and that they can penetrate from both above and
below the disk.  Their ionization rate is given by
\begin{equation}
\zeta_{CR} = \frac{1}{2} \zeta_0 [\exp(-\Sigma_1(z,R)/10^2) + 
\exp(-\Sigma_2(z,R)/10^2)]
\end{equation}
\citep{semenov04}
where $\Sigma_1$ is the surface density between the height above
the midplane, $z$, and the
top of the disk, and $\Sigma_2$ is the surface density between $z$ and the 
bottom of the disk.

The decay of radioactive nucleides is an additional source of ionization.
$^{26}$Al can decay to form excited $^{26}$Mg, which
in turn decays by either positron decay or by electron capture. 
The ionization rate of these processes is assumed to be 
$\zeta_\mathrm{Al}$ = 6.1 $\times$ 10$^{-19}$ s$^{-1}$ \citep{un81}.

A final source of ionization are X--rays of which  T Tauri stars
are strong emitters.  We use the approach of \cite{gh04}
to calculate the X--ray photoionization rates of atoms and
molecules.  This method assumes that the
X-ray photoionization rate of a molecule is the
sum of the rate for its constituent atoms and that
X-ray ionization leads to the loss of a single electron.
The ionization rate per atom 
depends on the incident X-ray flux and the cross-section of the atom.
We calculate fits to the energy dependent atomic cross-sections given in \cite{vy95}.
The attenuation depth of X--rays is very small
and X-ray ionization is most efficient in the surface layers
of the disk. For further details, see \citet{ww09}.

\subsection{Recombination of N$_2$H$^+$}
The recombination of N$_2$H$^+$ with electrons has
recently been the subject of study in the laboratory with
two different groups producing two different 
results.  There are two possible pathways for this reaction
\begin{eqnarray}
\hbox{\chem{N_2H^+}} + e^- & \longrightarrow & \hbox{NH} + \hbox{N} \label{eq:1}\\
& \longrightarrow & \hbox{N$_2$} + \hbox{H} \label{eq:2}
\end{eqnarray} 
The pathway followed by this reaction can influence the abundance of
nitrogen-bearing molecules in the gas phase.  If the primary product
is NH then in cold regions of the disk, NH will quickly
freezeout and be converted to NH$_3$ ice, removing nitrogen from the
gas phase.  N$_2$, on the other hand, does not react on the grains and
is relatively easy to desorb.  N$_2$ is also an important means of
controlling the degree of ionization \citep{cd05}.  The presence of
N$_2$ in the gas can prevent the transfer of deuteration along the chain of
isotopologues from \chem{H_3^+} to \chem{D_3^+} by destroying the less
deuterated isotopologues before they have a chance to form \chem{D_3^+}.

In W07 we took the results of \cite{geppert04} which
found that the NH route is most likely to happen, occurring 65\% of
the time.  However, more recent laboratory work from \cite{molek} has
cast some doubt on this result.  \cite{molek} find that the route
leading to N$_2$ dominates, with no significant NH being
formed.

Using the \citeauthor{geppert04}\ results leads to lower
abundances of \chem{N_2} in the midplane compared to models using
\citeauthor{molek}  This will translate into a higher 
deuteration of \chem{H_3^+} and lower abundances of gas phase
nitrogen molecules. 

Here we use the more recent results of \citeauthor{molek}
and assume that little NH is formed by the recombination of \chem{N_2H^+}.

\subsection{The initial conditions}

We assume that the material incorporated into the disk has first been
processed to some extent in the parent molecular cloud.  The input
abundances for the disk model are therefore taken from the output of a
molecular cloud model.  The latter uses the elemental abundances found
in Table~\ref{tab:cloud} and is allowed to run for 1 Myrs at a total
hydrogen density $n_H$ = 2 \x 10$^{4}$ cm$^{-3}$, temperature = 10 K
and visual extinction $A_V$ = 10 magnitudes.  All the chemical
processes included in the disk model are also included in the
molecular cloud model with the exception of ionization by the decay of
radioactive nuclei.  The input disk abundances are given in
Table~\ref{tab:input}.  The processes that drive deuteration chemistry
in molecular clouds have been well described by previous authors e.g.\
\cite{roberts04,rm00a,ctr97,bm89,tielens83}.

\begin{deluxetable}{ll}
\tablecolumns{2}
\tablewidth{0pt}
\tablecaption{\label{tab:cloud}Elemental abundances used in the model.
Abundances are given with respect to the total abundance of hydrogen
atoms, $n_\mathrm{H}$ = 2 $n$(H$_2$) + $n$(H).  Initially we assume that
1\% of the hydrogen is atomic, with the rest molecular, and that 
1\% of the deuterium is atomic, with the rest in HD.}
\tablehead{
\colhead{Element} & \colhead{Abundance}\\
& \colhead{$n(x)/n_\mathrm{H}$}
}
\startdata
H$_2$ & 0.495 \\
H & 0.01 \\
He & 0.14 \\
D & 1.6 \x\ 10$^{-5}$ \\
O & 1.76 \x\ 10$^{-4}$ \\
C & 7.3 \x\ 10$^{-5}$ \\
N & 2.14 \x 10$^{-5}$ \\
Fe & 3.0 \x 10$^{-9}$ \\
Mg & 7.0 \x 10$^{-9}$\\
\enddata
\end{deluxetable}

\begin{deluxetable}{lllllll}
\tabletypesize\footnotesize
\tablecolumns{7}
\tablewidth{0pt}
\tablecaption{\label{tab:input}Input fractional abundances 
with respect to total  number of hydrogen nuclei
for the disk model
as determined by a molecular cloud model at 1 Myrs.  The results for two
models are shown.  The numbers vary a little from those in Table 3 in
W07 because of the use of the \cite{molek} results for
N$_2$H$^+$ recombination instead of \cite{geppert04}.  The physical parameters
used were density, $n_H$ = 2 $\times$ 10$^4$ cm$^{-3}$, temperature = 10 K, and
$A_V$ = 10 magnitudes.  Freezeout, thermal desorption and cosmic
ray heating of grains were included.  Grain chemistry contributes to the formation
of some of the ice molecules e.g.\ water, whereas others e.g.\ CO form in the gas phase
and then freezeout but are not produced {\it in situ} on the grain surfaces.}
\tablehead{
\colhead{Molecule} & \multicolumn{2}{c}{Abundance}   & & \colhead{Molecule} & \multicolumn{2}{c}{Abundance} \\
\cline{2-3} \cline{6-7}\\
                   &  \colhead{Gas} &  \colhead{Grain}   & &                   & \colhead{Gas} & \colhead{Grain}
}
\startdata
H             & 2.8 (-5)  & \nodata   & & D             & 6.7 (-7)  & \nodata  \\
H$_3^+$       & 3.0 (-9)  & \nodata   & & H$_2$D$^+$    & 5.8 (-10) & \nodata  \\
HD$_2^+$      & 9.8 (-11) & \nodata   & & D$_3^+$       & 1.5 (-11) & \nodata  \\
CO            & 3.2 (-5)  & 2.7 (-6)  & & CO$_2$        & 5.8 (-8)  & 2.2 (-7)  \\
HCO$^+$       & 3.5 (-9)  & \nodata   & & DCO$^+$       & 3.1 (-10) & \nodata   \\
H$_2$CO       & 2.4 (-7)  & 1.6 (-6)  & & HDCO          & 1.6 (-8)  & 7.7 (-8)  \\
CH$_3$OH      & 1.6 (-10) & 4.0 (-8)  & & CH$_3$OD      & 9.3 (-12) & 9.1 (-10)  \\
CH$_2$DOH     & 1.5 (-11) & 3.2 (-9)  & & O             & 8.5 (-7)  & \nodata   \\
O$_2$         & 4.4 (-8)  & \nodata   & & OH            & 2.4 (-8)  & \nodata   \\
OD            & 4.1 (-8)  & \nodata   & & H$_2$O        & 3.8 (-8)  & 1.4 (-4)  \\
HDO           & 3.8 (-9)  & 1.2 (-6)  & & D$_2$O        & 4.2 (-11) & 1.3 (-8)  \\
N             & 1.0 (-7)  & \nodata   & & N$_2$         & 1.9 (-6)  & 6.5 (-8)  \\
NO            & 4.5 (-8)  & \nodata   & & CN            & 6.6 (-8)  & \nodata   \\
HCN           & 2.1 (-8)  & 2.4 (-6)  & & DCN           & 6.7 (-10) & 2.0 (-8)  \\
HNC           & 9.9 (-9)  & 1.1 (-7)  & & DNC           & 3.9 (-10) & 2.1 (-9)  \\
HC$_3$N       & 2.5 (-7)  & 9.9 (-7)  & & DC$_3$N       & 9.4 (-9)  & 3.3 (-8)  \\
NH$_3$        & 8.6 (-9)  & 1.3 (-5)  & & NH$_2$D       & 2.7 (-10) & 6.7 (-8)  \\
NHD$_2$       & 5.1 (-12) & 3.1 (-10) & & ND$_3$        & 8.7 (-13) & 2.1 (-11) \\
N$_2$H$^+$    & 2.8 (-10) & \nodata   & & N$_2$D$^+$    & 2.8 (-11) & \nodata   \\
C             & 3.9 (-7)  & \nodata   & & C$^+$         & 3.1 (-9)  & \nodata   \\
C$_2$H        & 6.4 (-9)  & \nodata   & & C$_2$H$_2$    & 1.1 (-7)  & 3.3 (-7)  \\
CH$_4$        & 3.1 (-6)  & 1.1 (-5)  & & CH$_3$D       & 1.6 (-7)  & 3.4 (-7)  \\
CH$_2$D$_2$   & 6.1 (-9)  & 1.8 (-8)  & & CHD$_3$       & 1.5 (-10) & 5.0 (-10) \\
CD$_4$        & 8.2 (-13) & 2.1 (-12) & & HD            & 1.2 (-5)  & \nodata  \\
\enddata
\end{deluxetable}

\section{Results}

In common with previous authors e.g.\ \cite{ah99,wl00,aikawa02,zadelhoff03} we find that the
disk can be divided vertically into three chemically distinct
layers. Grains are coldest in the midplane, leading to 
the existence of ices of at least some species at $R$ $>$ 2 AU.
Inside of 2 AU the grains are hot 
enough that all molecules are desorbed in the midplane.  Above
the midplane is a molecular layer where thermal desorption is efficient
and the disk is sufficiently optically thick that molecules can survive in 
the gas phase. This leads to high gas phase molecular abundances.
In the surface layer UV and X-rays can penetrate, dissociating 
molecules into their constituent atoms and ions.  The distinction
between the midplane and the molecular layer is not as
clear as that seen in outer disk models, since the grains are warm
enough even in the midplane for some volatile species (e.g.\ CO 
and N$_2$) to desorb.  We use these three layers in the discussions
below.

\begin{figure}
\caption{\label{fig:abund_0.01}Fractional abundances in the inner disk
  for Model 1 ($\alpha$ = 0.01).  Note change of radial scale for
  molecules that are only abundant in the inner few AU
  i.e.\ \chem{H_2O}, HDO, \chem{NH_3} and \chem{NH_2D}.  The dashed
  line indicates the surface of the disk (at 6 
pressure scale heights above the midplane).  Also
  marked are optical depths, $\tau$, of 0.1, 1 and 10. }
\includegraphics{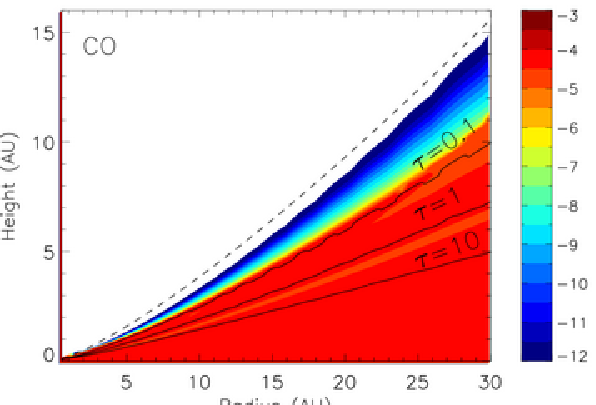}
\includegraphics{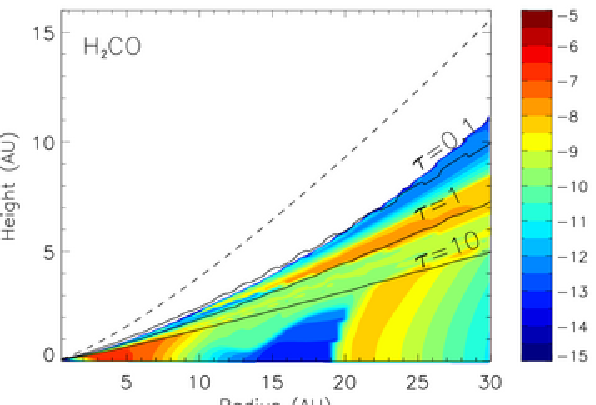}\\
\includegraphics{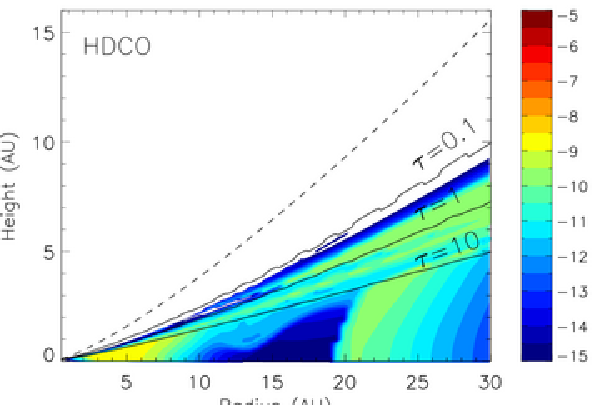}
\includegraphics{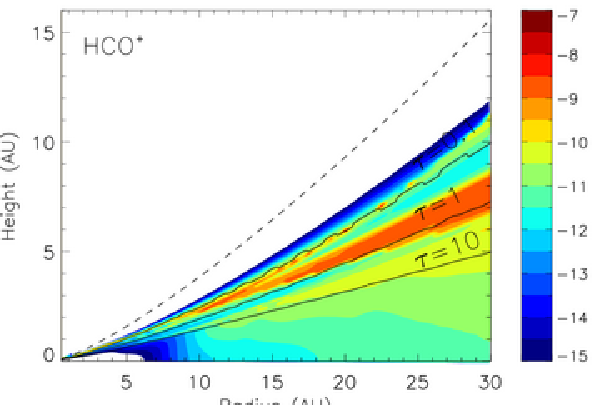}\\ 
\includegraphics{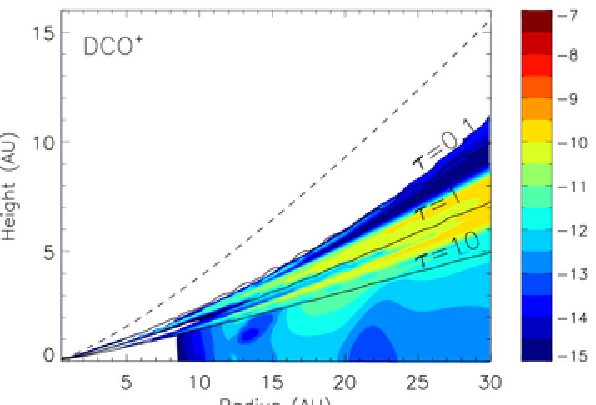}
\includegraphics{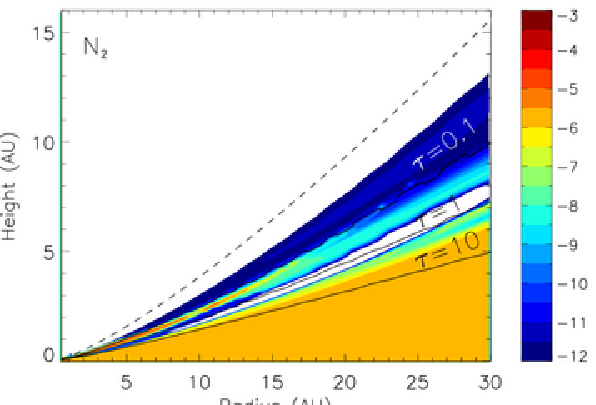}\\
\end{figure}

Figure~\ref{fig:abund_0.01} shows the fractional abundance
distributions in the inner disk.  The abundance distribution is a
combination of the effects of the formation and destruction processes
at a given radius, together with the transport to that radius of
molecules formed further out in the disk.  Many molecules desorb as
their sublimation temperature is reached and then show little
variation in their gas phase abundance with radius e.g.\ N$_2$ and CO
show no significant change in abundance across the inner 30 AU of the
disk below the surface photodissociation layer.  The upper extent of
these molecules is determined by photodissociation.  N$_2$ shows a
more complicated behavior than CO, with a low abundance region (where
$x$(N$_2$) $<$ 10$^{-12}$).  Here photodissociation is still efficient
and the resulting nitrogen atoms are incorporated into other
nitrogen--bearing molecules e.g.\ HCN.  HCN has a high binding energy
of 4173 K and the grain temperature in this region is low enough that
once HCN freezes out it is not able to thermally desorb.  This leads
to a loss of nitrogen atoms from the gas and hence N$_2$ cannot
reform.  Above the low abundance layer, HCN still forms from nitrogen
atoms, but the difference is that the grains are warm enough for
desorption of HCN to be efficient.  Hence nitrogen atoms can cycle
between N$_2$ and HCN.  Below the low abundance region N$_2$ does not
photodissociate and has a high abundance.

\addtocounter{figure}{-1}
\begin{figure}
\caption{{\it cont}}
\includegraphics{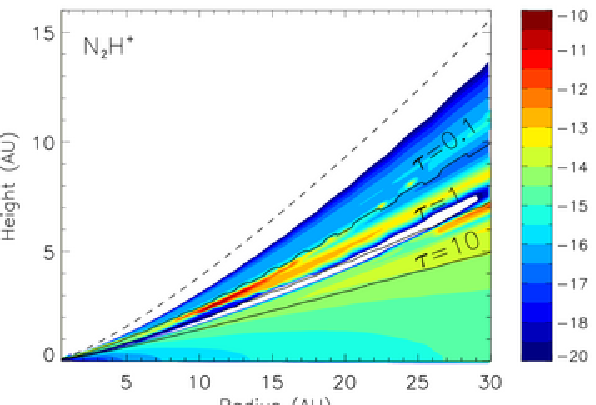}
\includegraphics{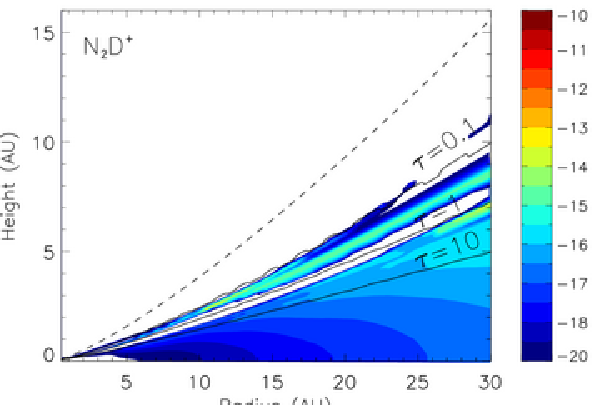}\\ 
\includegraphics{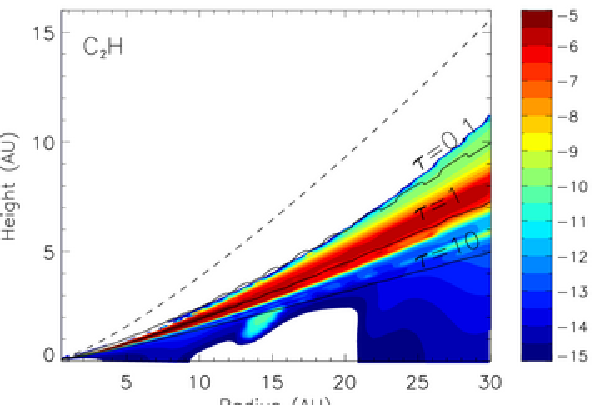}
\includegraphics{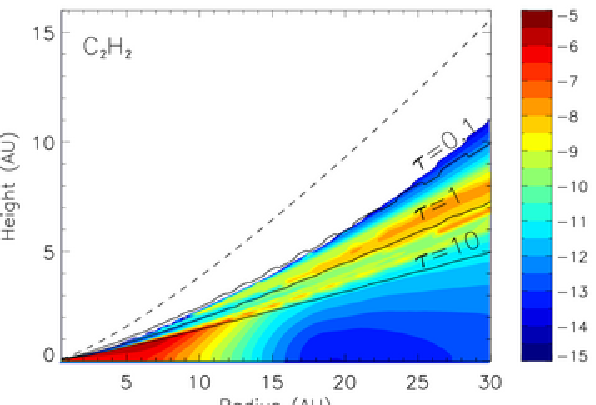}\\
\includegraphics{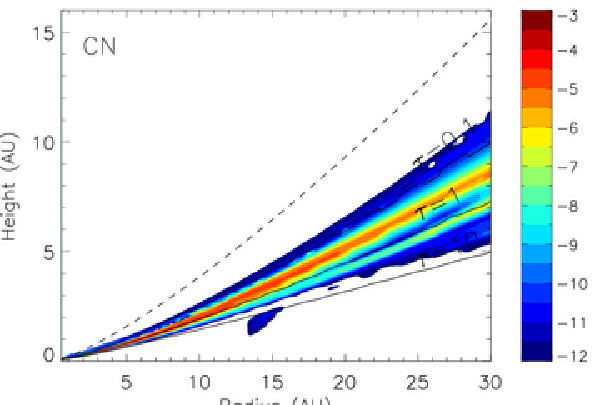}
\includegraphics{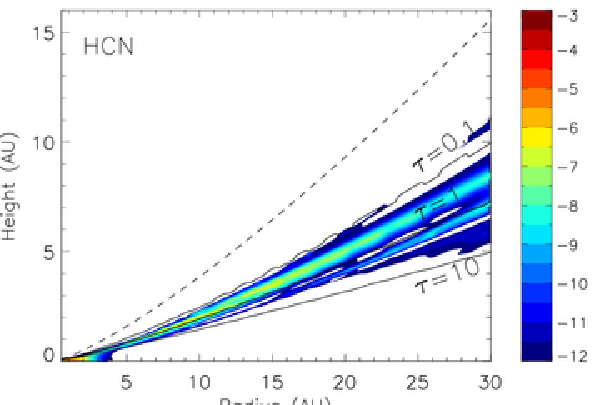}\\ 
\includegraphics{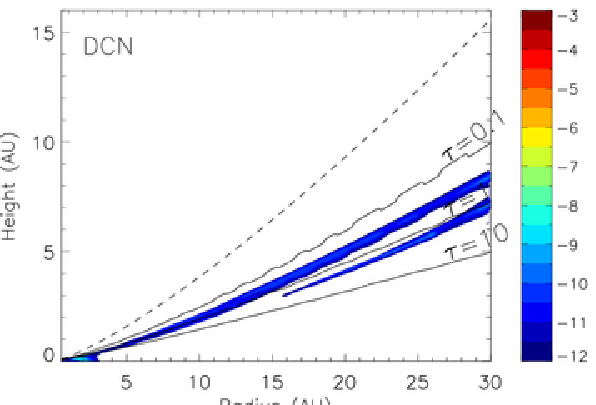}
\includegraphics{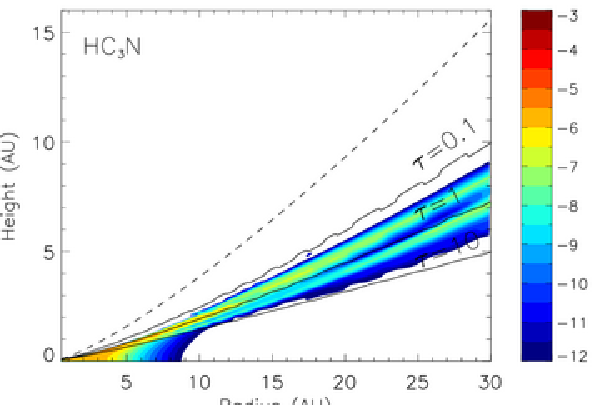}\\
\end{figure}

\addtocounter{figure}{-1}
\begin{figure}
\caption{{\it cont}}
\includegraphics{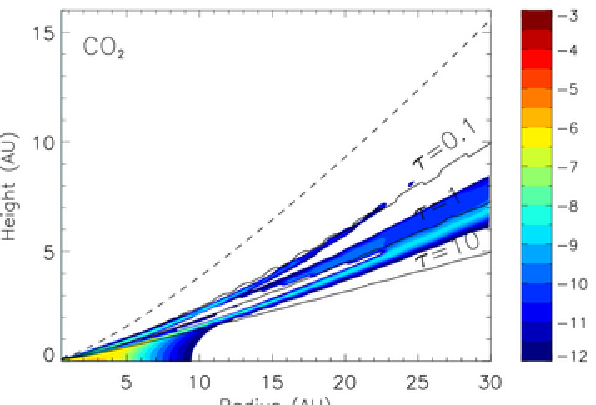}
\includegraphics{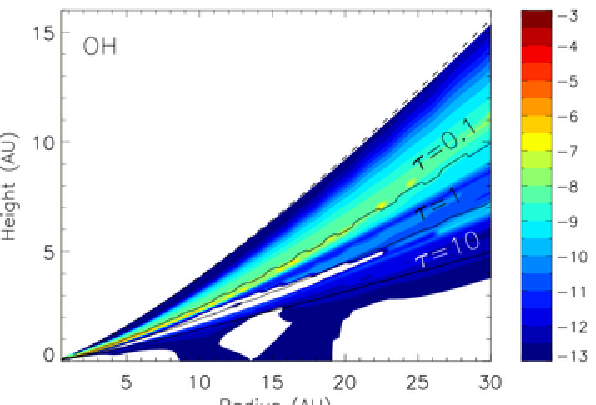}\\ 
\includegraphics{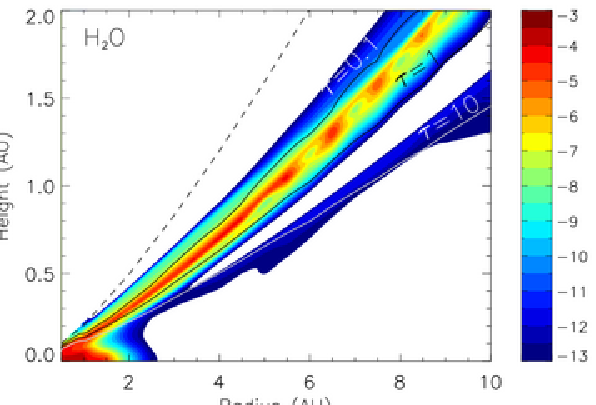}
\includegraphics{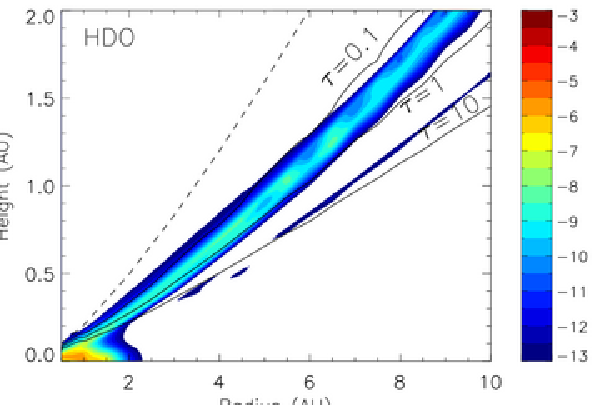}\\
\includegraphics{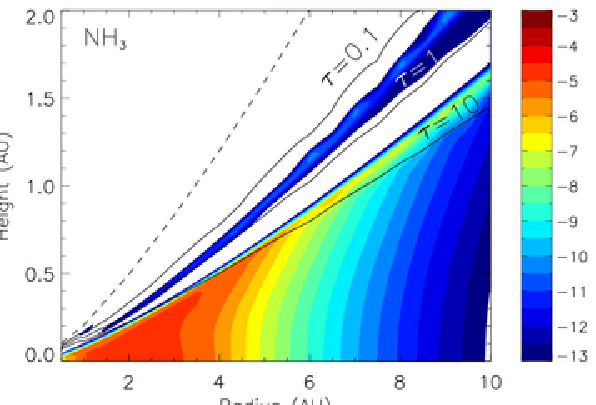}
\includegraphics{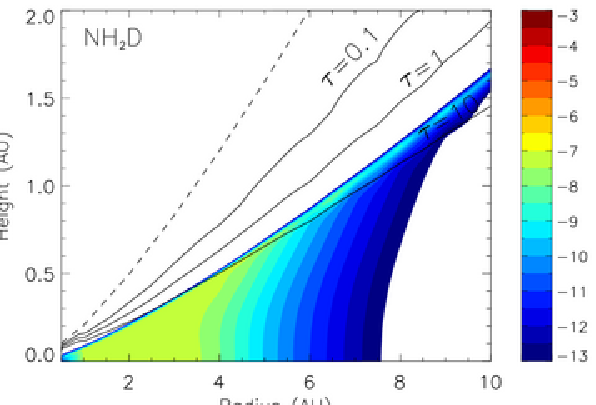}\\
\end{figure}

The effects of an increase in temperature towards the disk surface and
towards the star can be seen in the increase in abundance of less
volatile species such as H$_2$O and HCN in these regions.  Ammonia has
a lower binding energy and is present for a larger radial extent than
either H$_2$O or HCN, but it does not survive in the surface layers
where it is quickly
destroyed by reaction with abundant CN, producing HCN.  The HCN
freezes out and is retained on the grains and NH$_3$ is not reformed.
NH$_3$ only exists in regions shielded from UV radiation.

\chem{H_2CO} shows two abundance peaks in the midplane, as
well as a peak in the molecular layer above it.  The outermost
midplane peak (at $R$ = 20-23 AU) and the molecular layer peak are
due to thermal desorption.  Between 7 and 20 AU H$_2$CO 
is quickly destroyed by gas phase reactions with HCO$^+$ and DCO$^+$
followed by neutralization of the resulting ions (H$_3$CO$^+$ and H$_2$DCO$^+$)
either with electrons or on grain surfaces.  While some of the ions
will reform H$_2$CO, 1/3 will instead form CO, leading to a removal
of H$_2$CO from the gas.

The second peak at $R$ $<$ 8 AU is due to formation in the
gas phase by 
\begin{equation}
\hbox{\chem{O}} + \hbox{\chem{CH_3}} \longrightarrow \hbox{\chem{H_2CO}} + 
\hbox{H}
\end{equation}
The oxygen atoms are supplied by the destruction of CO by reaction
with \chem{He^+}.  The \chem{CH_3} is supplied by the breakdown of
\chem{C_3H_4} (which is desorbed at $\sim$ 7 AU). HDCO forms by
similar processes.

C$_2$H exists in the boundary between the molecular and
surface layers as a photodissociation product of C$_4$H$_2$ at
$R$ $>$ 10 AU and of HC$_3$N inside of this radius.

The abundances of molecular ices do not vary much across
the disk, until their desorption temperature is reached.
The ice abundances are set by reactions in the molecular
cloud and are not altered by grain chemistry in the disk.

\subsection{Gas phase column densities}

\begin{figure}
\caption{\label{fig:alp0.01_cd}The calculated column densities in Model 1 of
some molecules in the inner 30 AU. }
\plottwo{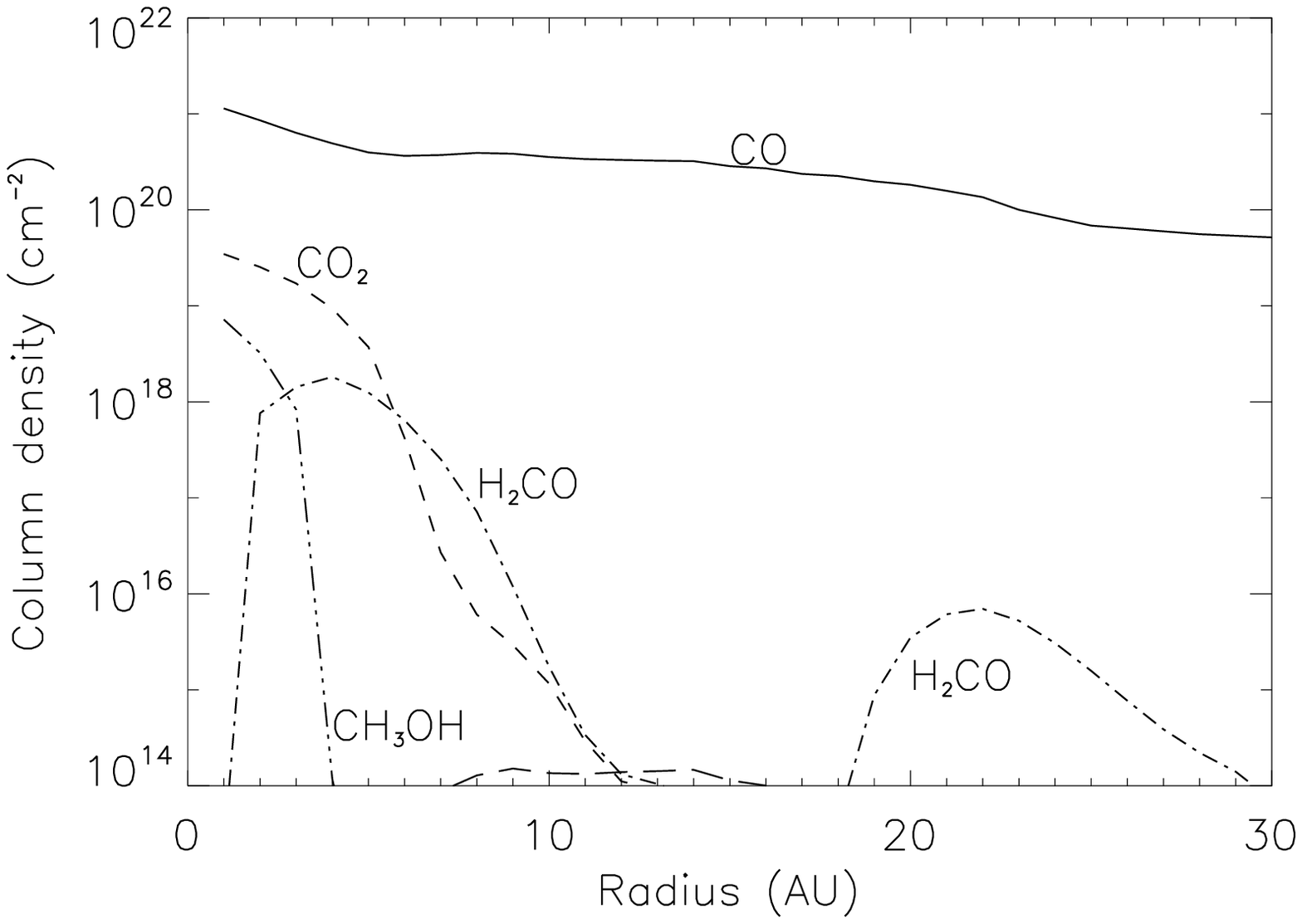}{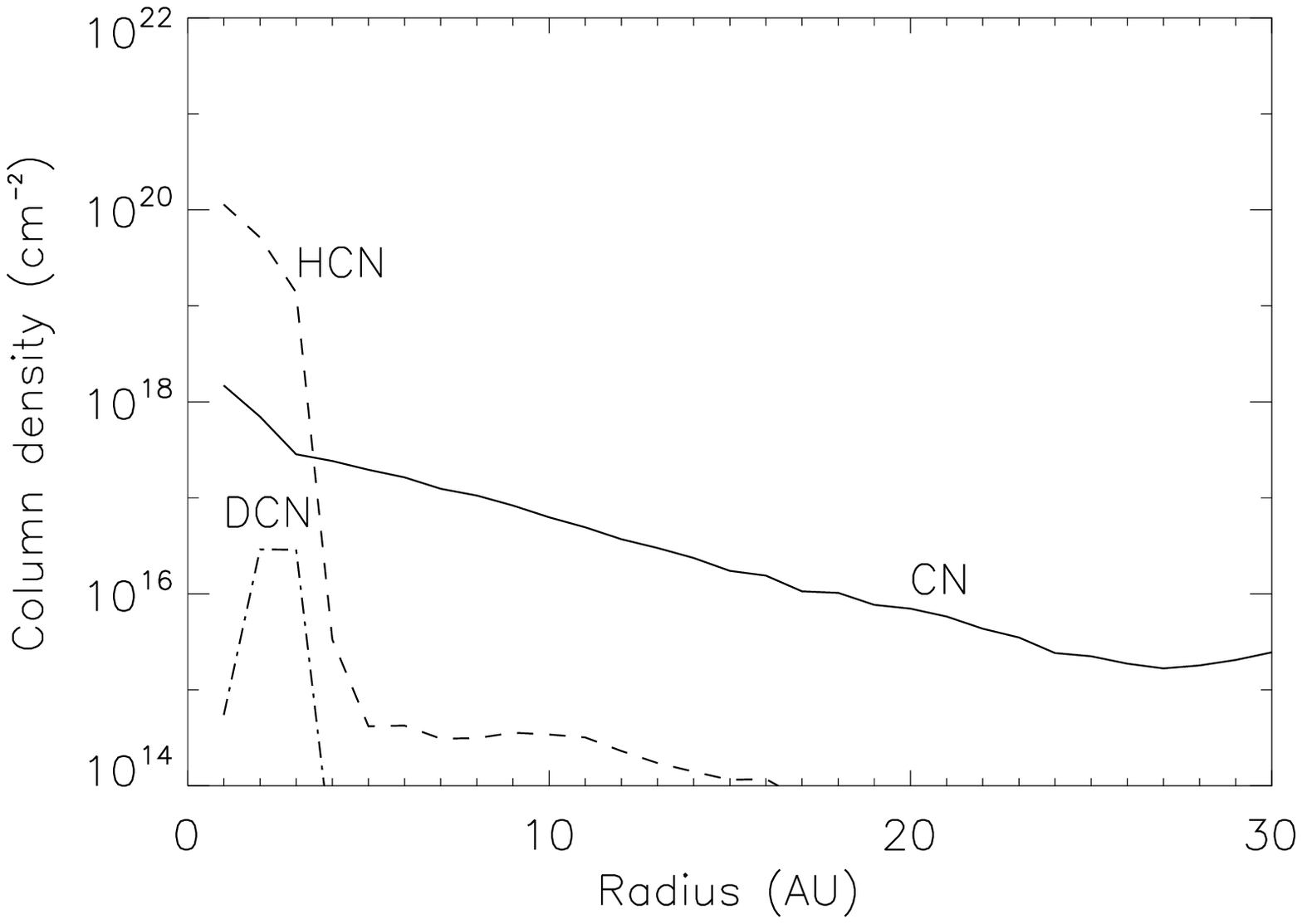}\\
\plottwo{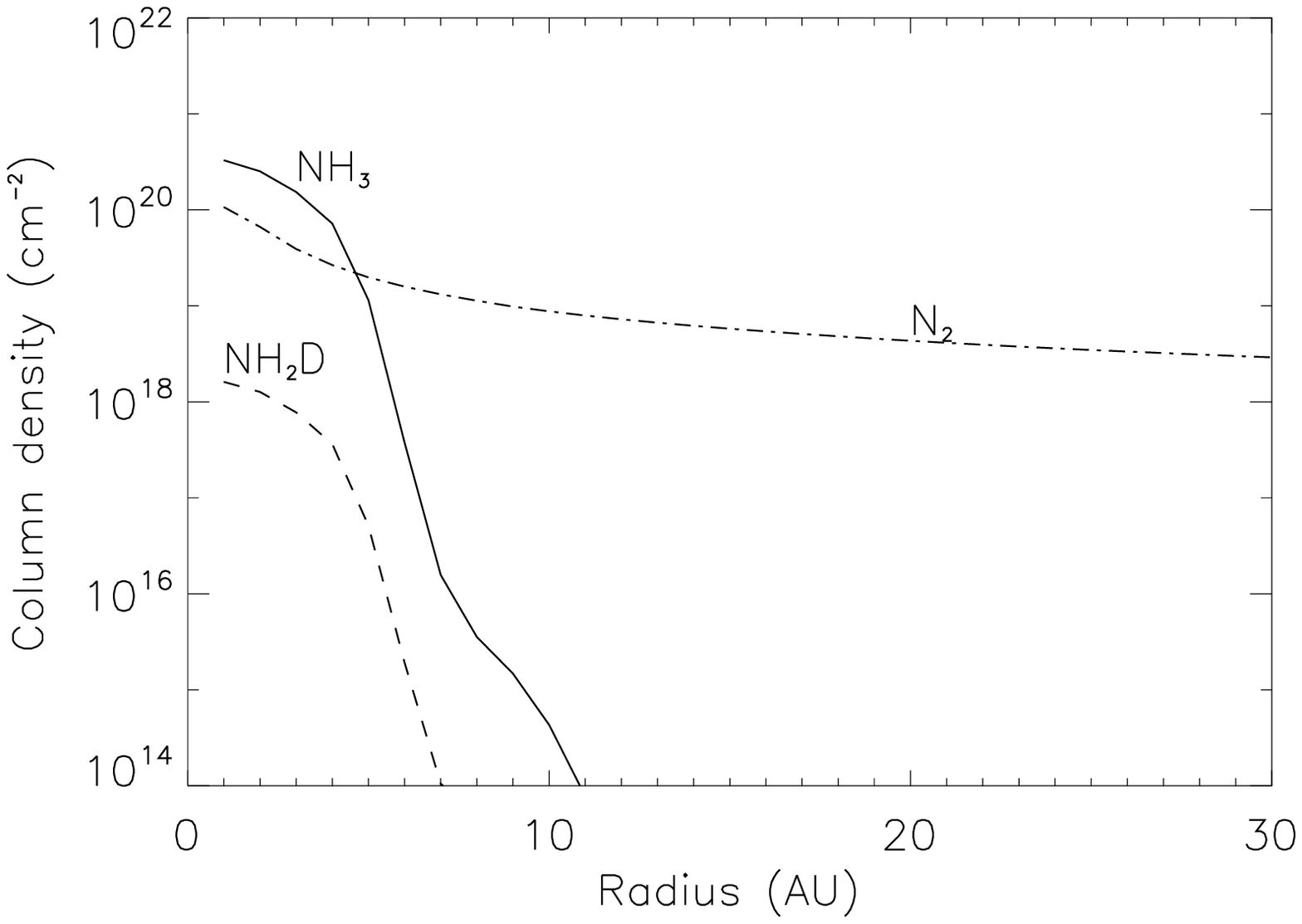}{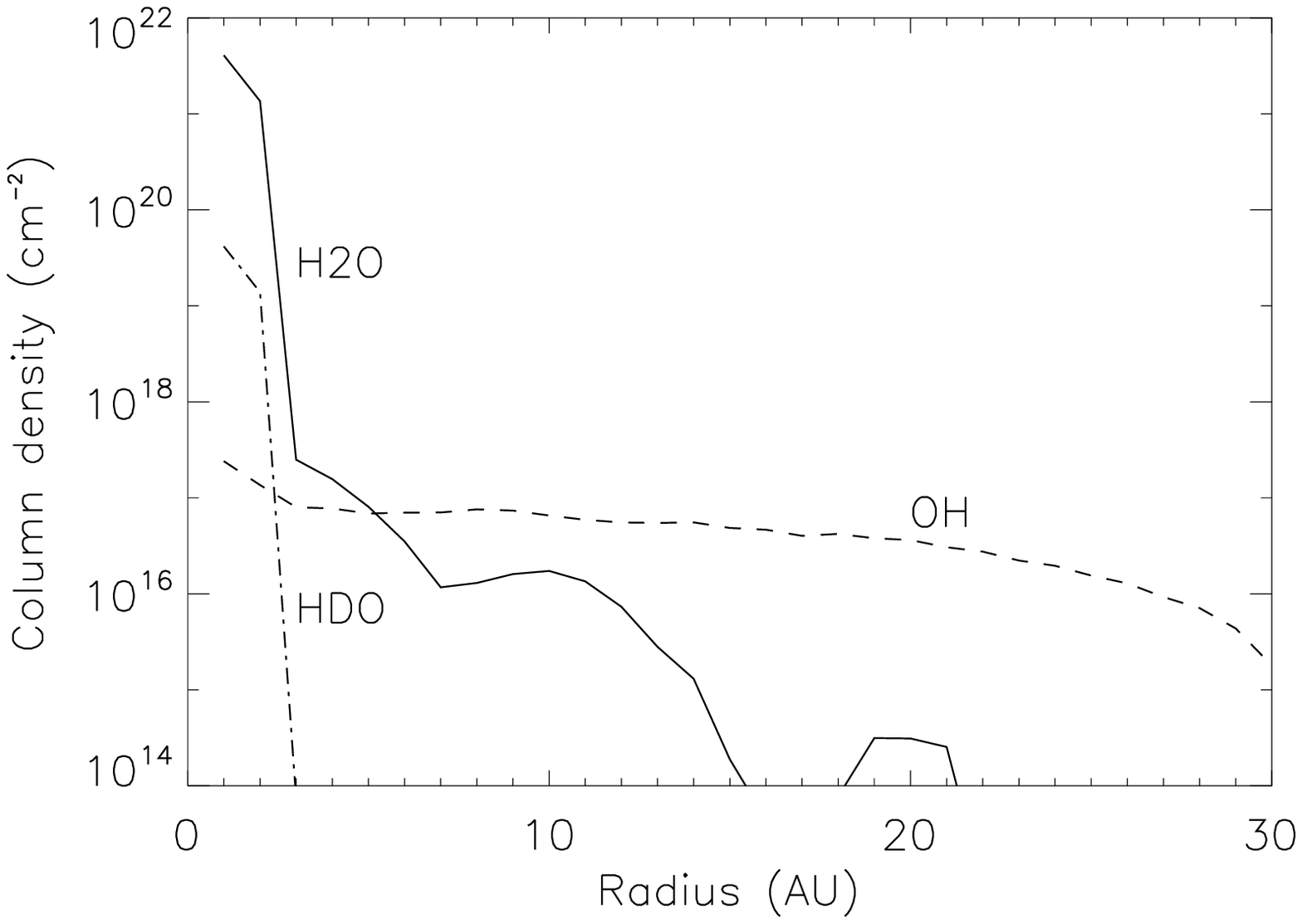}\\
\plottwo{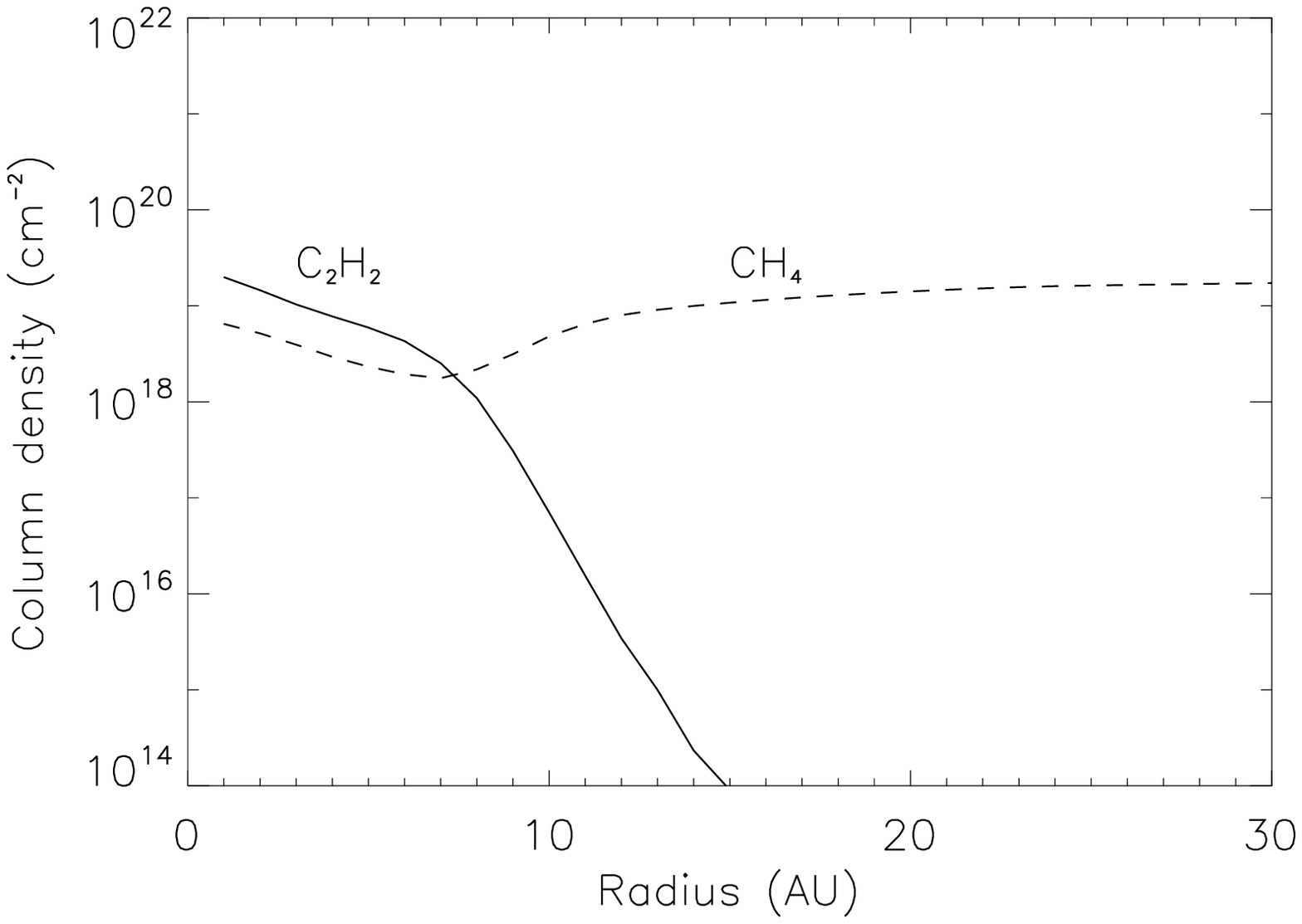}{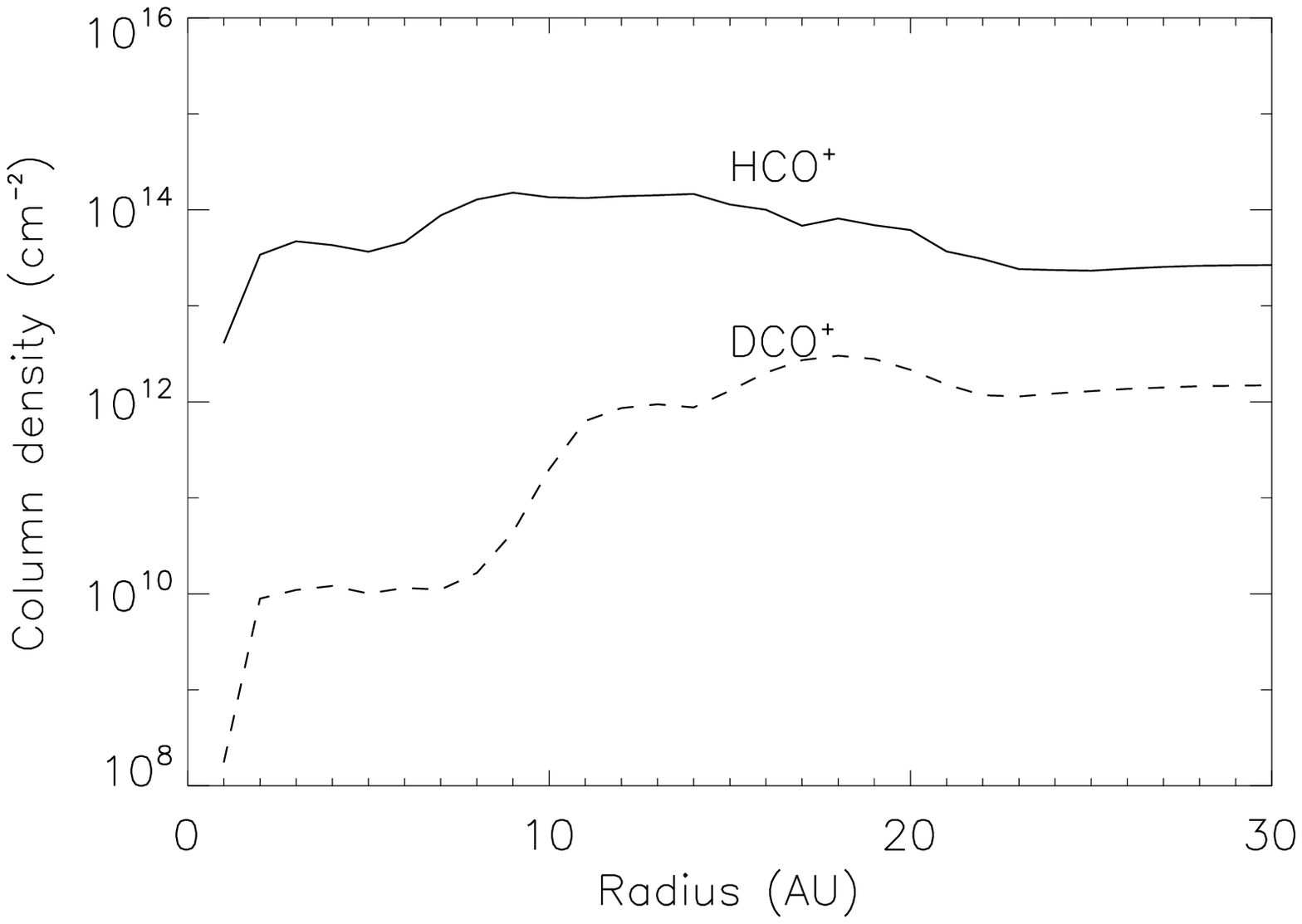}
\end{figure}

Figure~\ref{fig:alp0.01_cd} shows the radial variation in column
density measured vertically from the midplane for some
molecules as calculated in Model 1.  Most molecules show a sharp
increase in column density due to thermal desorption as the radius
decreases  e.g.\ HCN, CO$_2$, NH$_3$.  A corresponding increase in
fractional abundance can be seen for these molecules in
Figure~\ref{fig:abund_0.01}.  The increase in the column density
of the equivalent deuterated molecules occur at slightly smaller radii
because of their slightly higher binding energies.  The fall in DCN at
$R$ $<$ 1.5 AU is because of its reaction with hydrogen atoms to form
HCN.  The reverse reaction \mbox{(D + HCN $\longrightarrow$ DCN + H)} can
also occur, and at the same rate, but DCN does not reform because the
atomic D/H ratio is low at small radii.  $N$(CN) also increases
closer to the star.  It is a photodissociation product of DCN and HCN
and therefore exists mainly in the surface layers.

H$_2$CO shows two column density peaks: one around 23 AU and one
around 3 -- 4 AU corresponding to the fractional abundance peaks in
the midplane seen in Figure~\ref{fig:abund_0.01}.

\begin{deluxetable}{lllllllll}
\tablecolumns{8}
\tablewidth{0pt}
\tablecaption{\label{tab:cd}Calculated vertical column densities from midplane to the disk surface
for some gas phase species in Models 1 and 2.}
\tablehead{
\colhead{Molecule} & \multicolumn{3}{c}{Model 1 ($\alpha$ = 0.01)} & & \multicolumn{3}{c}{Model 2 ($\alpha$ = 0.025)} \\
\cline{2-4} \cline{6-8}\\
& \colhead{1 AU} & \colhead{5 AU} & \colhead{10 AU} & & \colhead{1 AU} & \colhead{5 AU} & \colhead{10 AU} 
}
\startdata
H$_2$       & 1.6 (25) & 2.4 (24) & 2.3 (24) & & 4.9 (25) & 1.9 (25) & 9.2 (24) \\
CO          & 1.1 (21) & 3.8 (20) & 3.6 (20) & & 1.9 (21) & 1.9 (21) & 6.0 (20) \\
\chem{CO_2} & 3.5 (19) & 1.2 (18) & 7.0 (14) & & 7.3 (19) & 1.7 (19) & 6.8 (13) \\ 
\chem{H_2O} & 4.1 (21) & 8.2 (16) & 2.5 (16) & & 1.4 (22) & 1.4 (17) & 7.1 (15) \\
HDO         & 4.2 (19) & 6.4 (12) & 8.7 (11) & & 1.7 (20) & 8.7 (12) & 2.6 (11) \\
OH          & 2.4 (17) & 7.6 (16) & 6.8 (16) & & 1.2 (17) & 1.1 (17) & 8.3 (16) \\
HCN         & 1.1 (20) & 6.6 (14) & 4.2 (14) & & 4.1 (20) & 1.8 (14) & 3.6 (14) \\
DCN         & 5.0 (14) & 2.4 (11) & 4.3 (10) & & 2.6 (15) & 1.4 (10) & 2.6 (10) \\
CN          & 1.5 (18) & 2.2 (17) & 8.6 (16) & & 3.4 (18) & 1.9 (17) & 8.3 (16) \\
\chem{C_2H_2} & 2.0 (19) & 6.0 (18) & 3.6 (16) & & 3.3 (19) & 2.2 (19) & 2.5 (17) \\
\chem{CH_4}   & 6.5 (18) & 2.3 (18) & 4.5 (18) & & 2.8 (18) & 4.8 (18) & 9.5 (17) \\
\enddata
\end{deluxetable}

\clearpage

\subsection{Deuteration in the inner disk}

In low temperature regions such as molecular clouds enhancements
in molecular deuteration are driven by the reactions of
\chem{H_2D^+}, \chem{CH_2D^+} and \chem{C_2HD}.  These
form by
\begin{eqnarray}
\hbox{H$_3^+$} + \hbox{HD} & \leftrightharpoons & \hbox{H$_2$D$^+$} + \hbox{H$_2$} + \Delta E_1 \nonumber\\
\hbox{CH$_3^+$} + \hbox{HD} & \leftrightharpoons & \hbox{CH$_2$D$^+$} + \hbox{H$_2$} + \Delta E_2\\
\hbox{C$_2$H$_2^+$} + \hbox{HD} & \leftrightharpoons & \hbox{C$_2$HD$^+$} + \hbox{H$_2$} + \Delta E_3 \nonumber \\
\end{eqnarray}
where $\Delta E_1$, $\Delta E_2$ and $\Delta E_3$ are activation
barriers, which inhibit the reverse reactions at low temperatures.
($\Delta E_1$ = 220 K at low temperatures, falling to $\sim$ 130 K as 
the temperature increases, $\Delta E_2$ $\sim$ 370 K and $\Delta E_3$ 
$\sim$ 530 K. 
This allows deuteration to be transferred to other molecules
via ion--molecule reactions.  \cite{roberts03} and \cite{roberts04} showed
that multiply deuterated forms of \chem{H_3^+}, \chem{CH_3^+} and 
\chem{C_2H_2^+} also play an important role in determining the deuteration 
in cold regions.  Grain surface reactions can also efficiently increase 
deuteration in ices e.g.\ \cite{tielens83,bm89}.  For example, the
high abundances of deuterated formaldehyde and methanol observed in
star forming regions \citep{parise02,loinard02,bacmann03} have
been attributed to the reaction of deuterium atoms on grains
during an earlier low temperature phase of evolution.

\begin{figure}
\caption{\label{fig:h3+}The fractional abundance distributions of the
  isotopologues of \chem{H_3^+} in a model with $\alpha$ =
  0.01. }
\includegraphics[scale=0.5]{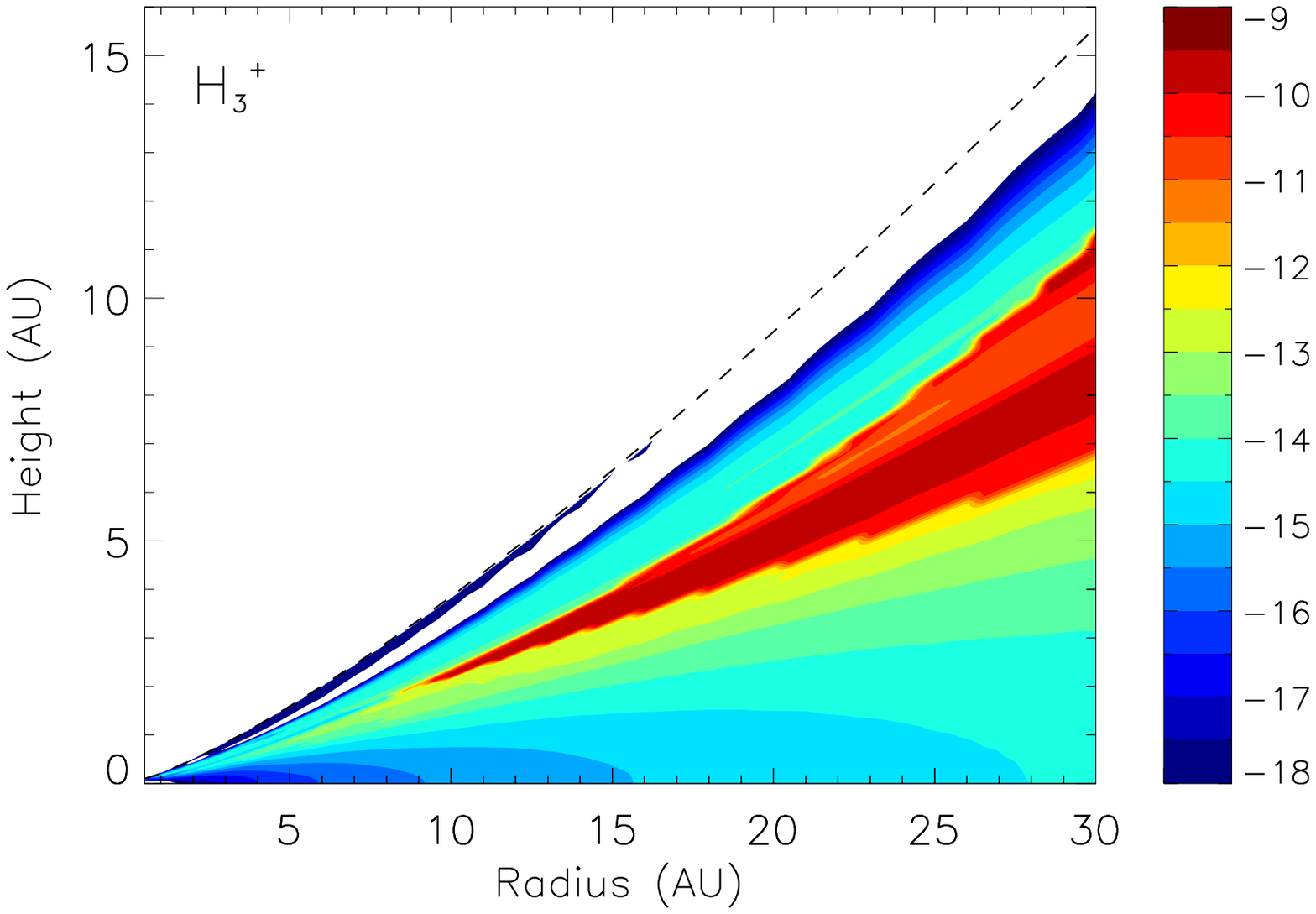}
\includegraphics[scale=0.5]{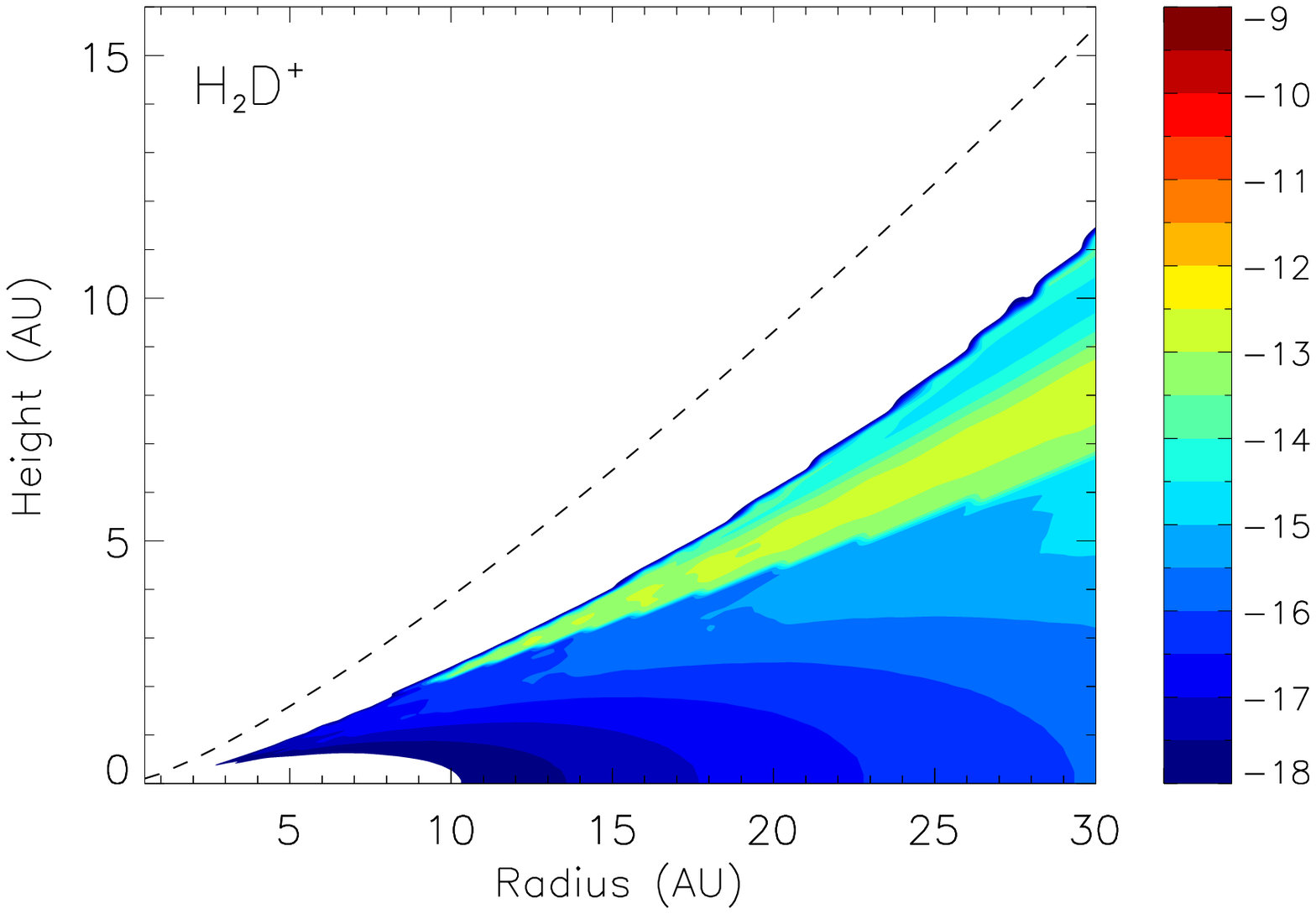}\\
\includegraphics[scale=0.5]{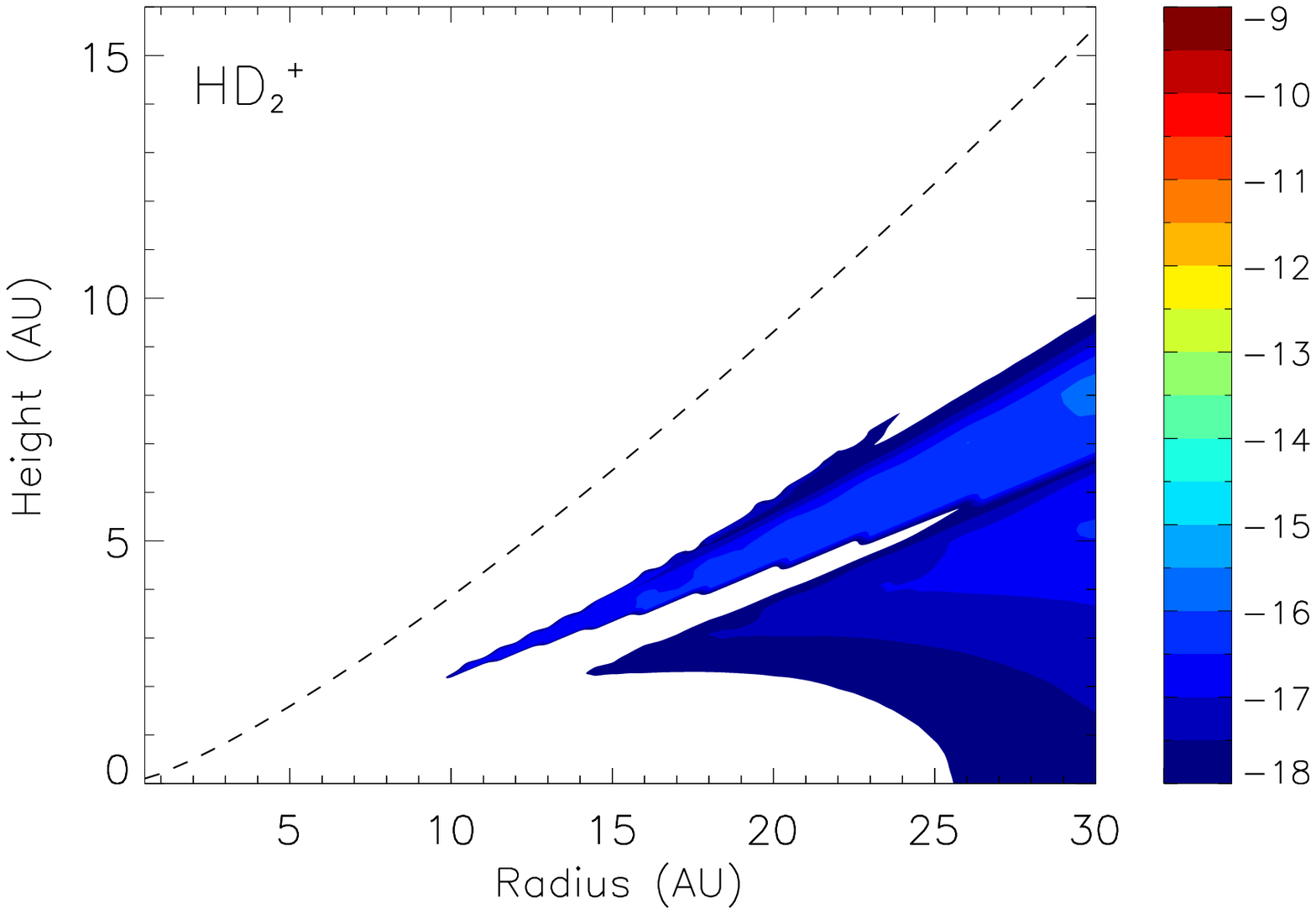}
\includegraphics[scale=0.5]{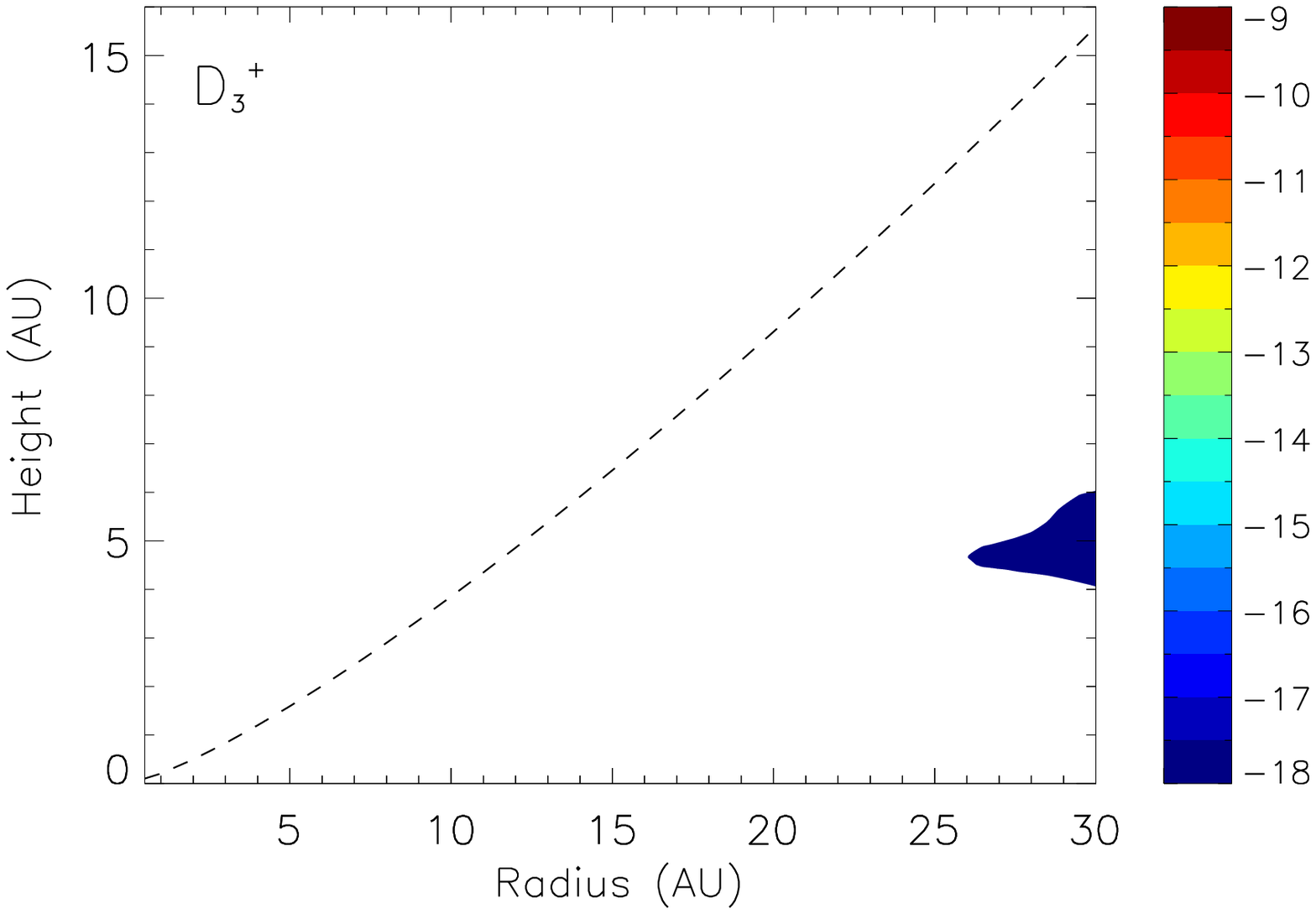}
\end{figure}

\begin{figure}
\caption{\label{fig:DH_h3+}The D/H ratios of the isotopologues of \chem{H_3^+} in Model 1.}

\includegraphics[scale=0.5]{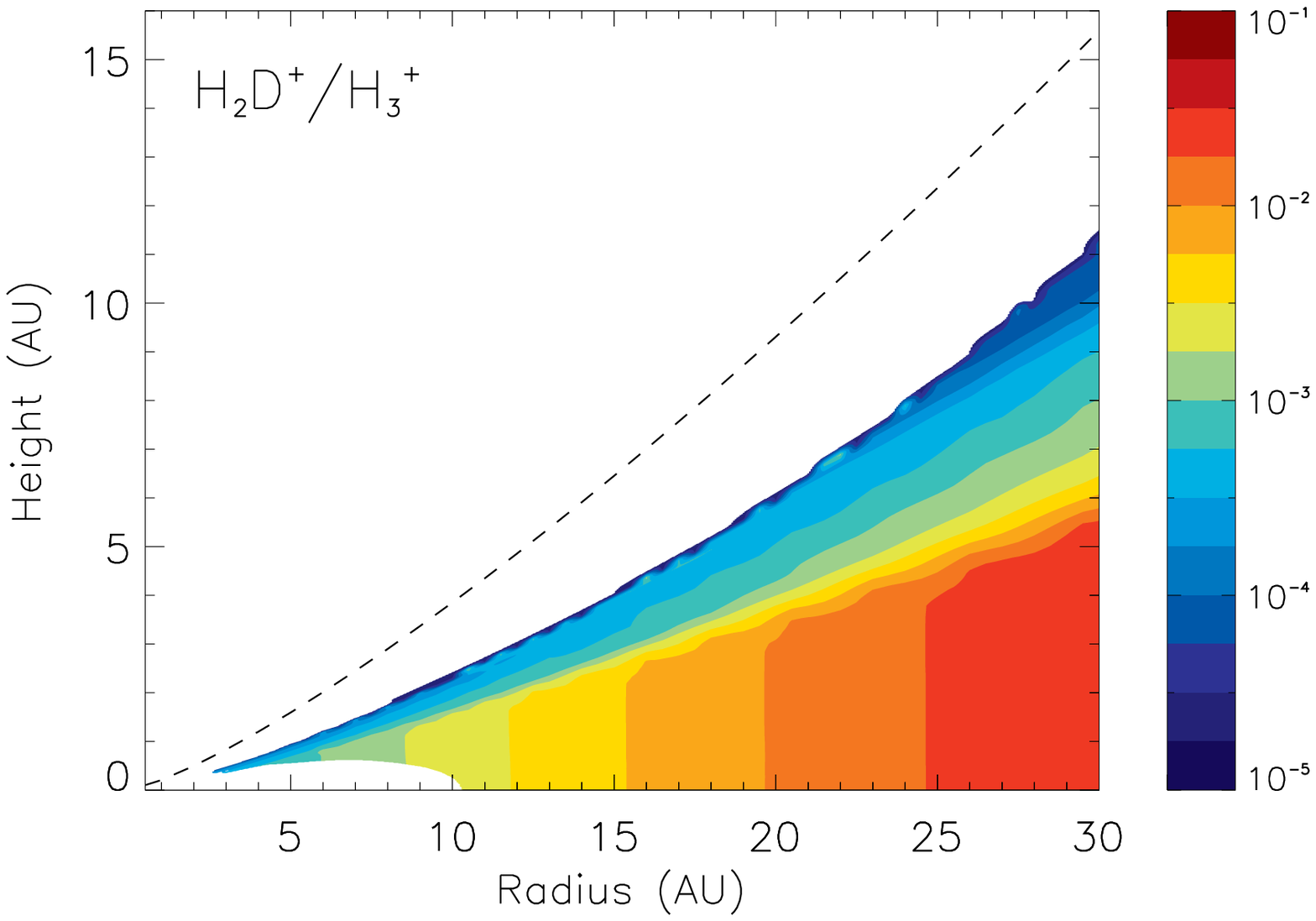}
\includegraphics[scale=0.5]{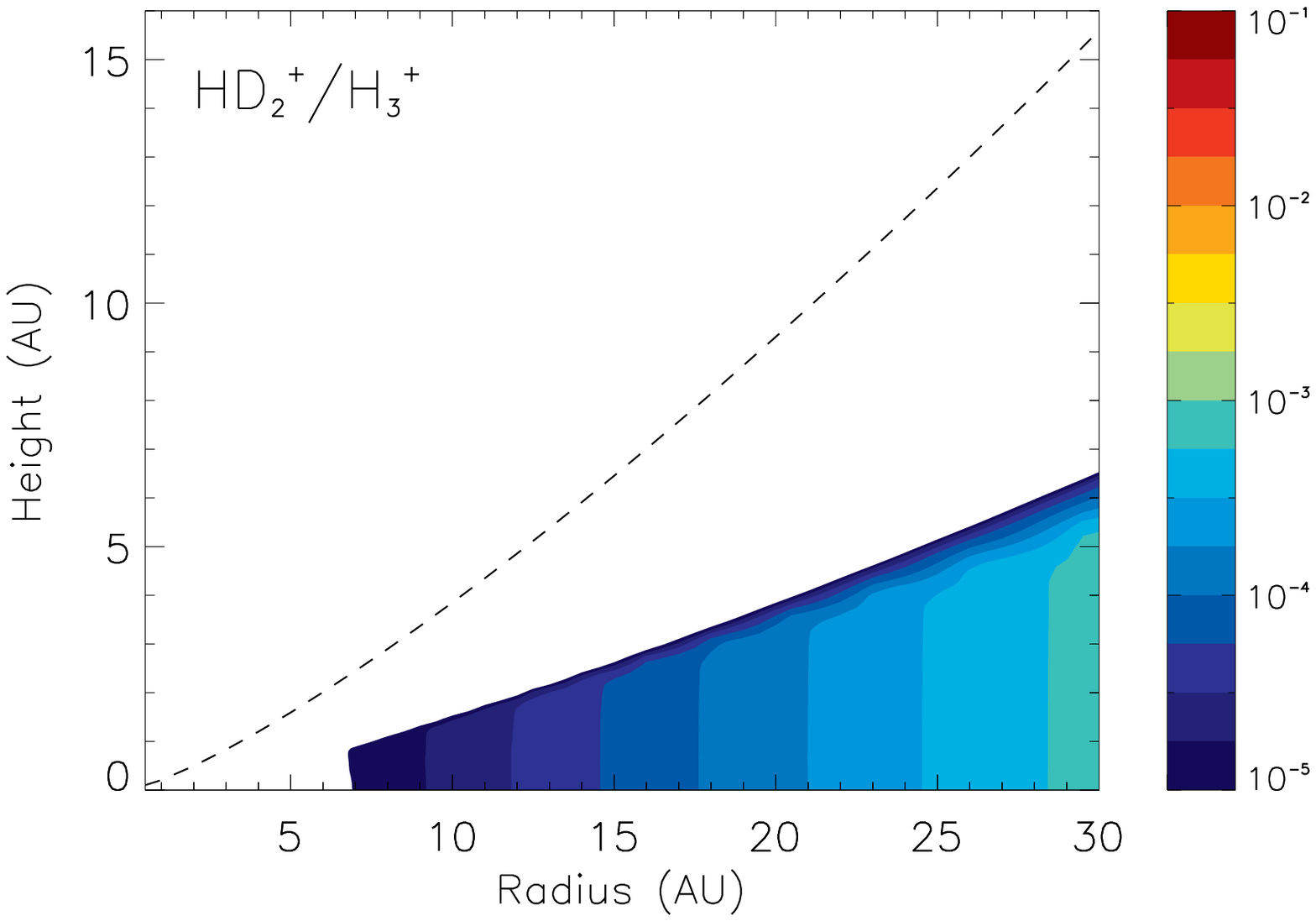}
\includegraphics[scale=0.5]{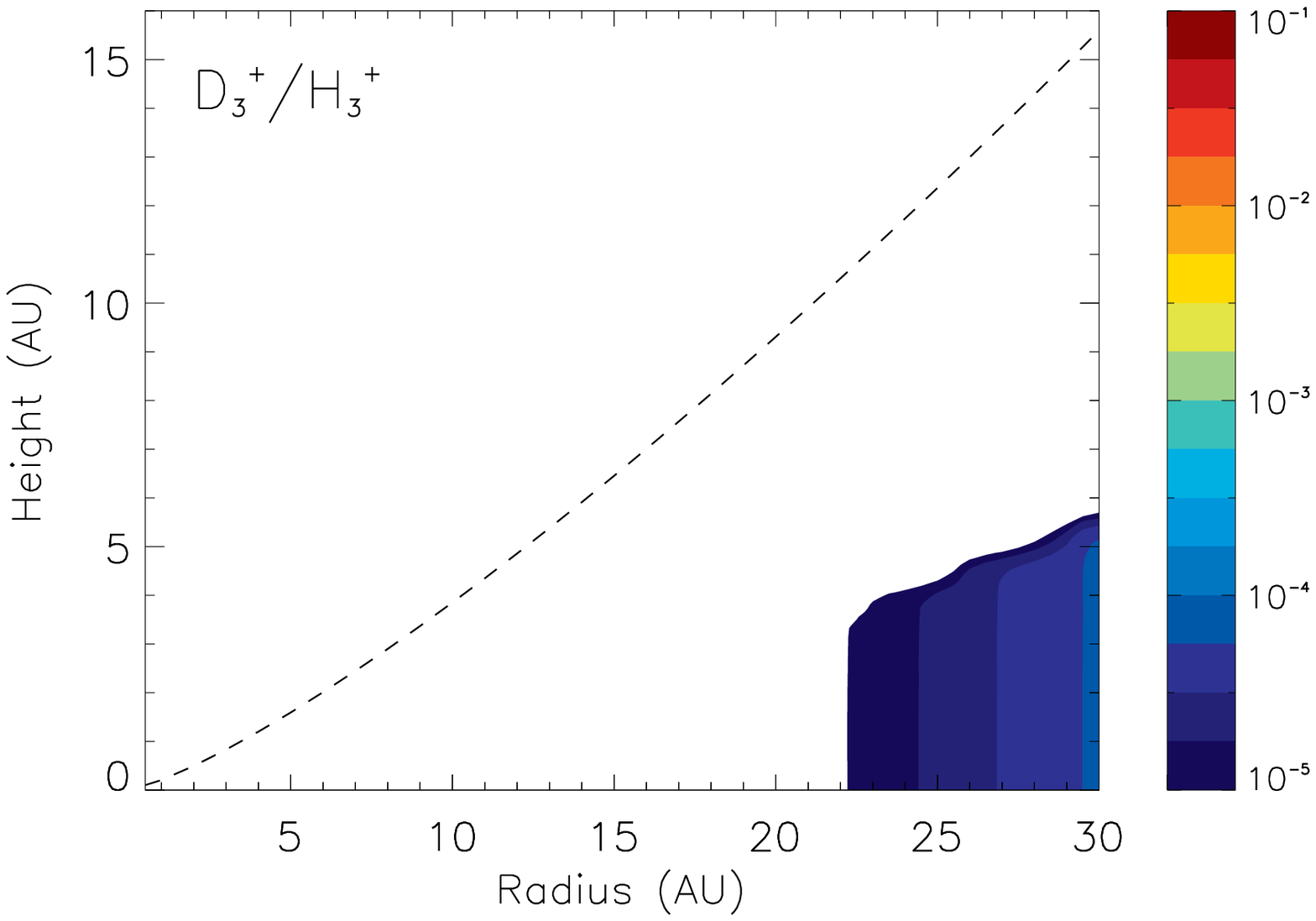}
\end{figure}

In our disk model we find that the
isotopologues of \chem{H_3^+} and \chem{CH_3^+} are the main
drivers of deuteration with some contribution
from the reaction of deuterium atoms.  \chem{C_2HD^+} plays only a minor
role.  Figure~\ref{fig:h3+} shows the
fractional abundances of \chem{H_3^+} and its isotopologues, and the
level to which this molecular ion is deuterated is shown in
Figure~\ref{fig:DH_h3+}.  The abundance of \chem{H_3^+} peaks above
the midplane where cosmic ray ionization is efficient and hydrogen is
mainly atomic.  But here the temperatures are relatively high, and the
deuteration is low.  In the midplane, the D/H ratios fall off rapidly
with decreasing radius, and the more highly deuterated forms are
present only at larger radii (Fig.~\ref{fig:DH_h3+}).  In the outer
disk ($R$ $>$ 100 AU), the models of W07 and
\cite{cd05} found that \chem{HD_2^+} and \chem{D_3^+} are produced
with significant abundances and therefore play an important role in
the deuterium chemistry of this region.  In the inner disk the
temperature is higher and the abundances of these isotopologues are much
lower.  Deuteration is therefore driven only by H$_2$D$^+$.

In common with \chem{H_3^+}, the deuteration of CH$_3^+$ and atomic
hydrogen peak in the midplane at R $>$ 15 AU.  The abundance of
CH$_3^+$ is very small in this region, however, and its 
isotopologues play
little part in driving the deuteration.  Deuteration by reaction with
atomic D is most efficient at $R$ $>$ 17 AU, and inside of 5 AU.  The
abundance of atomic D and H in the midplane is relatively high because
of the high grain temperature.  This reduces the residence time of the
atoms on the grains so that they are more likely to desorb than to
react with other atoms or radicals.

Other species do not necessarily show a decrease in D/H ratio with
decreasing radius (and increasing temperature).  Many molecules retain
the D/H ratio that they had when they formed in the cold molecular
cloud or in the colder regions of the disk, as they are transported
inwards.  The transport process occurs on a shorter timescale than the
chemical reactions that would destroy the deuteration.  For example,
the deuteration of H$_2$CO and NH$_3$ reflects that set in the ices
during the molecular cloud phase and is not affected by chemistry in
the disk. 

\begin{figure}
\caption{\label{fig:deut_disk}The molecular D/H ratios
in the inner disk as calculated in Model 1.  The D/H ratios are only shown
in regions where the fractional abundance of the non--deuterated
species is greater than 10$^{-12}$ (the exception is CH$_3^+$
for which D/H ratios are shown across all of the disk
regardless of the molecular abundances).  Those species which
are abundant at $R$ $>$ 10 AU, have radial scales running
from 0.5 to 30 AU.  Other species are only shown for the inner
10 AU.  The atomic D/H ratio and the \chem{CH_2D^+}/\chem{CH_3^+}
ratios are important for driving molecular deuteration of
some molecules.  Relatively high molecular deuteration
is retained even in warm regions, since the ratios are
set in the cooler outer regions of the disk and the
material is transported inwards at a faster rate than
the chemical reactions that would destroy the deuteration.}

\includegraphics[scale=0.4]{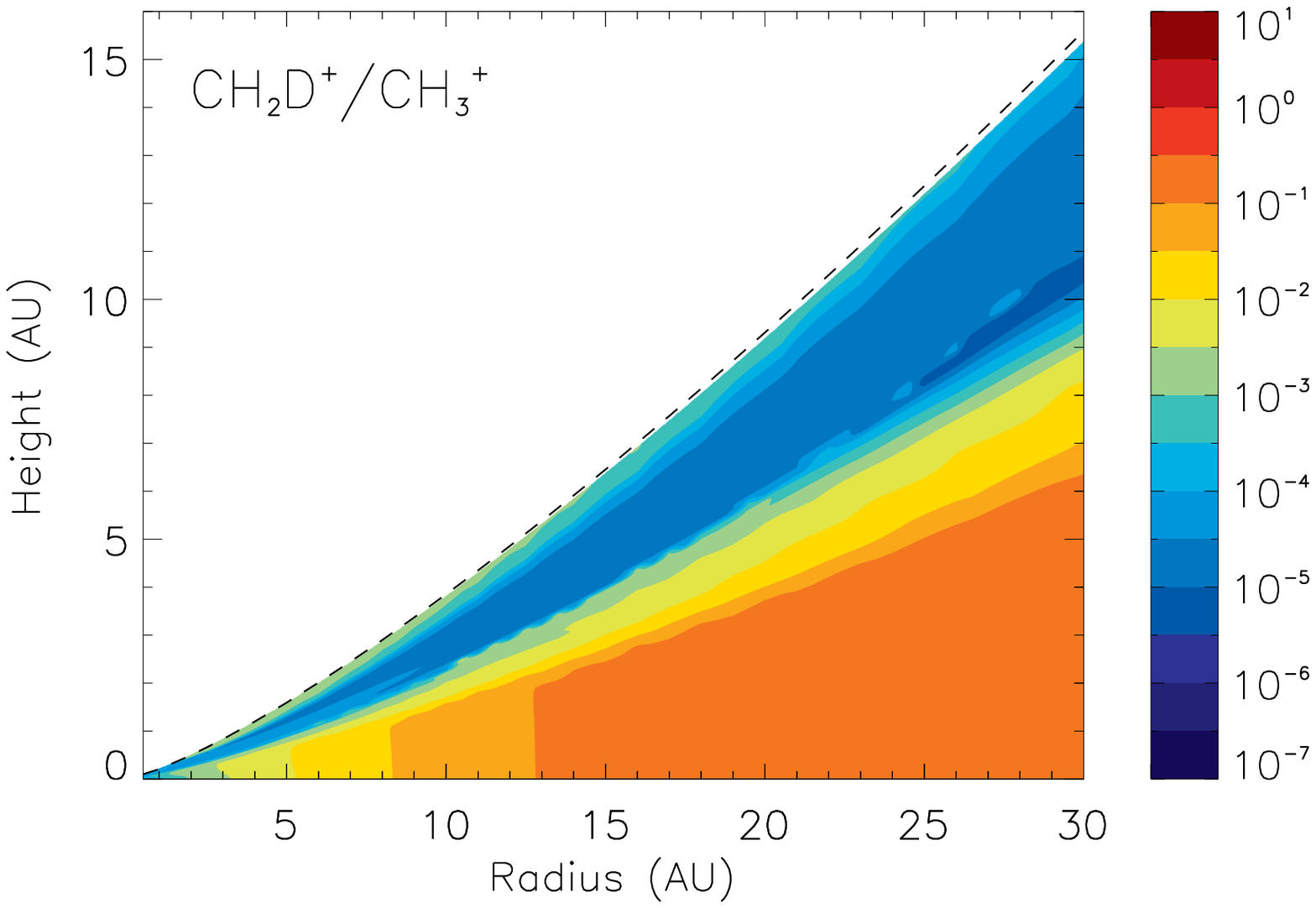}
\includegraphics[scale=0.4]{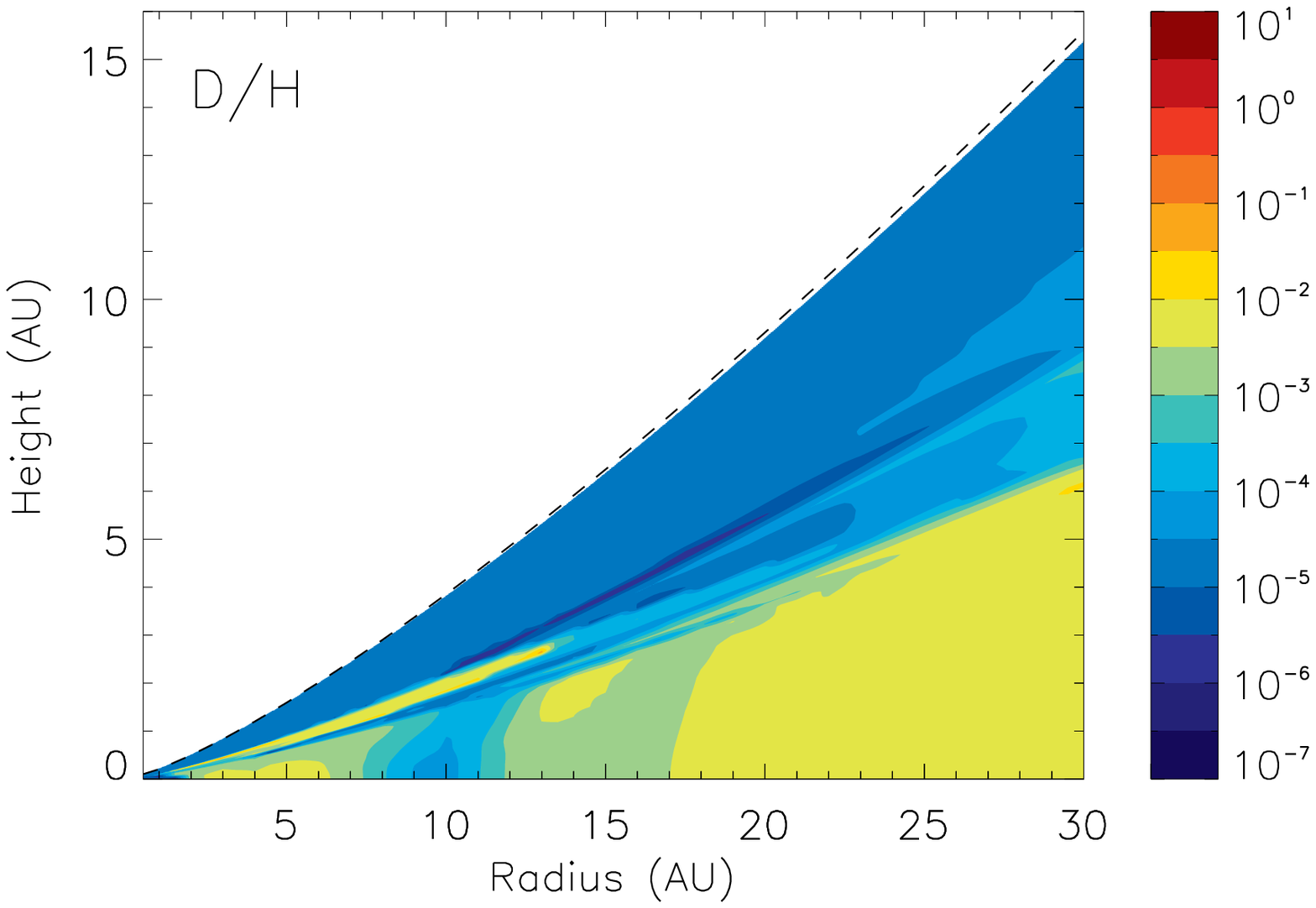}\\
\includegraphics[scale=0.4]{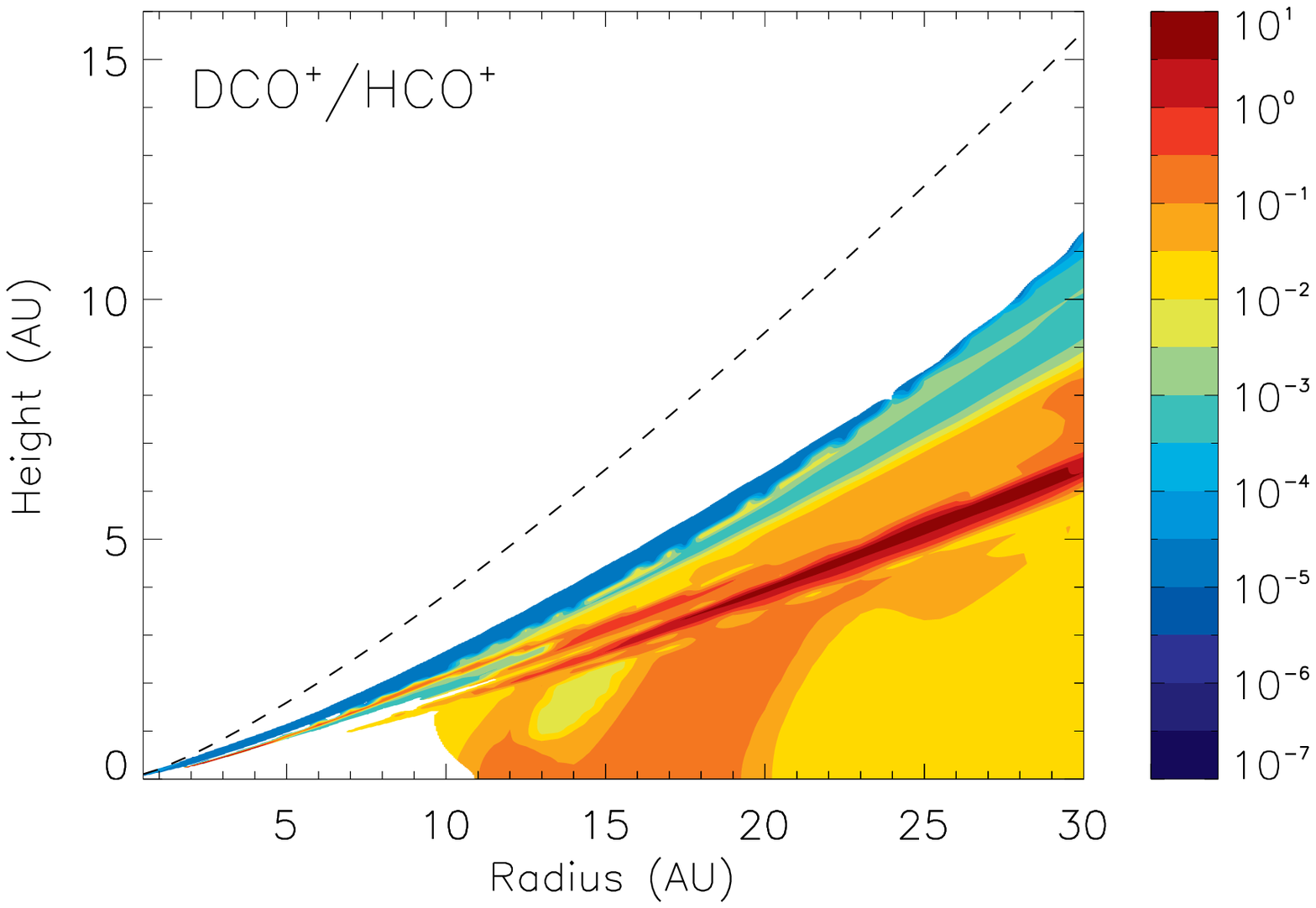}
\includegraphics[scale=0.4]{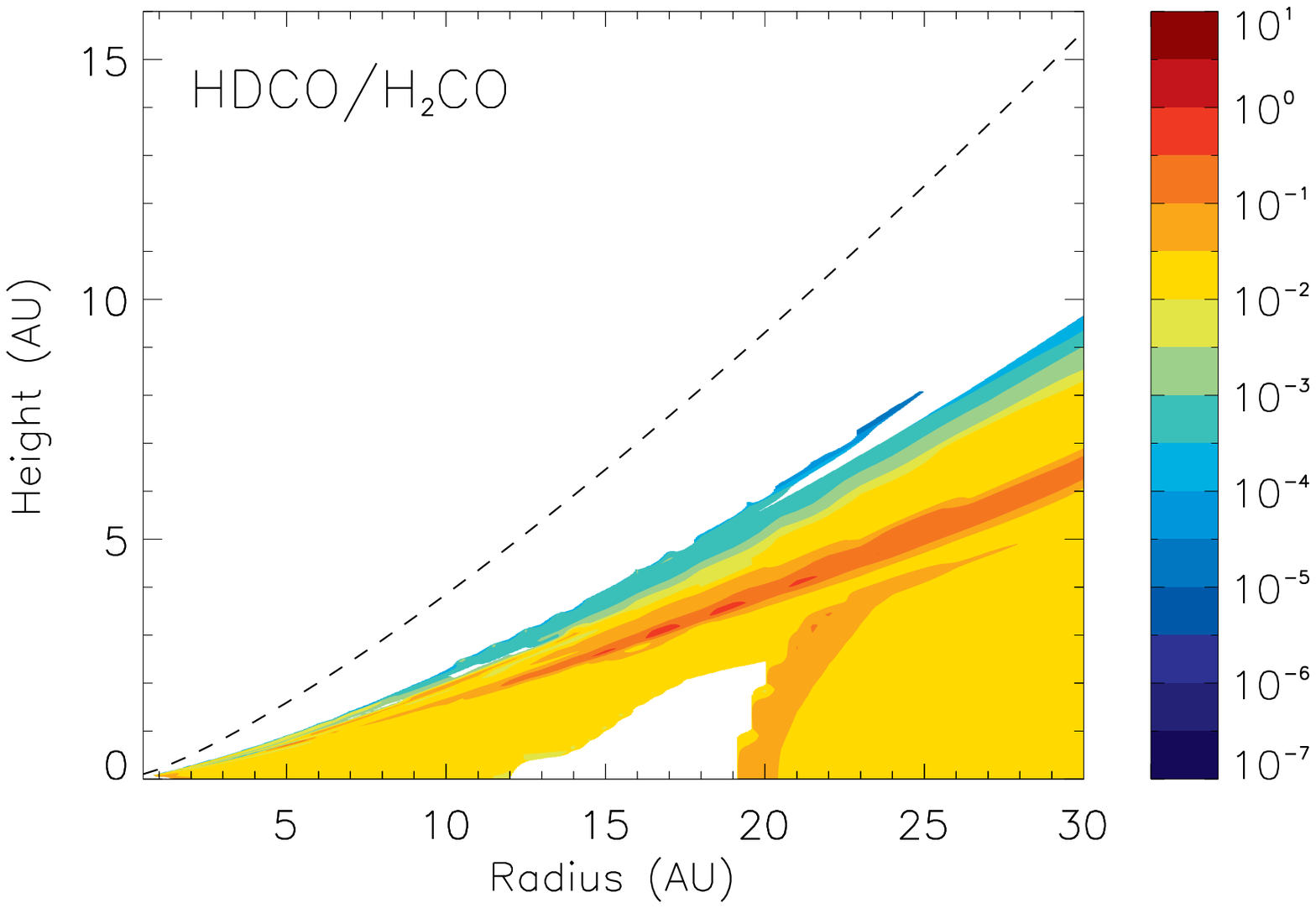}\\
\end{figure}

The deuteration of gas phase water is less than that seen in the ice.
Water is desorbed in two regions, in the surface layers and at $R$ $<$
2 AU.  At high $z$ HDO is formed by
\begin{equation}
\hbox{OD} + \hbox{H$_2$} \longrightarrow \hbox{HDO} + \hbox{H}
\end{equation}
and destroyed by
\begin{equation}
\hbox{HDO} + \hbox{H}\longrightarrow \hbox{OH} + \hbox{HD}
\end{equation}
Similar reactions affect H$_2$O, but since the destruction product is
OH, this is able to cycle back to form more water by reaction with
H$_2$. Hence a decrease in HDO/H$_2$O is seen at high $z$.  Below this
the ratio is higher (yellow stripe in Fig.~\ref{fig:deut_disk})
but it is still lower than that set in the grain mantles, again due to
the formation of OH from the destruction of HDO.

\addtocounter{figure}{-1}
\begin{figure}
\caption{{\it cont}}
\includegraphics[scale=0.4]{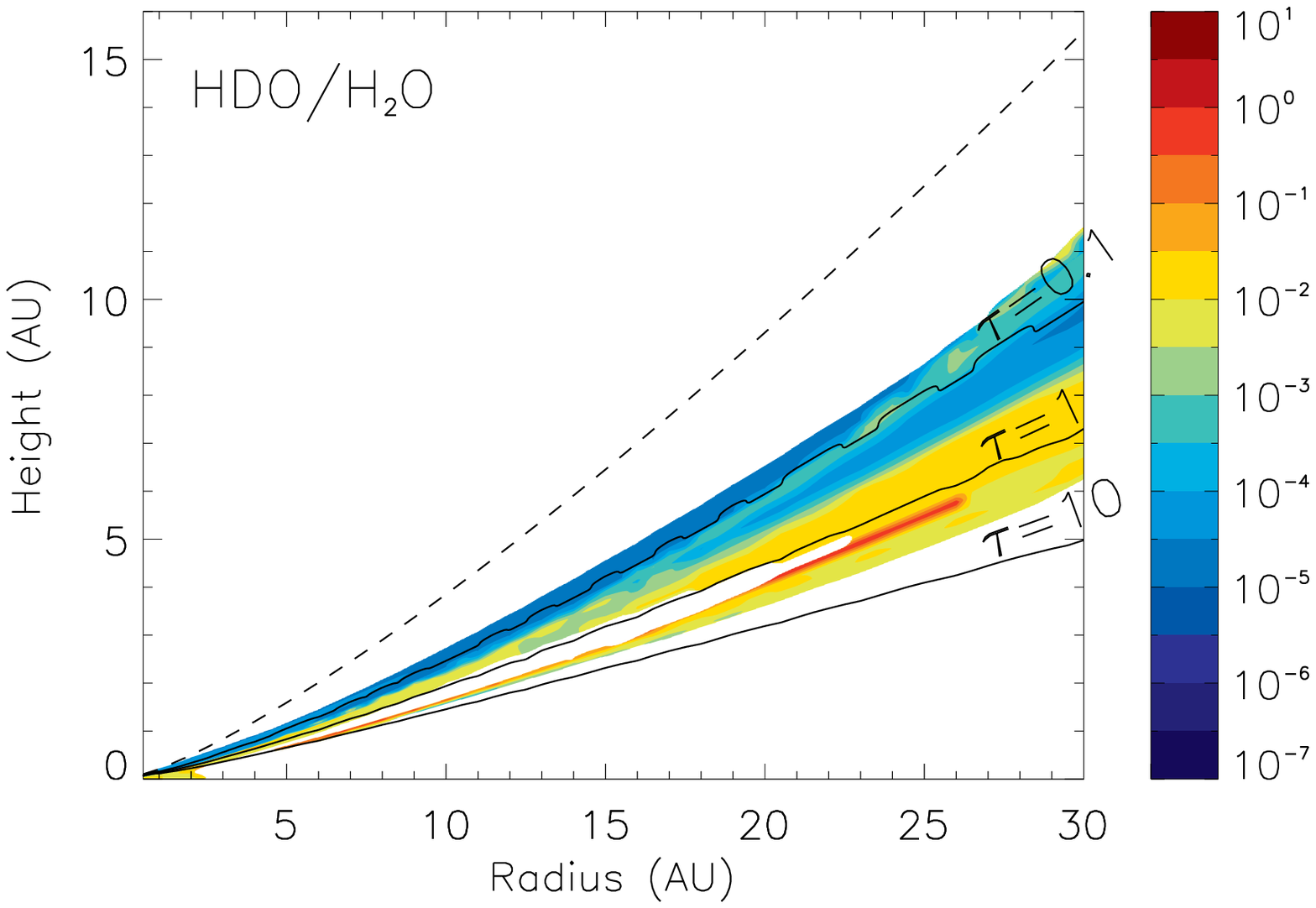}
\includegraphics[scale=0.4]{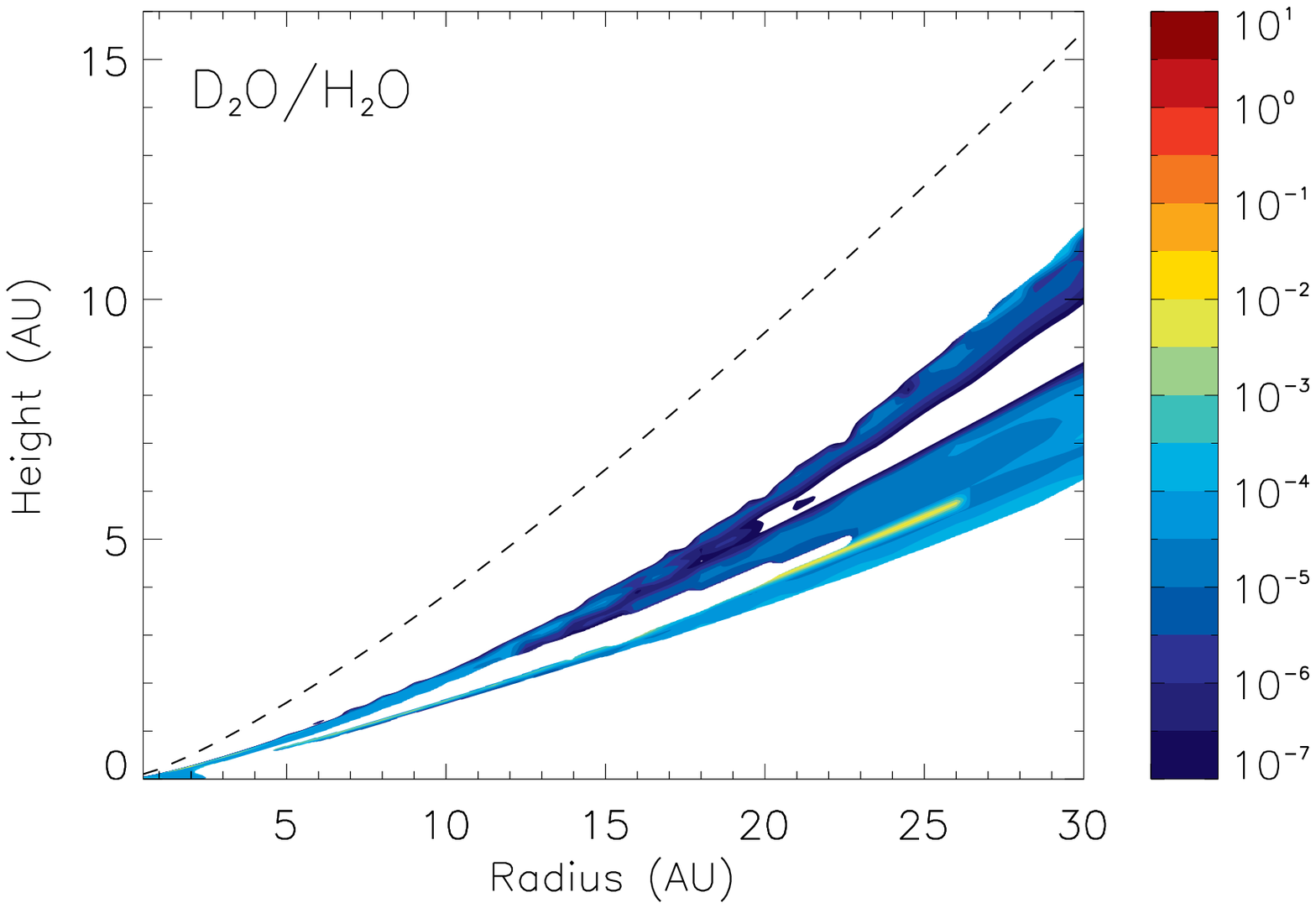}\\
\includegraphics[scale=0.4]{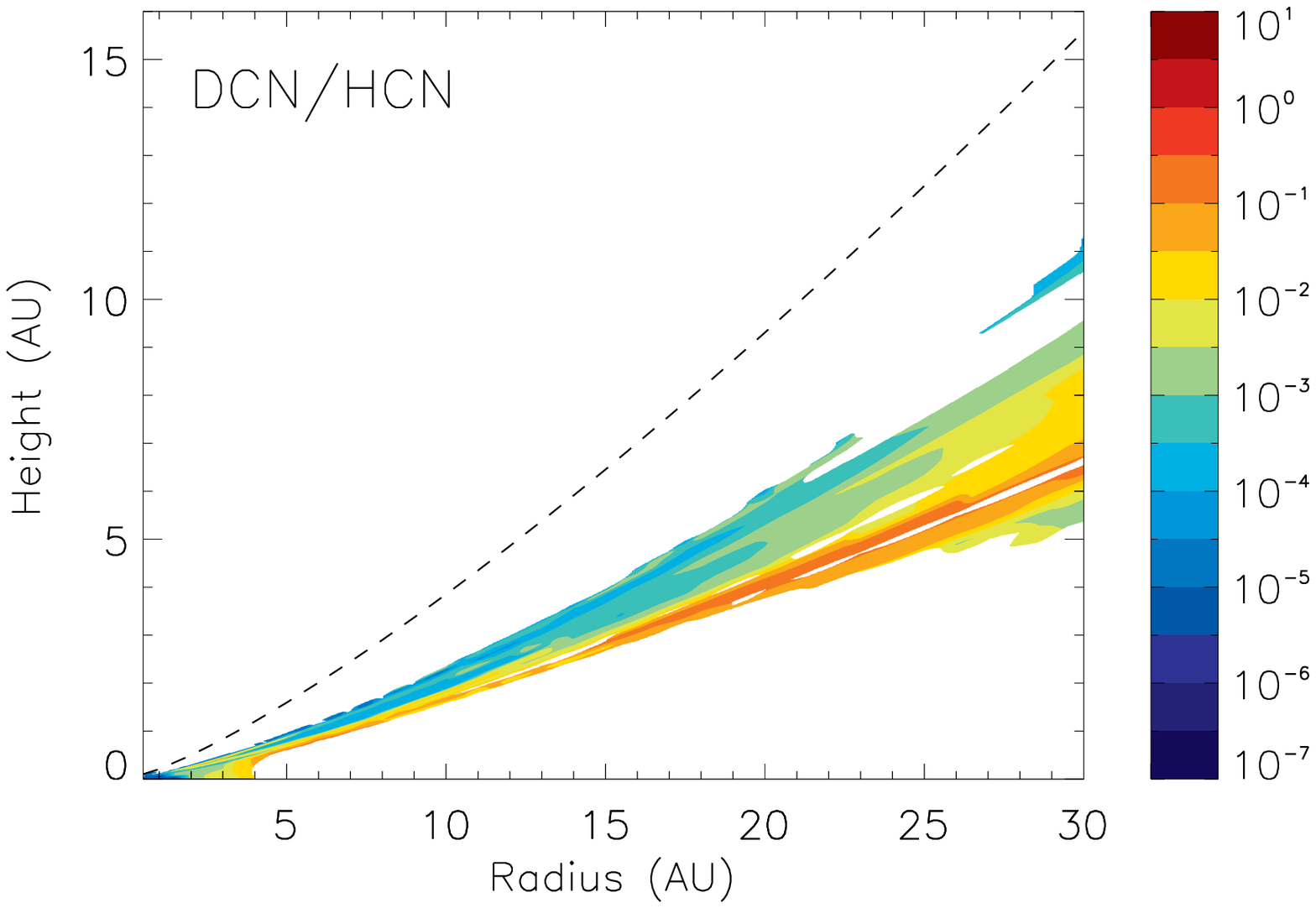}
\includegraphics[scale=0.4]{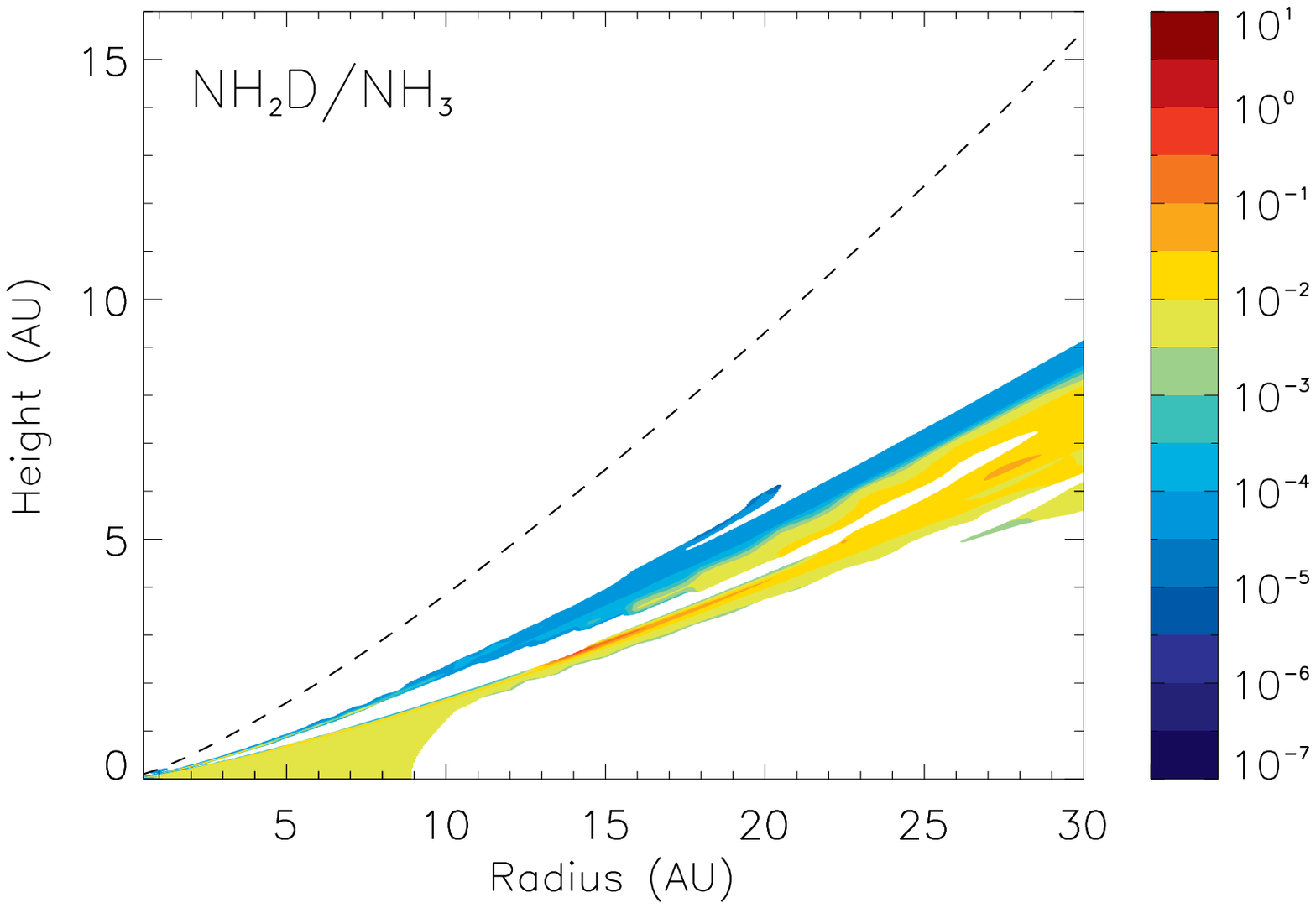}
\end{figure}

The increase in DCO$^+$/HCO$^+$ in the midplane between 20 and 10 AU 
is due to the increase in grain temperature which reduces
the efficiency of H$_2$, HD and D$_2$ formation on grains, and
therefore increases the abundance of deuterium atoms in the gas phase.
The deuterium atoms react with HCO$^+$ to form DCO$^+$.  Inside of 
10 AU the abundances of both DCO$^+$ and HCO$^+$ are very small
because they are destroyed by reaction with hydrocarbons
such as C$_3$H$_4$ and C$_2$H$_2$.

\begin{deluxetable}{llllll}
\tablecolumns{6}
\tablewidth{0pt}
\tablecaption{\label{tab:dhratio}The molecular
D/H ratios in the midplane for Model 1. \nodata indicates that
the molecular fractional abundances are very low ($<$ 10$^{-15}$).
X:gr indicates a grain mantle species}
\tablehead{
\colhead{Molecule} & \colhead{Input} & \colhead{25 AU} & \colhead{10 AU} & 
\colhead{5 AU} & \colhead{1 AU} 
}
\startdata
D/H                  & 2.00 (-2)  & 5.25 (-3) & 6.70 (-5) & 5.50 (-3) & 5.50 (-6) \\
DCO$^+$/HCO$^+$      & 9.00 (-2)  & 1.89 (-2) & 2.30 (-2) & \nodata   & \nodata   \\
CH$_2$D/CH$_3^+$     & 1.10 (-1)  &  \nodata  & \nodata   & \nodata   & \nodata   \\
H$_2$D$^+$/H$_3^+$   & 2.00 (-1)  & \nodata   & \nodata   & \nodata   & \nodata   \\
HDCO/H$_2$CO         & 7.00 (-2)  & 1.69 (-2) & 1.07 (-2) & 1.60 (-2) & \nodata   \\
HDO/H$_2$O           & 1.00 (-1)  & \nodata   & \nodata   & \nodata   & 1.00 (-2) \\
D$_2$O/H$_2$O        & 1.00 (-3)  & \nodata   & \nodata   & \nodata   & 1.06 (-4) \\
DCN/HCN              & 3.20 (-2)  & \nodata   & \nodata   & \nodata   & 5.53 (-6) \\
NH$_2$D/NH$_3$       & 3.10 (-2)  & \nodata   & \nodata   & 4.09 (-3) & 4.50 (-3) \\
HDO:gr/H$_2$O:gr     & 9.00 (-3)  & 9.34 (-3) & 9.50 (-3) & 9.60 (-3) & 1.07 (-2) \\
DCN:gr/HCN:gr        & 9.00 (-3)  & 3.00 (-2) & 2.96 (-2) & 2.99 (-3) & \nodata   \\
HDCO:gr/H$_2$CO:gr   & 4.80 (-2)  & 2.96 (-2) & 1.54 (-2) & 2.17 (-3) & \nodata   \\
NH$_2$D:gr/NH$_3$:gr & 5.13 (-3)  & 5.34 (-3) & 5.26 (-3) & 5.26 (-3) & \nodata   \\
\enddata    
\end{deluxetable}

\clearpage

\subsection{Ionization fraction}

Disks around young stars provide a means of dissipating angular
momentum and they regulate the rate at which mass accretes onto the
protostar.  The evolution of the star-disk system is controlled by
angular momentum transport, but the mechanism by which this is
achieved is unclear.  It is likely to involve turbulence
\cite[see][and references therein]{mv84} since molecular viscosity is
too low to have much effect.  The mechanism driving the turbulence has
not yet been identified but several processes have been suggested
e.g.\ \cite{lp80,rg82,dubrulle93,
  bhs96,bh91,li00,kb03,klahr04,dubrulle05,sr05}.  Magneto-rotational
instability (MRI) is widely regarded as the most promising mechanism
for driving turbulence \citep{bh91,hb91,hawley96}.  In MRI, magnetic
field lines linking gas at different distances from the star are
stretched due to the decrease in orbital frequency with
radius. Magnetic tension forces decelerate the inner gas, which
spirals inwards and accelerates the outer gas, which spirals outwards
so that the bend in the field line grows with time.  Instability
occurs if the field lines are frozen in the gas.  MRI requires the
ionization fraction $x$(e) to be above a minimum value ($>$ 10$^{-12}$
at 1 AU in the midplane of the MMSN
;\cite{is05}).  There is some debate as to whether MRI is
applicable in disks, where high densities and low temperatures lead to
low ionization levels \citep{bb94, gammie96}.  Many models have found
that active regions (where the ionization level is high enough to
drive MRI) can exist in the disk along with dead zones where the
magnetic field is not coupled to the gas
e.g.\ \cite{sano00,fromang02,mp03,semenov04, in06a,in06b,in06c}.  The
size and location of the dead zone varies with the model, but most
cover the region of terrestrial planet formation and it is thought
that dead zones may also play a role in halting the inward migration
of planets \citep{matsumura07}.

\begin{figure}
\caption{\label{fig:alp0.01_xe}{\it Left:} (a) The ionization level in 
disk of Model 1.  A dead zone (defined as the region
where $x(e)$ $<$ 10$^{-12}$) exists in the midplane
at $R$ $<$ 10 AU; {\it Right:} (b) The ionization level in the
higher mass disk of Model 2, showing an increase in size
of the dead zone with increasing disk mass}.
\plottwo{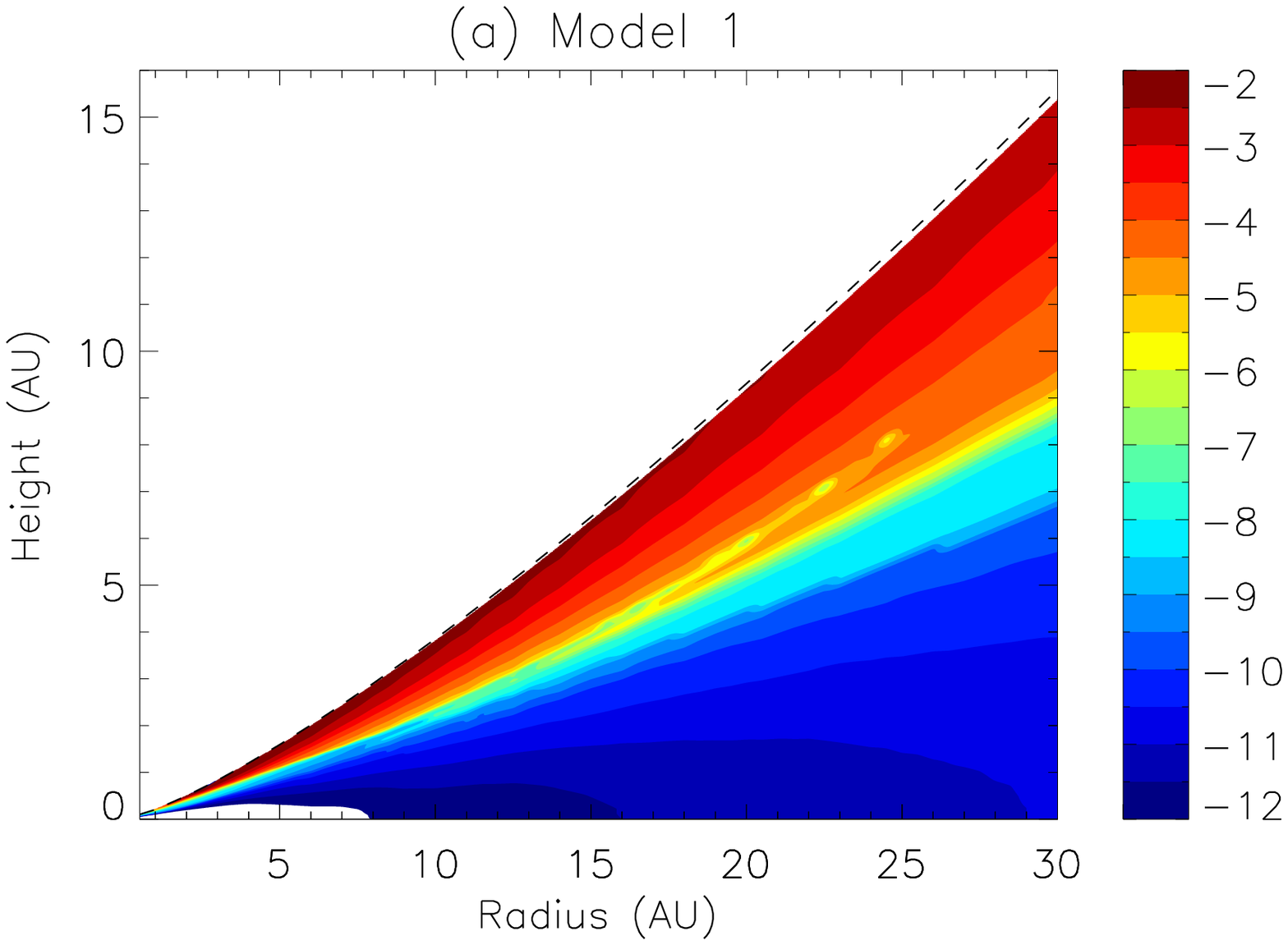}{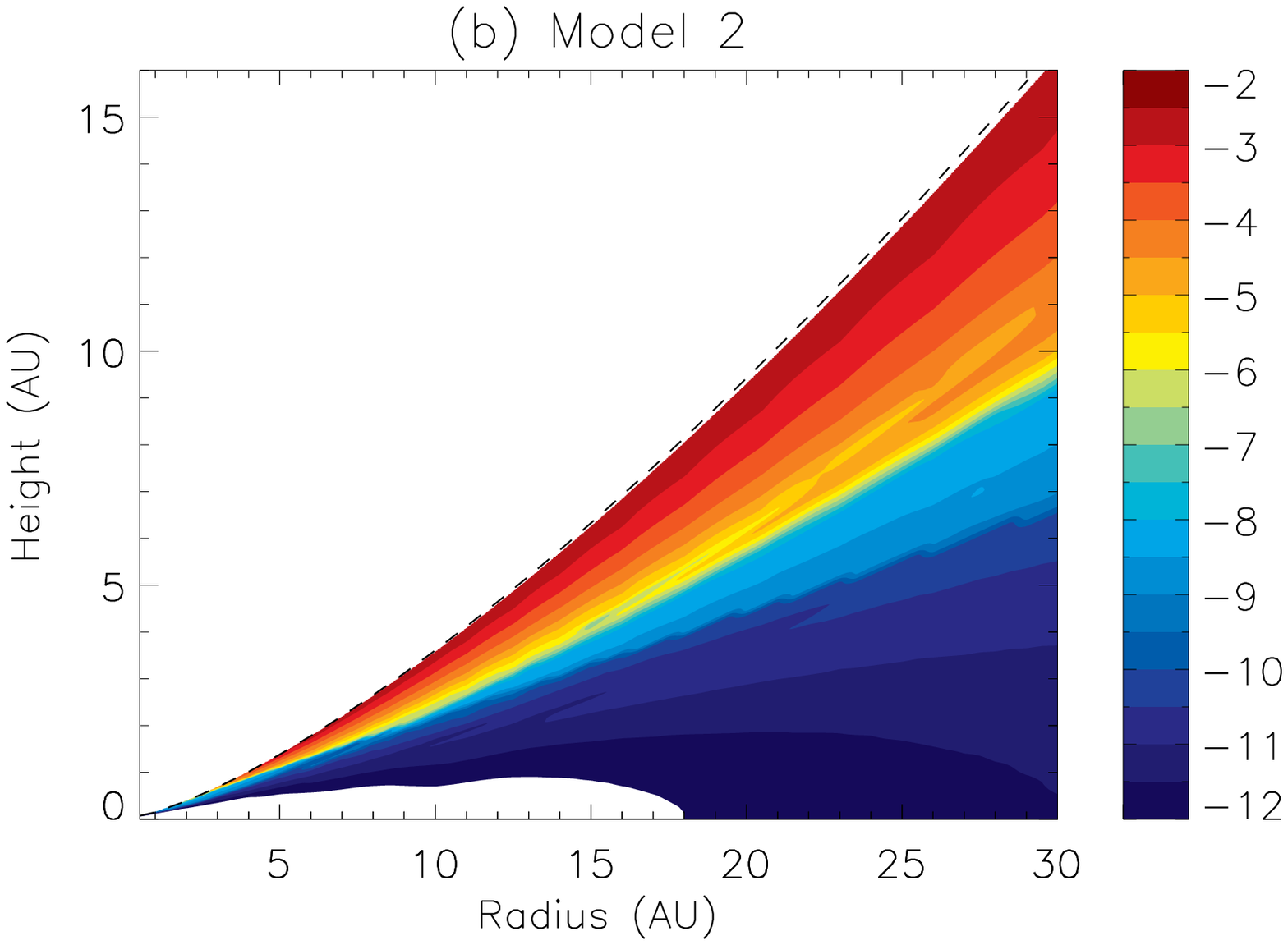}
\end{figure}

The total ionization level in Model 1 is shown in
Figure~\ref{fig:alp0.01_xe}(a).  A dead zone (defined as a region
where $x(e)$ $<$ 10$^{-12}$) is present in the midplane inside of
$\sim$ 10 AU.  The extent of the dead zone is sensitive to the mass of
the disk -- increasing the mass by increasing the value of $\alpha$ to
0.025 (Model 2) extends the dead zone out to 18 AU in the midplane
(Fig.~\ref{fig:alp0.01_xe}(b)).  This is a consequence of the
higher densities in Model 2, which both increase the rate at which ions and
electrons collide and neutralize and reduce the cosmic ray
ionization rate in the midplane.  The effects of an increase in disk
mass on the abundance distributions of other molecules are discussed
in Section~\ref{sec:diskmass}.

\begin{figure}
\caption{\label{fig:ion}The distribution of ions in Model 1.  HCO$^+$
  is the dominant ion at $R$ $>$ 8 AU, CH$_3$CO$^+$ and hydrocarbon
  chain ions dominating inwards of this.  At the surface the main ion
  is H$^+$, with a layer of C$^+$ below.  Hydrocarbon ions exist in a
  layer below the C$^+$. A similar distribution is seen in Model
  2. }
\plotone{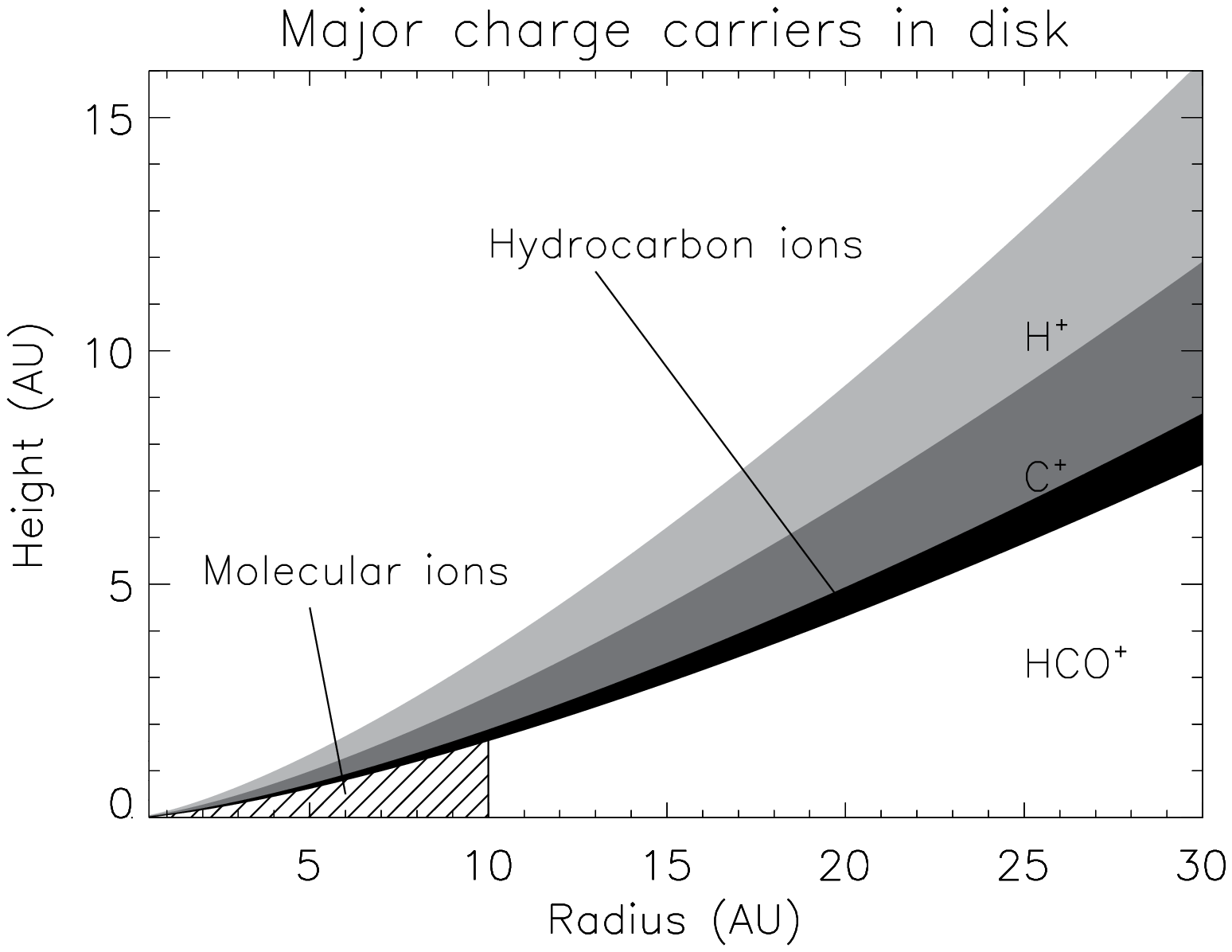}
\end{figure}

The dominant ion varies with position in the disk.  
Fig~\ref{fig:ion} illustrates this for Model 1.
At the surface, H$^+$ is the main ion at
all radii.  It is formed efficiently by X-ray ionization of H atoms.
Below this is a layer of C$^+$ which lies above a thin layer of
hydrocarbon ions e.g.\ C$_4$H$_2^+$ and C$_3$H$_3^+$.  The hydrocarbon
ions dominate in a region where there are still some photons and where
the high abundance of C$^+$ means that hydrocarbons and their ions can
form efficiently.  HCO$^+$ is the main charge carrier below the
hydrocarbon layer for $R$ $>$ 8 AU.  It is destroyed at $R$ $<$ 8 AU
by reaction with hydrocarbon chains e.g.\ C$_3$H$_4$ and C$_2$H$_2$
producing hydrocarbon ions that are the main contributors to the
charge in the central regions.  HCO$^+$ also contributes to the
production of another abundant ion in the inner regions --
CH$_3$CO$^+$ -- by reacting with CH$_2$CO.

\clearpage

\subsection{\label{sec:diskmass}Effect of disk mass} 

Model 1 described above has $\alpha$ = 0.01 and a total disk mass of
0.032 M$_\odot$ inside of 300AU.  The surface
density is 38.8 g cm$^{-2}$ at 5 AU.  This is somewhat lower than that
of the MMSN model and we have therefore also considered
a higher mass model, again provided to us by Dr.\ P.\ d'Alessio.  Model
2 has a higher value of $\alpha$ (= 0.025) resulting in a higher surface
density (150 g cm$^{-2}$ at 5 AU).

\begin{figure}
\caption{\label{fig:model2}Fractional abundances in the inner disk
for Model 2 ($\alpha$ = 0.025).  The first few figures, from CO to 
CN, show those
molecules which are present throughout the disk, the later ones show
those molecules which are only present in the gas phase in the
inner 10 AU. }
\includegraphics{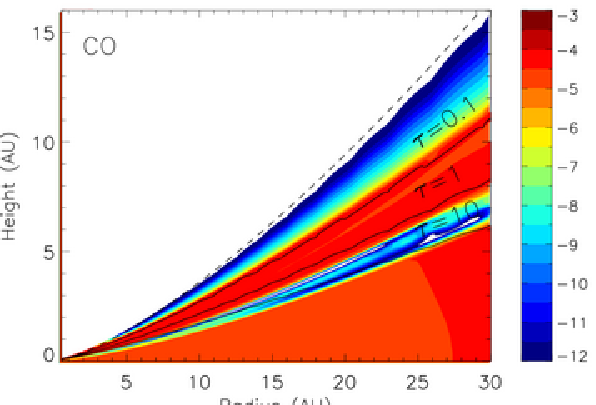}
\includegraphics{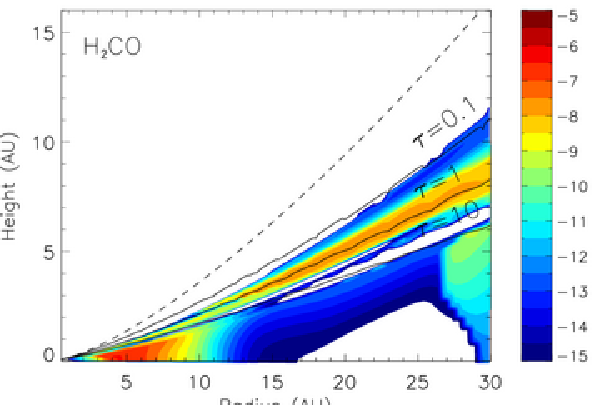}\\
\includegraphics{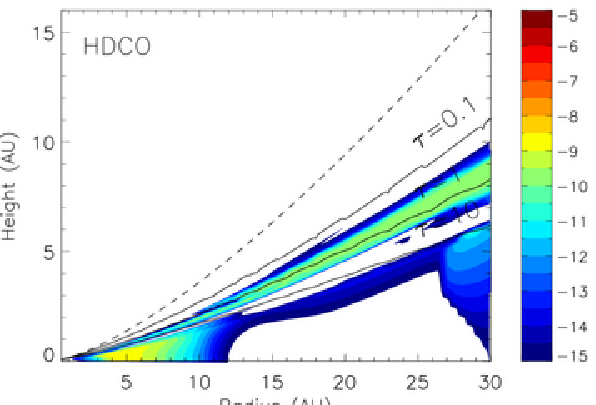}
\includegraphics{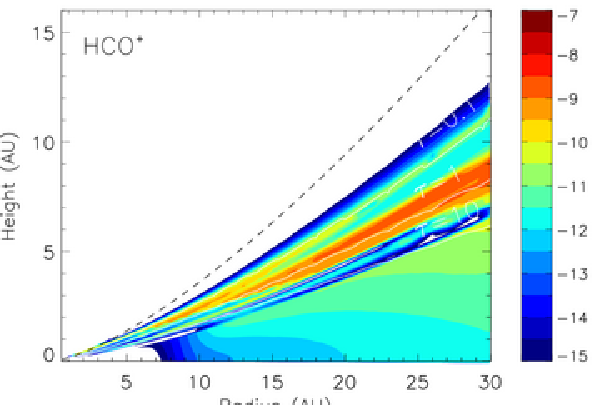}\\
\includegraphics{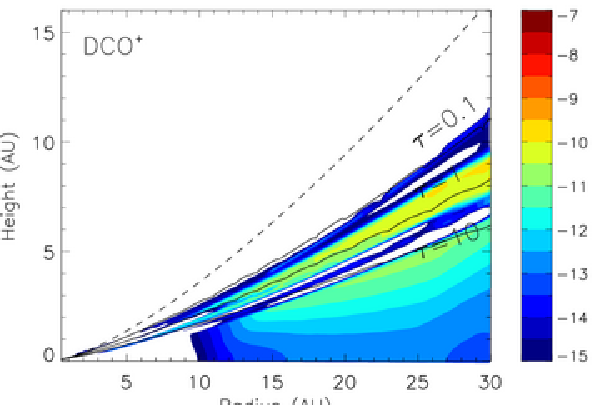}
\includegraphics{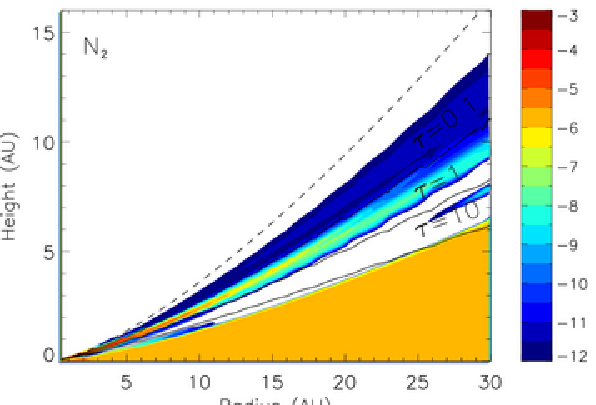}\\
\end{figure}

We find that the mass of the disk does not greatly affect
the molecular distributions in the inner disk (Fig.~\ref{fig:model2}).
The higher mass does result in a narrower surface layer of molecules e.g.\
for CN, H$_2$O etc, and the slightly higher midplane grain temperature
in Model 2 means that molecules desorb at slightly larger radii.

The deuteration of some molecules varies with the disk mass.  
Figure~\ref{fig:comp_deut_col} shows the radial variation of 
column density deuteration for both models.  HDO/H$_2$O and
DCO$^+$/HCO$^+$ are similar in the two models, but for
DCN/HCN and HDCO/H$_2$CO there is a drop in deuteration with the
higher disk mass. DCN and HCN exist in a layer above the midplane
in both models, with the Model 2 layer being at slightly higher $z$
due to the increase in optical depth of this model.
In both models the abundance of DCN is mainly dependent on the
ratio of D atoms which react with HNC to form DCN, and H atoms
which destroy DCN forming HCN.  The
ratio of atomic D/H is higher in Model 1 than Model 2, leading to
a higher DCN/HCN ratio.

\addtocounter{figure}{-1}
\begin{figure}
\caption{{\it cont}}
\includegraphics{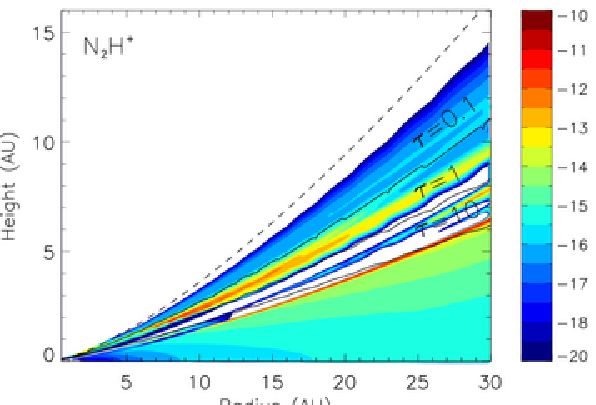}
\includegraphics{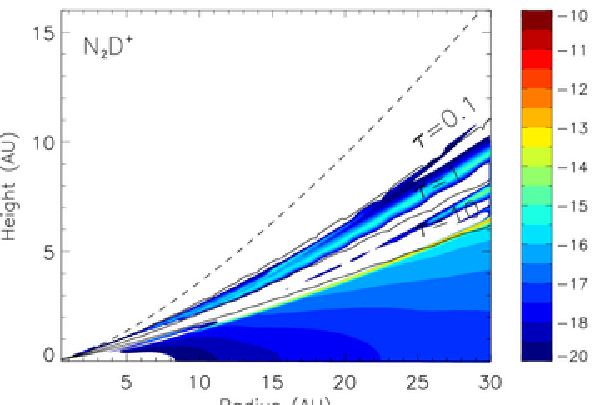}\\
\includegraphics{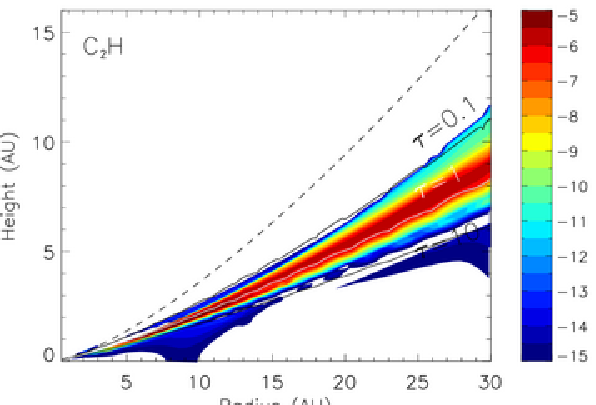}
\includegraphics{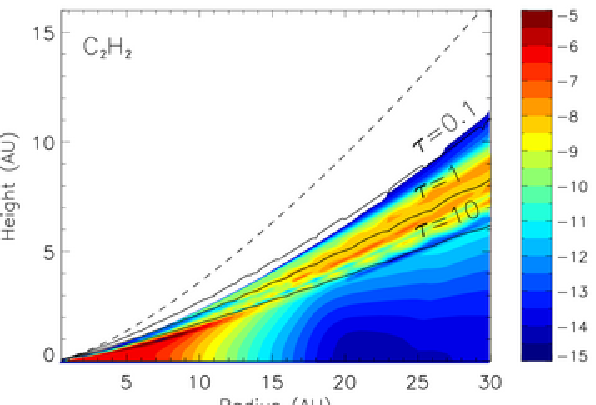}\\
\includegraphics{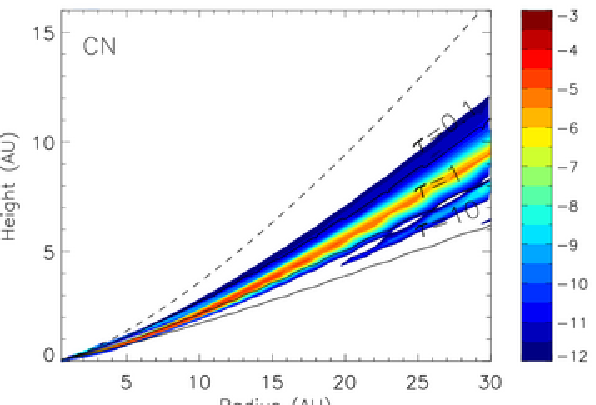}
\includegraphics{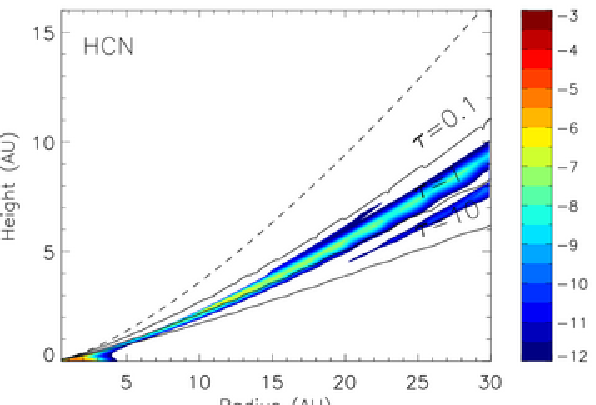}\\
\includegraphics{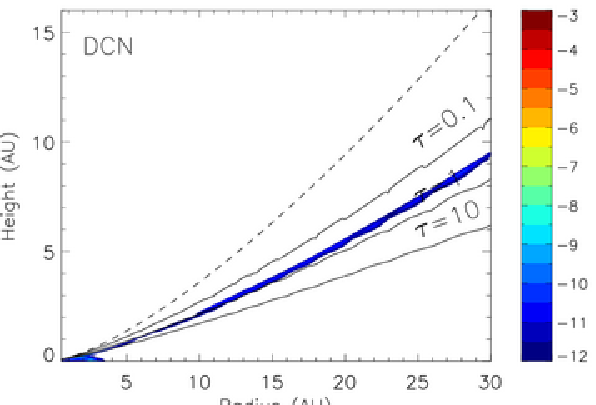}
\includegraphics{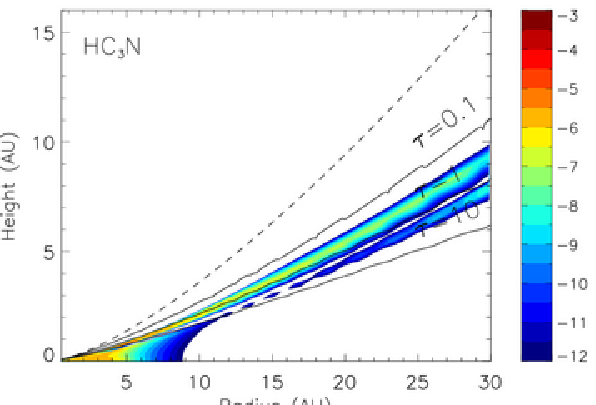}\\
\end{figure}

\addtocounter{figure}{-1}
\begin{figure}
\caption{{\it cont}}
\includegraphics{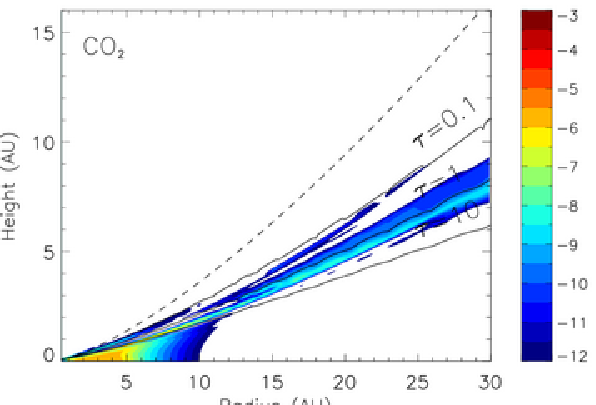}
\includegraphics{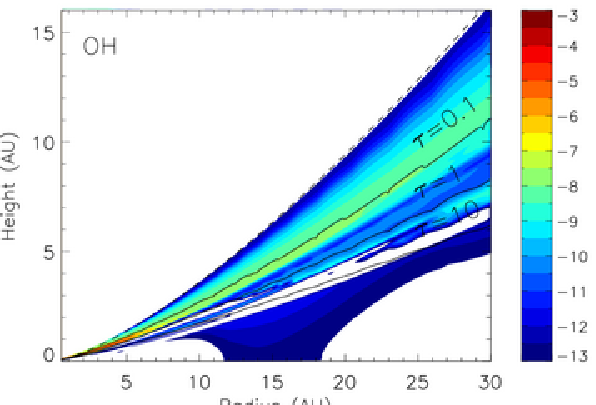}\\
\includegraphics{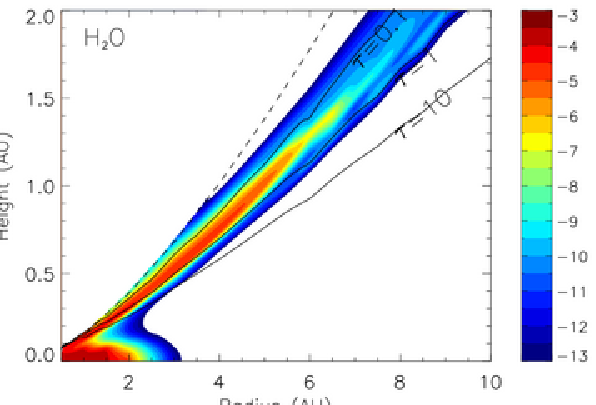}
\includegraphics{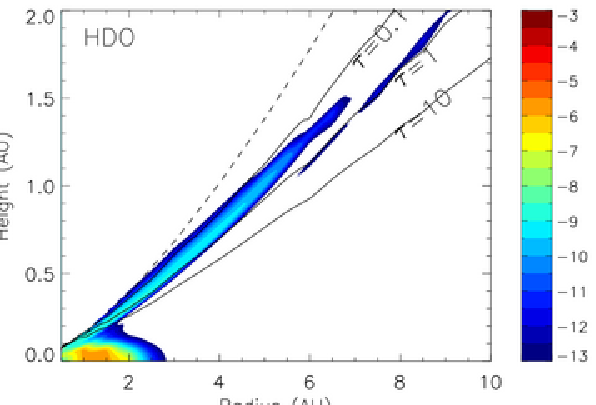}\\
\includegraphics{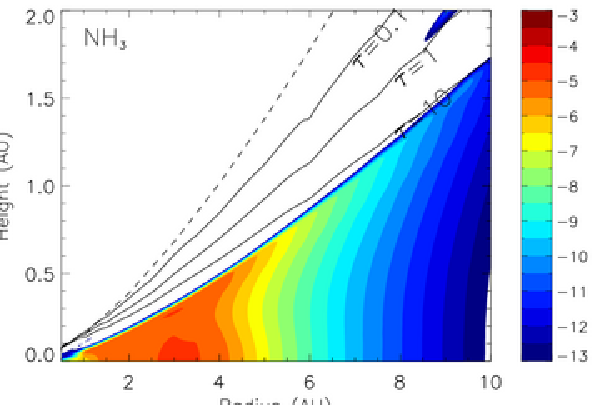}
\includegraphics{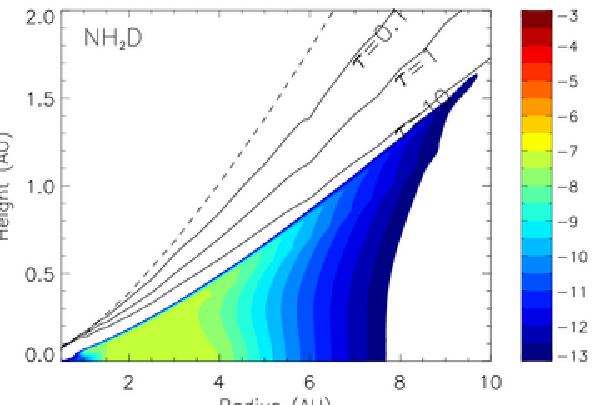}
\end{figure}

\begin{figure}
\caption{\label{fig:comp_deut_col}The radial variation of
column density deuteration (i.e.\ $N$(XD)/$N$(XH)) for 
Model 1 (dashed line) and 2 (solid line).}
\plottwo{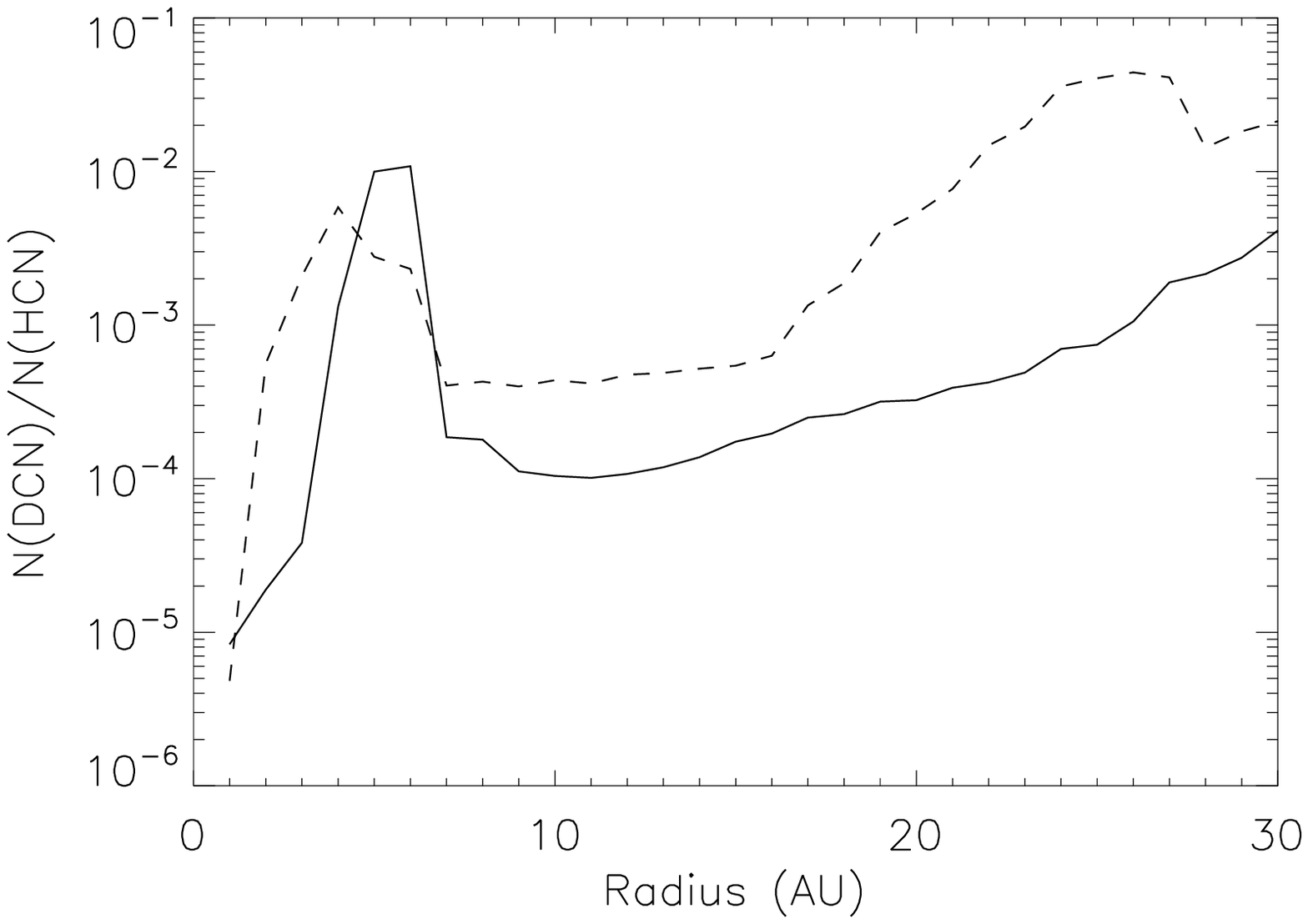}{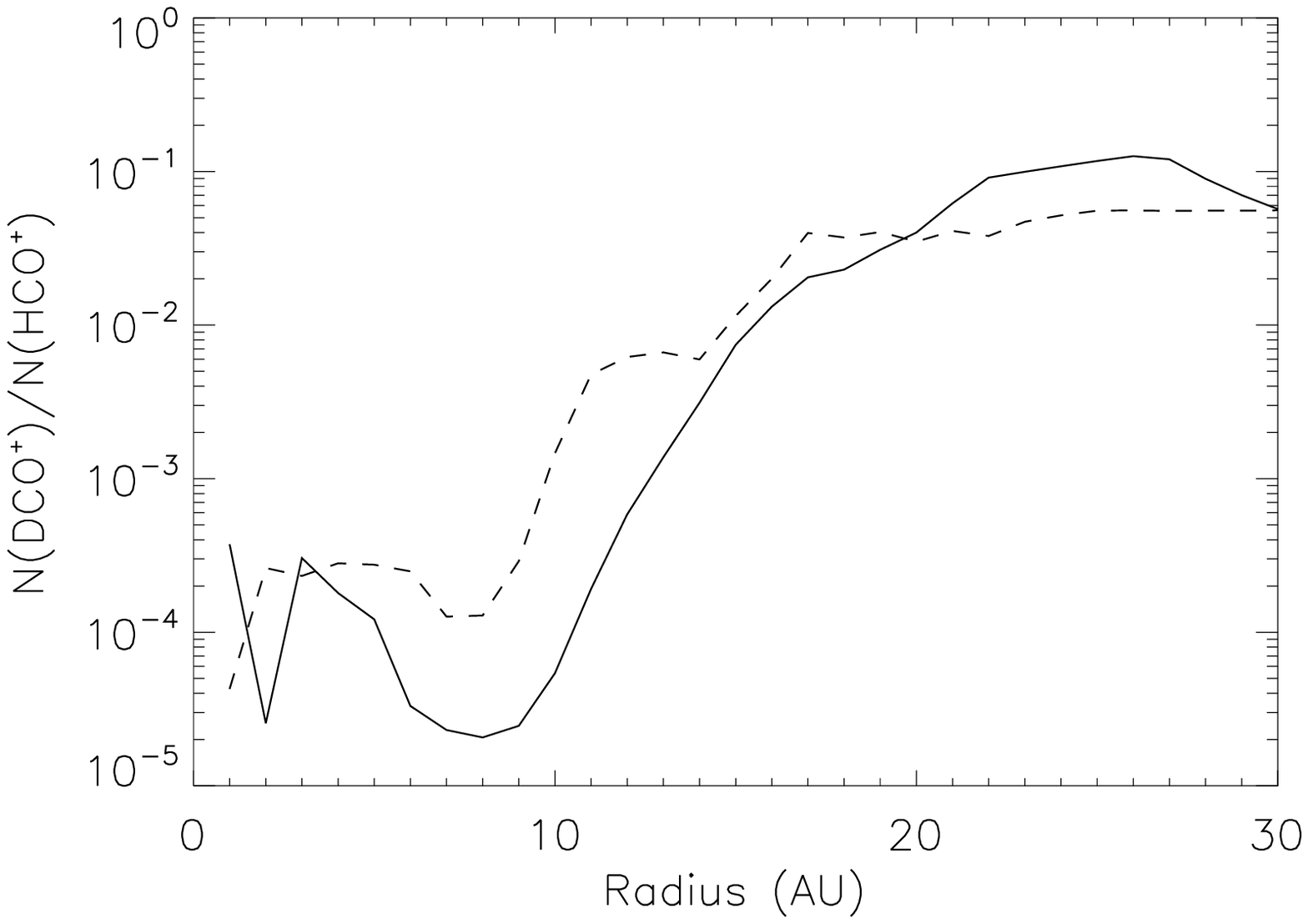}\\
\plottwo{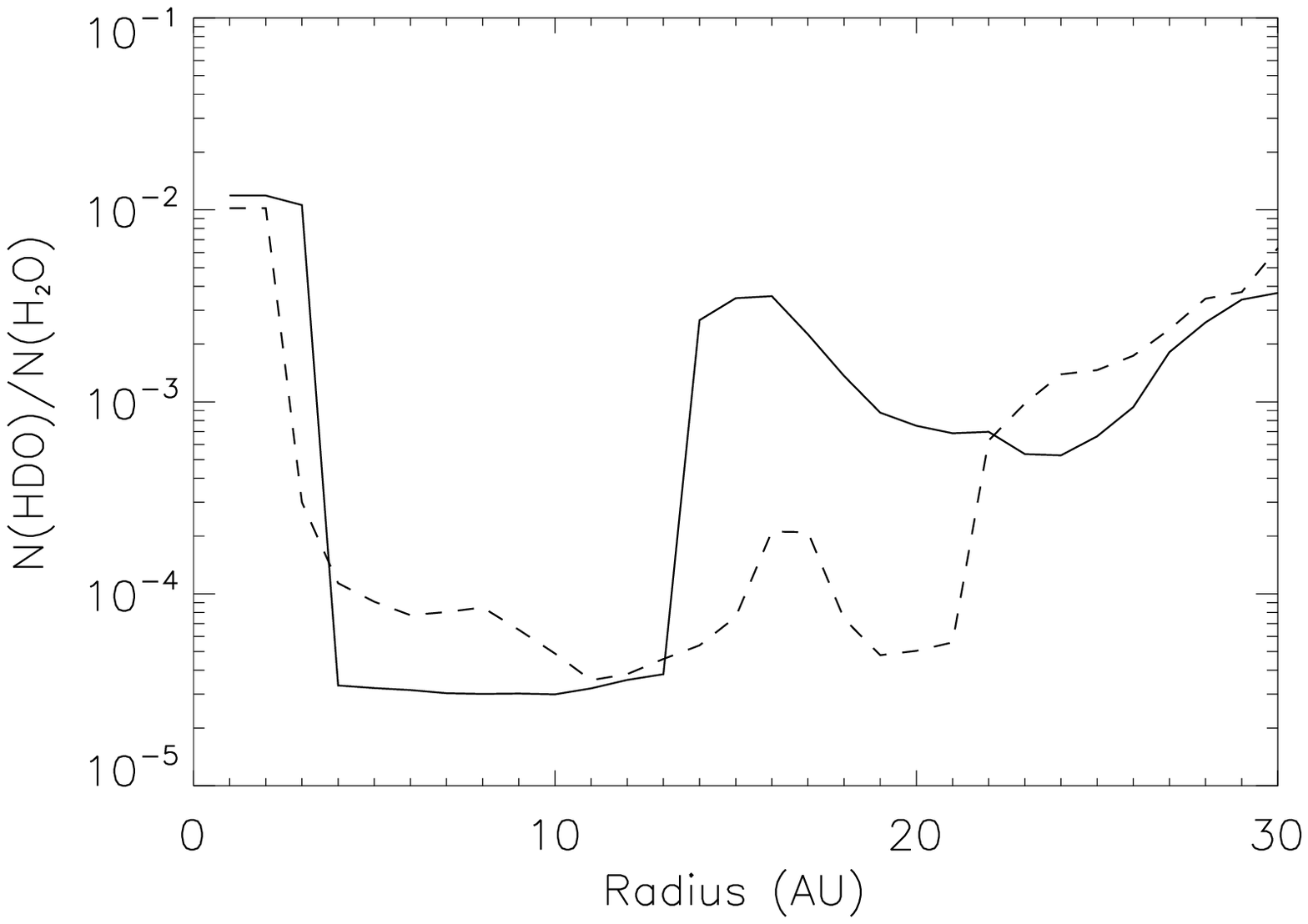}{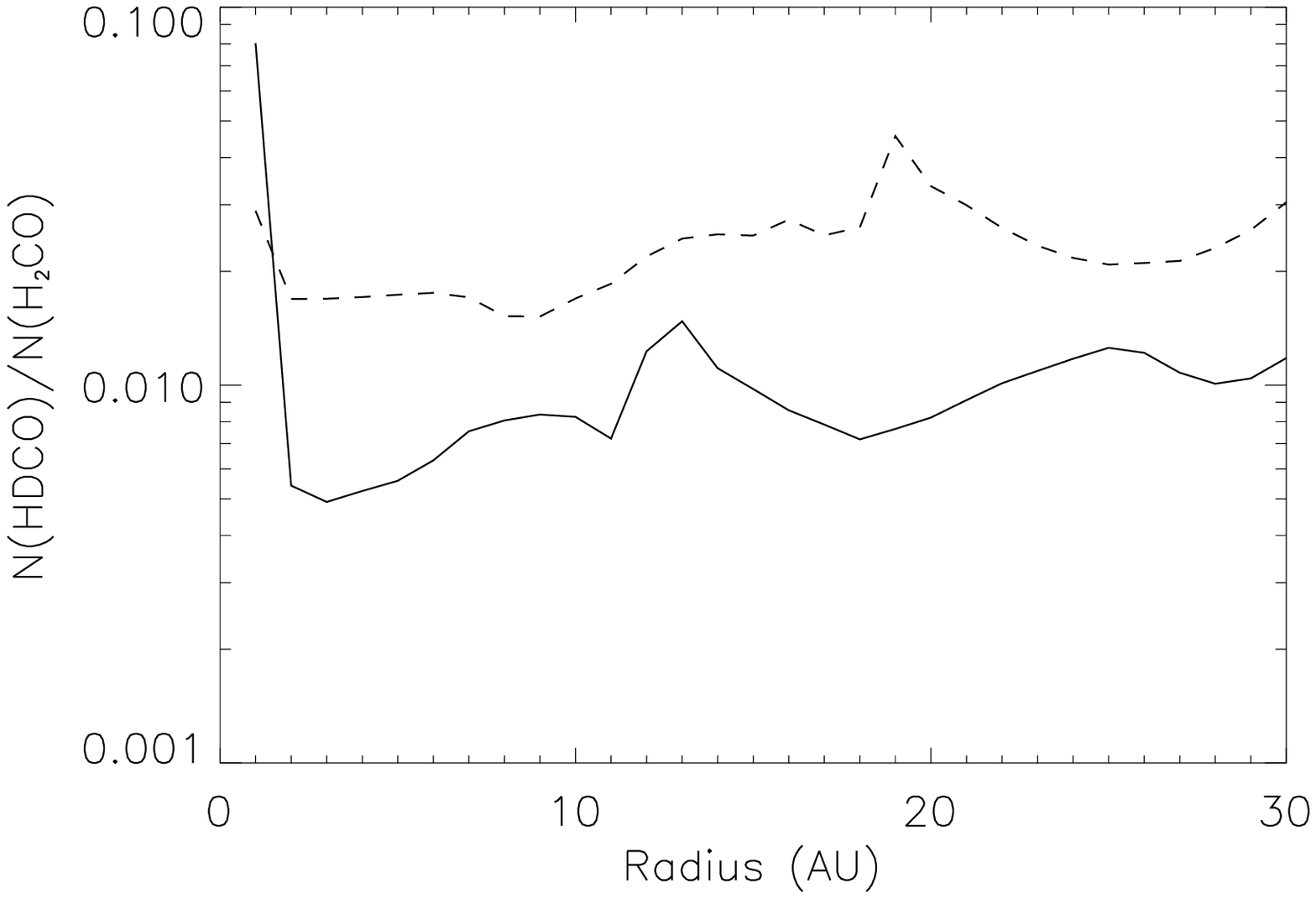}
\end{figure}

\clearpage

\subsection{Comparison with previous inner disk models}

\cite{markwick02} use a similar, but undeuterated, chemical network to
model the chemistry within 10 AU of a protostar.  However there are
major differences between our disk model and theirs 
which result in very different
temperature structures and hence different calculated abundances.
Although our Model 1 and that used by \citeauthor{markwick02}\
have similar grain temperatures in the midplane at 10 AU, their
model has a more rapid increase in midplane temperature as the radius
decreases.  They also find a vertical temperature inversion at all
radii ($<$ 10 AU) so that the surface layers are cooler than the
midplane. The relative surface and midplane temperatures are
controlled in part by the surface density of the disk -- the higher
the surface density, the thicker the disk and the warmer the midplane
is relative to the surface layers.  The \citeauthor{markwick02}\ disk
has a mass accretion rate that is 10\x\ higher than ours, leading to a
higher surface density and hence a colder surface temperature.

The grain temperature is very important since it governs
the location at which molecules can desorb.  
\citeauthor{markwick02}\
find that at 10 AU almost all the molecules are 
accreted onto grains in the surface layer.  Because of 
our higher grain temperature in this region we
find most molecules are in the gas phase, with only
water and its isotopologues remaining as ice.  We also
find that grain mantles persist at smaller radii compared
to Markwick, again because of the difference in temperature
distribution assumed. 

More recently the chemistry in the
disk within 3 AU of the star has been studied by \cite{agundez08}.
These authors concentrated on the warm, UV irradiated, surface
regions and found that photodissociation of CO provided a pool of
carbon atoms that leads to the formation of simple organics such
as \chem{C_2H_2}, HCN and CH$_4$.  The formation of \chem{C_2H_2}
depends on the reaction of H$_2$ with either C$_2$ or C$_2$H --
reactions that are not generally included in astrochemical models
since they have activation barriers of $\sim$ 1400 K.  However
in the warm inner disk the gas temperatures are high enough for them
to make a significant contribution to the carbon chemistry.  These authors
find good agreement with observations and their results (apart
from C$_2$H$_2$) are in broad agreement with those presented here.

Deuterium chemistry has previously been studied in the inner disk
by \cite{ah99} who follow the evolution of material into a radius of
30 AU.  Their model differs from ours in several ways.  They include a
description of the effects of collapse on the input abundances.
Although they include gas--grain interactions (freezeout and thermal
desorption) they do not consider the effects of reactions on the grain
surface (except for the formation of H$_2$ and for recombination of
ions and electrons).  This leads to a number of important differences
between their conclusions and ours.  \citeauthor{ah99} find that gas
phase chemistry is important for determining the deuteration of some
molecules.  Most notably this includes the DCN/HCN ratio observed in
comets.  They find that DCN and HCN are formed by CHD + N
$\longrightarrow$ DCN + C and CH$_2$ + N $\longrightarrow$ HCN + C
respectively.  The derived ratio can be affected by both the cosmic
ray ionization rate (lowering $\zeta_{CR}$ results in a lower D/H
ratio) and by whether or not HCO$^+$ recombines dissociatively with
electrons on grain surfaces.  In contrast we find that to a large
extent the DCN/HCN ratio calculated in ices at the comet formation
radius in our disk is driven by what happened in the molecular cloud,
where grain surface chemistry plays an important role in determining
the deuteration.

We know from observations that the amount of water ice seen in
molecular clouds can only be produced by grain chemistry.  This alone
demonstrates the importance of considering grain chemistry in models,
but the mechanism for including it is still a matter for debate.
Another issue that remains unresolved is the nature of non--thermal
desorption processes that prevent the complete removal of molecules
from the gas in regions where thermal desorption is inefficient.
Other observations of star forming regions \citep{parise02,loinard02,bacmann03}
also indicate that grain chemistry is important in the formation
of molecules such as deuterated formaldehyde and methanol and hence
cannot be ignored when modeling deuteration effects.  Equally our model
falls short in not considering the collapse phase and the effects
on the chemistry of the incorporation
of cloud material into the protostellar disk.  
Neither our model, nor that of \cite{ah99} provides a complete picture
of the chemical history of disks.

\clearpage

\section{Comparison to observations}
\subsection{Cometary ices}

The comets we see today are the remains of icy planetesimals that
formed early in the history of the solar system.  Since their
formation they have been kept mainly at cold temperatures
which restrict further chemical evolution.  Consequently their
chemical make up reflects to first order, the composition
of the protostellar disk at their time of formation. 
Simple species such as H$_2$O, CO, CO$_2$, HCN, CH$_3$OH, H$_2$CO
have been observed in the comae of many comets, but 
D/H ratios have been measured in only four comets --
Halley (HDO/H$_2$O), Hyakutake (HDO/H$_2$O), Hale--Bopp (DCN/HCN,
HDO/H$_2$O) and C/2002 T7 (LINEAR).  
These are all long--period comets, and so formed
inside of the Trans-Neptunian region, after which their
orbits were perturbed by interaction with the giant planets and
they were ejected outwards towards the Oort cloud.  The chemical 
composition of the comets should reflect the temperature,
density and ionization state of the region in which they formed.
Measurements of the D/H ratios of cometary ices could also be used
to constrain the degree of processing of the interstellar ices
incorporated into the protosolar disk, and place limits on 
the amount of radial mixing experienced by material in the disk.

The observed molecular deuteration is similar for all four comets.
In Hale--Bopp, DCN/HCN = 2.3 \x 10$^{-3}$ \citep{meier98a}.
(D/H)$_{\hbox{H$_2$O}}$ is measured as 3.3 \x 10$^{-4}$ in Hale--Bopp \citep{meier98a},
2.9 \x 10$^{-4}$ in Hyakutake \citep{bm98} and  $\sim$ 3.1 \x 10$^{-4}$ 
\citep{eberhardt95,balsiger95} in Comet/P Halley.  In
C/2002 T7 (LINEAR)  (D/H)$_{\hbox{H$_2$O}}$ is 2.5 \x 10$^{-4}$ 
\citep{hut08}. 
(These measurements
correspond to x(HDO)/x(H$_2$O) ratios of 6.6 \x 10$^{-4}$, 5.8 \x 10$^{-4}$,
6.2 \x 10$^{-4}$ and 5.0 \x 10$^{-4}$ for Hale-Bopp, Hyakutake, 
Halley and C/2002 T7 (LINEAR)
respectively.)

\begin{figure}
\caption{\label{fig:gdcn}The deuteration of HCN and H$_2$O ices in
Model 2.}
\plottwo{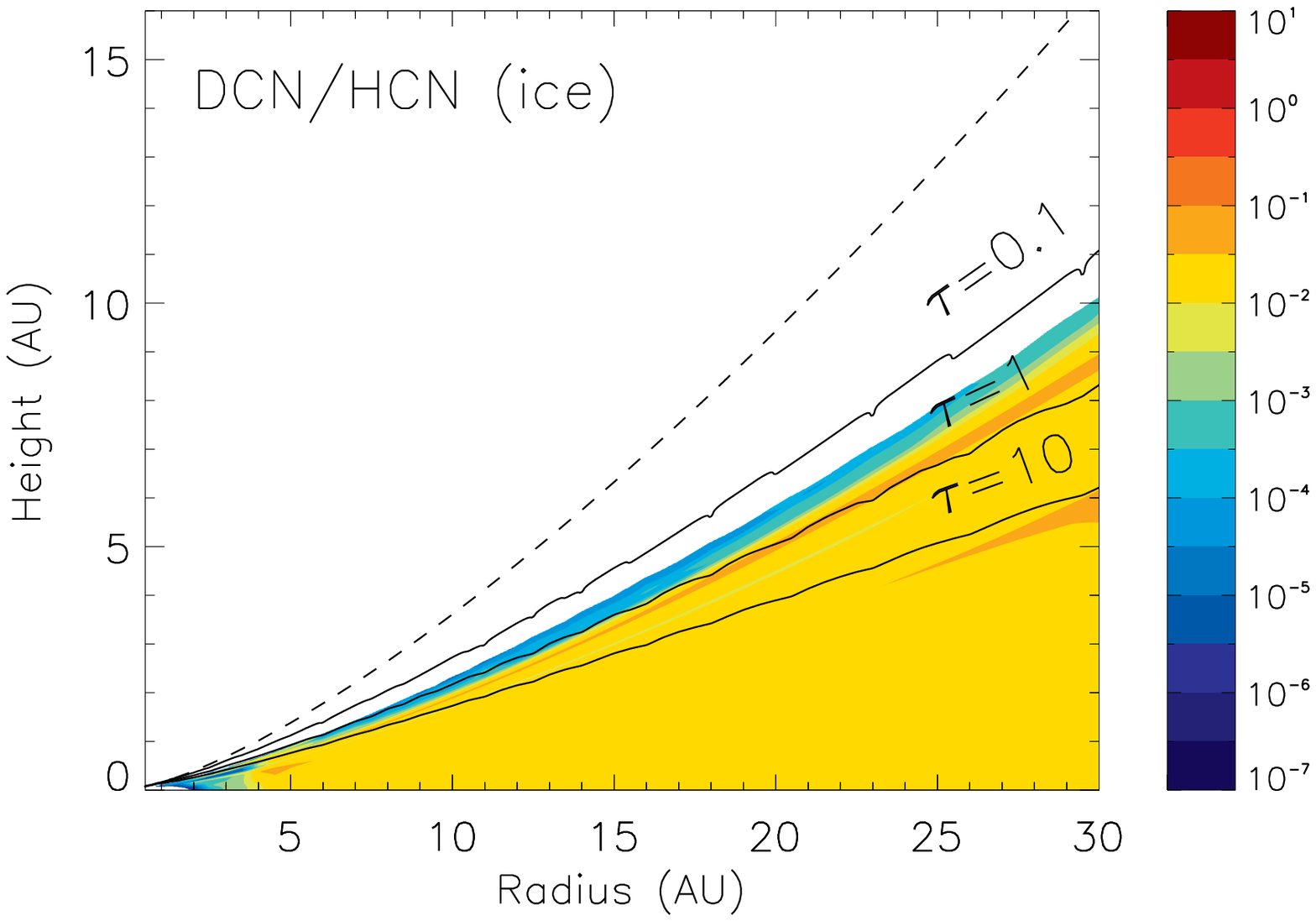}{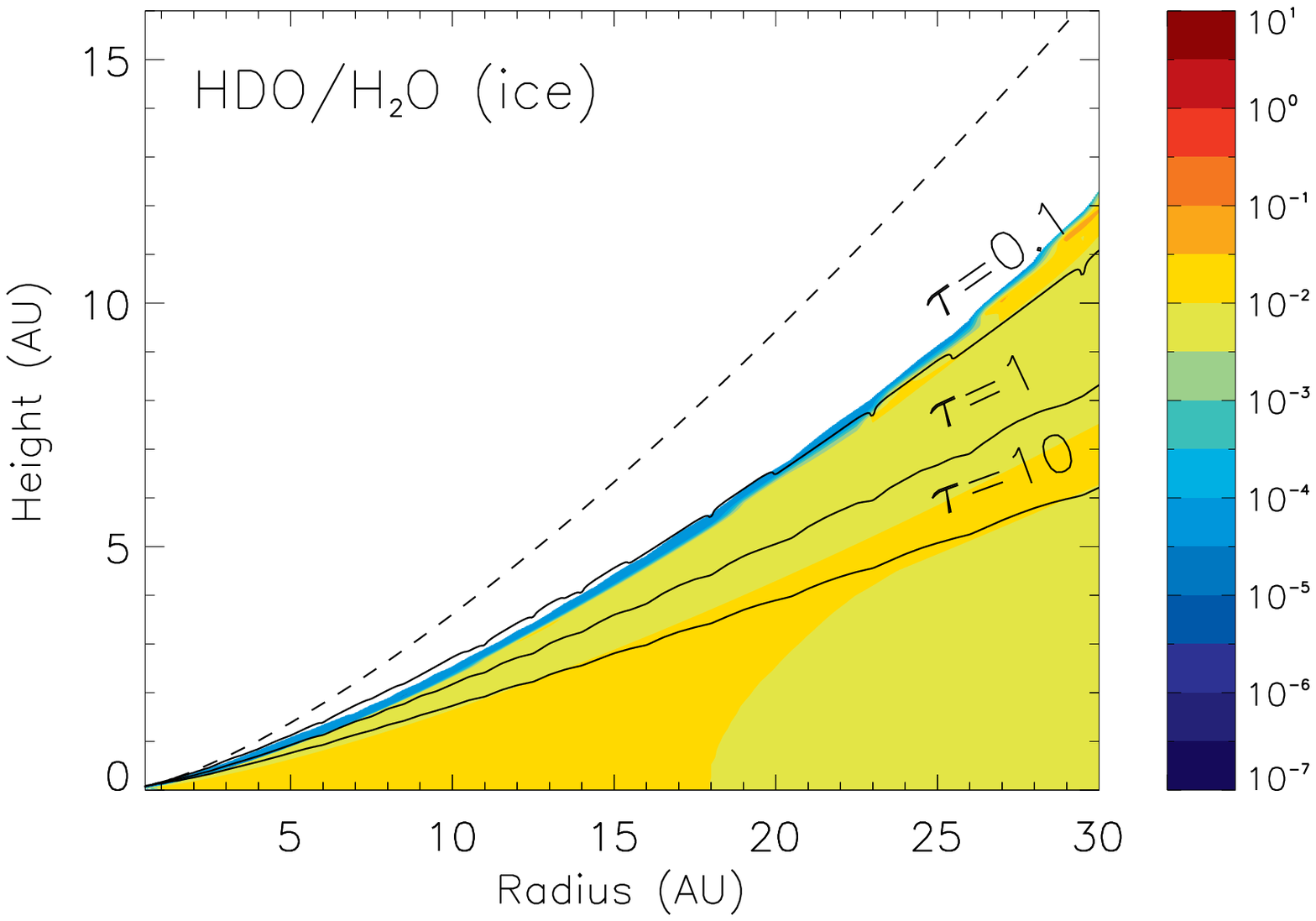}
\end{figure}

Figure~\ref{fig:gdcn} shows the variation in the x(DCN)/x(HCN)
and x(HDO)/x(H$_2$O) ratios in the ice mantles over the radius range
from 0.5 to 10 AU.  Our model shows very little variation in the
ratios in the midplane in this region.  The x(HDO)/x(H$_2$O) ratio is
set in the molecular cloud phase and very little processing of the
water ice occurs in the disk.  Hence in the midplane, the
x(HDO)/x(H$_2$O) ratio in ices has a constant value of 0.022 from 30
to 2 AU.  DCN/HCN shows some slight variation but the value is still
roughly 0.018 from 5 - 10 AU.  For both molecules the model
deuteration ratios are much higher than observed.  The model DCN/HCN
is approximately 8 times higher than observed, and the HDO/H$_2$O
ratio is 33 -- 44 times higher than observed.  Hence, although comets
have to contain ices accumulated at low temperatures they must also
include some ices that were formed in a higher temperature phase.

There are several possible explanations for the discrepancy between
our model and the observations.  Since the disk chemistry
does not change the HDO/H$_2$O ratio from the value set in the 
molecular cloud it may be that the problem lies in our
molecular cloud model, and that HDO ice is formed too efficiently.
In our cloud model the ratio of HDO/H$_2$O in the ice is 0.9 \%, compared
to an upper limit of $\sim$ 0.5 -- 2 \% towards low mass
protostars \citep{parise03}. This upper limit is much less
than the deuteration ratios seen in methanol and formaldehyde
which are also expected to form on grain surfaces.
If all three molecules form at
the same time then their D/H ratios should be similar and 
should reflect the gas phase atomic D/H ratio at the time of
formation.  Observations of methanol and formaldehyde in low
mass star formation regions find much higher D/H ratios that
the upper limit derived for HDO/H$_2$O ice, suggesting that there
are still unresolved issues in our understanding of the formation
of these molecules.

Another possibility is that the ices are removed from the
grains during infall through the accretion shock at the surface
of the disk.
Models suggest that most of the water ice on infalling
grains can be removed within
30 AU of a solar type star \citep[e.g.][]{lunine91,nh94}.
Recent Spitzer observations  \citep{watson07} have
found water vapor emission in a Class 0 object (NGC 1333-IRAS 4B), 
which has
been attributed to emission from a surface accretion shock.
Water would therefore enter the disk as vapor rather than as ice.
This could affect the D/H ratio
of the ices that reform in the disk, since water
formed in the warm post-shock region
would have a lower D/H ratio than water formed in the cold
molecular cloud and
this ratio would be transferred to the ices when the
molecules were redeposited.  However it is unclear how
widespread or long--lived this phenomenon is since
water emission was detected in only one object out of a sample of 30.

Although vertically averaged hydrodynamical disk models
find that material is advected radially inwards e.g.\
\cite{pringle81, lp85},with outward flow occurring only 
in the cold outer regions of the disk, more complex models
have shown that large scale meridional flow patterns develop
in disks e.g.\ \cite{urpin84, kl92, roz94, rg02,tg07}.  In 
these models material flows inwards at the surface and outwards
in the midplane, leading to a mixing of
warm and cold material.  Observational evidence of the
importance of mixing in disks comes from comets.  Silicates
are thought to have undergone significant high temperature
processing, which converted amorphous silicates into
the crystalline form, before they were incorporated into comet nuclei
\citep{hanner99}.  One way to do this would be by
radial mixing where the silicates were heated to high
temperatures in the inner disk before being mixed outwards
to the comet formation region.  Additional evidence of mixing
in the history of comets has been seen in the STARDUST samples
\citep{brownlee06,zolensky06}.  If grains can be
mixed then gases could be too, and this would
have the effect of changing the deuteration ratios by
bringing together molecules formed in the cold outer disk
or in the molecular cloud
with those formed in warmer regions of the disk.
Overall, mixing could be expected to reduce the D/H ratios
seen in cometary ices.

Diffusion has been included in models of the deuterium 
chemistry in the midplane of disks by \cite{drouart99} and 
\cite{mousis00}.  \cite{drouart99} assume that the
water is initially gaseous with (D/H)$_{\hbox{H${_2}$O}}$ = (D/H)$_{\hbox{H}_2}$
and they follow the evolution of (D/H)$_{\hbox{H${_2}$O}}$ over time in 
a model that includes turbulent diffusion.  \cite{mousis00}
take a similar approach and modeled the DCN/HCN ratio.  These authors
concluded that cometary ices condense in the nebula, rather than
being retained during the disk and star formation, and are composed
of material that is partially reprocessed in the disk. \cite{horner07}
also found a relationship between the D/H ratios in ices and the distance
from the Sun at which they condense.

Our model is therefore probably too simplistic in its assumptions and
further work is required to study how the accretion shock alters
the D/H ratios and the survivability of the ices in the star
formation process, and to consider the effects of mixing.
\cite{ilgner04,semenov06} and \cite{wlab06} have all shown that
mixing in the vertical direction 
can greatly affect the abundances in the disk, but
isotopes have not yet been studied with these models.  Also,
as shown by \cite{drouart99,mousis00} and \cite{horner07}, radial
mixing alters D/H ratios in ices, bringing cold material into the
inner disk where it can desorb and returning material to the colder
regions where it can reaccrete.  Both radial and vertical diffusion
alter the molecular D/H ratios by combining material formed under
different temperature regimes.

\subsection{Gas phase observations of extrasolar disks}
Our basic model (excluding isotopes) is the same as
used in \cite{ww09} who considered the chemistry of carbon isotopes.  
That paper includes a
comparison of the result to recent infrared observations of
the inner regions of protostellar disks.  Here we merely
summarize our findings for completeness, but the reader
is referred to \cite{ww09} for more details.  

Several molecules e.g. CO, \chem{C_2H_2}, HCN, OH, H$_2$O and
CO$_2$ have been observed in AA Tau \citep{cn08}, 
DR Tau and AS 205 A \citep{salyk08}, GV Tau \citep{gibb04,
gibb07} and IRS 46 \citep{lahuis06}.  The derived rotational temperatures 
suggest that 
the molecular emission comes from a region above the midplane.
In our models, these temperatures are reached in the region
where thermal desorption is efficient, and the gas and grain
temperatures are decoupled.  
We find that in Model 1, CO, H$_2$O and OH have peak abundances in
regions of the disk with 
similar temperatures to the observed rotation temperatures
for these molecules.  The
surface peak of CO$_2$ also coincides with the observed rotation
temperature, although the peak abundance of this molecule is closer to
the midplane.  For C$_2$H$_2$ the observations lie above the modeled
distribution, and for HCN they are at the upper edge of the model. 

\begin{deluxetable}{lrrrrrr}
\tablecolumns{7}
\tablewidth{0pt}
\tablecaption{\label{tab:obs}The observed column densities in the inner regions
of protostellar disks.  The values for Model 1 are calculated at 1AU for the surface
layers of the disk where T$_{gas}$ $>$ 150 K.}
\tablehead{
\colhead{Molecule} & \multicolumn{6}{c}{Column Density (10$^{16}$ cm$^{-2}$)}\\
                   & \colhead{GV Tau$^1$} & \colhead{IRS46$^2$} & \colhead{AA Tau$^3$} & \colhead{AS 205A$^4$} & \colhead{DR Tau$^4$} & \colhead{Model 1}\\
}
\startdata
H$_2$O     & \nodata & \nodata & 65.0 & 60.0     & 80.0    & 1280.0    \\
OH         & \nodata & \nodata & 8.1  & 20.0     & 20.0    & 23.0      \\
HCN        & 3.7     & 5.0     & 6.5  & \nodata  & \nodata & 4.4       \\
C$_2$H$_2$ & 7.3     & 3.0     & 0.8  & \nodata  & \nodata & 9.0 (-8)  \\
CO$_2$     & \nodata & 10.0    & 0.2  & \nodata  & \nodata & 13.0      \\
CO         & 590.0   & 200.0   & 49.0 & 60.0     & 70.0    & 1600.0    \\
\enddata
\tablerefs{(1) \cite{gibb07}, (2) \cite{lahuis06}, (3) \cite{cn08}
(4) \cite{salyk08}}
\end{deluxetable}

We can also compare our results with the column densities observed
(Table~\ref{tab:obs}).  
The calculated column densities at 1 AU for several molecules
are given in Table~\ref{tab:cd} and are much higher than observed, but
these values are calculated from the disk midplane
to its surface and therefore include material that is much
colder than observed.
If we instead calculate the column densities only in the upper regions of the 
disk where the gas temperature is warmer than 150 K, 
the agreement is better.  The calculated column densities are
N(HCN) = 4.4 \x 10$^{16}$ cm$^{-2}$, N(CO) = 1.6 \x 10$^{19}$ cm$^{-2}$, 
and N(CO$_2$) = 1.3 \x 10$^{17}$ cm$^{-2}$ for Model 1.  N(C$_2$H$_2$) is
far too low (9 \x 10$^8$ cm$^{-2}$).  This low abundance
may be due to the
exclusion of reactions involving H$_2$ with C$_2$ or C$_2$H
that were found to be important in the formation of C$_2$H$_2$ in disks
by \cite{agundez08} but are not included in our reaction scheme.  The
calculated N(H$_2$O) is also too high compared to observations, but
N(OH) gives good agreement with the abundances measured in DR Tau
and AS 205A \cite{salyk08}.

A further test of the models is to compare the CH$_4$/CO ratio.
\cite{gibb07} found an upper limit of 0.0035, similar to the
upper limit \cite{gibb04} found for this ratio in HL Tau ($<$ 0.005).
The calculated ratio at 1 AU of 0.006 is consistent with 
both these observational measurements.

\cite{qi08} observed deuterated molecules in TW Hya and found a 
gradient in the DCO$^+$/HCO$^+$ ratio.  The ratio increases from 
0.01 at 30 AU to 0.1 at $\sim$ 70 AU.  There is a steep fall--off in
N(DCO$^+$) at $R$ $>$ 90 AU.  Our current model only extends out to 
35 AU but when we consider it together with 
our outer disk model \citep{willacy07}
we can see if the same trend shows up in the chemical modeling data.
Table~\ref{tab:dco+} shows the model column densities for DCO$^+$ and
HCO$^+$, as well as their ratio for radii between 30 and 150 AU. Although
we do not find the same steep fall off in N(DCO$^+$) at $R$ $>$ 90 AU, 
we do see a similar increase in DCO$^+$/HCO$^+$ from 30 -- 100 AU followed
by a fall at larger radii.  However our D/H ratio at 100 AU is $\sim$ 
10 \x larger than inferred from the observational data.

\begin{deluxetable}{lcccc}
\tablecolumns{5}
\tablewidth{0pt}
\tablecaption{\label{tab:dco+}The calculated values of N(DCO$^+$), 
N(HCO$^+$) and DCO$^+$/HCO$^+$.  The data for $R$ $>$ 30 AU
is taken from \cite{willacy07}.}
\tablehead{
 & \colhead{30 AU} & \colhead{50 AU} & \colhead{100 AU} & \colhead{150 AU}\\
}
\startdata
N(DCO$^+$) & 1.9 (12) & 1.5 (12) & 6.6 (12) & 2.3 (12) \\
N(HCO$^+$) & 3.3 (13) & 1.2 (13) & 3.9 (12) & 2.8 (12) \\
DCO$^+$/HCO$^+$ & 0.05 & 0.125 & 1.7 & 0.8\\
\enddata
\end{deluxetable}

\clearpage
\section{Conclusions}

We have presented the results of calculations of the chemical abundances in the
inner regions of protostellar disks including deuterium chemistry.
We find that for many molecules the deuteration is set by cold
temperature chemistry in the parent molecular cloud, and that the
chemistry in the disk itself has little effect.  In particular the
calculated D/H ratios in ices in the region of cometary formation are
found to be high relative to the (limited) cometary observations.
This suggests that our model is not complete, and additional
processing of ices is required after the molecular cloud stage,
beyond the chemistry included in our models here.  
Our carbon isotope modeling leads to a similar conclusion \citep{ww09}.
The processing could be
achieved in the accretion shock, and/or by mixing in the disk, both of
which could lead to lower D/H ratios.  Additional work is
required to fully understand the process by which interstellar grains
and ices are incorporated into planets and other planetary bodies.

When a comparison is made with the observations of external disks
\citep{cn08,gibb07,gibb04} good agreement is found between the
location of the molecules in our disks and that derived from the
observations.  The molecular emission comes from a region above the
midplane, where thermal desorption is efficient.

We find major differences in the results of our calculations and those
of other models, mainly as a consequence of either the physical disk
model assumed, or because of the chemical processes included.
We find that the disk chemistry has less effect on D/H
ratios than \cite{ah99} because we allow chemistry to occur on the
surfaces of dust grains and hence gas phase chemistry is less
important in determining deuteration.  The temperature structure in
our models is very different to that assumed by \cite{markwick02}, so
that, despite using similar chemical networks, we get very different
abundance distributions in the inner 10 AU of the disk.  This
highlights the effect that disk structure can have on the molecular
distributions in disks.  As observations improve in resolution, for
example with the advent of ALMA this may enable the differences
between models to be used to distinguish among them observationally.

\acknowledgements This research was conducted at the Jet Propulsion
Laboratory, California Institute of Technology under contract with the
National Aeronautics and Space Administration.  Partial support was
provided to KW by a grant from the NASA TPF Foundation Science
program. A portion of this research was carried out while
PMW was supported by an appointment to the NASA
Postdoctoral Program at JPL, administered by ORAU through a contract
with NASA.  We also wish to 
acknowledge the helpful comments of Dr.\ Bockel{\'e}e-Morvan 
concerning the measurements of the D/H ratio in cometary water.

\bibliography{/home/willacy/papers/disk_turb/disk}

\end{document}